%% file: main.tex
\documentclass[aps,prd,onecolumn,notitlepage,superscriptaddress,showpacs,floatfix,preprintnumbers,nofootinbib]{revtex4-1}
\usepackage[utf8]{inputenc}
\usepackage{mathrsfs}
\usepackage{amssymb}	
\usepackage{amsmath}
\usepackage{graphicx}
\usepackage{subdepth}
\usepackage{physics}
\usepackage{slashed}
\usepackage{xcolor}

\definecolor{lcolor}{rgb}{0.5,0,0}
\definecolor{citcolor}{rgb}{0,0.3,0.0}

\usepackage[caption=false]{subfig}

\newcounter{diag}
\newcounter{subdiag}[diag]
\newcommand{\newdiag}[1]{\refstepcounter{diag} \label{#1}}
\newcommand{\namediag}[1]{\refstepcounter{diag} \thediag \label{#1}}

\renewcommand{\thediag}{(\alph{diag})}

\usepackage[breaklinks,colorlinks,urlcolor=blue,citecolor=citcolor,linkcolor=lcolor]{hyperref}
\usepackage{subdepth}
\usepackage{mathtools}

\usepackage{tikz}
\usepackage[capitalise]{cleveref}

\input{defs.tex}

\begin{document}

\title{Diffractive deep inelastic scattering at NLO in the dipole picture}

\preprint{HIP-2024-XXX/TH}
\author{Guillaume Beuf}
\email{guillaume.beuf@ncbj.gov.pl}
\affiliation{ Theoretical Physics Division, National Centre for Nuclear Research, Pasteura 7, Warsaw 02-093, Poland}
\author{Tuomas Lappi}
\email{tuomas.v.v.lappi@jyu.fi}
\affiliation{ Department of Physics, University of Jyv\"askyl\"a,
P.O. Box 35, 40014 University of Jyv\"askyl\"a, Finland}
\affiliation{ Helsinki Institute of Physics, P.O. Box 64, 00014 University of Helsinki, Finland}

\author{Heikki Mäntysaari}
\email{heikki.mantysaari@jyu.fi  }
\affiliation{ Department of Physics, University of Jyv\"askyl\"a,
P.O. Box 35, 40014 University of Jyv\"askyl\"a, Finland}
\affiliation{ Helsinki Institute of Physics, P.O. Box 64, 00014 University of Helsinki, Finland}

\author{Risto Paatelainen}
\email{risto.paatelainen@helsinki.fi}
\affiliation{ Helsinki Institute of Physics, P.O. Box 64, 00014 University of Helsinki, Finland}

\author{Jani Penttala}
\email{janipenttala@physics.ucla.edu}
\affiliation{
Department of Physics and Astronomy, University of California, Los Angeles, CA 90095, USA
}
\affiliation{Mani L. Bhaumik Institute for Theoretical Physics, University of California, Los Angeles, California 90095, USA}
\affiliation{ Department of Physics, University of Jyv\"askyl\"a,
P.O. Box 35, 40014 University of Jyv\"askyl\"a, Finland}
\affiliation{ Helsinki Institute of Physics, P.O. Box 64, 00014 University of Helsinki, Finland}

\begin{abstract}
We compute the transverse and longitudinal  diffractive structure functions to full next-to-leading order accuracy in the dipole picture of deep inelastic scattering. Our calculation uses the standard light-cone perturbation theory method for the partonic content of the virtual photon, together with the Color Glass Condensate description of the target color field.  Our result includes as a subset the $q\bar{q}g$ contribution calculated earlier. We show that there is  a rapidity divergence that can be factorized into the BK/JIMWLK evolution of the target Wilson lines, and that all other divergences cancel. 
\end{abstract}

\maketitle 

\newpage
\tableofcontents

\section{Introduction}

At high collision energies QCD scattering is believed to be sensitive to gluon saturation where nonlinear effects generate an emergent transverse-momentum scale $\qs$. At high enough energy, the scale $\qs$ should grow until it can be treated as a weak-coupling momentum scale.  The physics of gluon saturation can appear in many forms, depending on the gauge and the Lorentz frame used in the calculation. For processes where a dilute probe collides with a dense, saturated target, the most convenient form for many calculations is the dipole picture of QCD scattering~\cite{Nikolaev:1990ja,Nikolaev:1991et,Mueller:1993rr,Mueller:1994jq,Mueller:1994gb}. In this picture one factorizes the scattering process into, on the one hand, the vacuum evolution of the probe into a  partonic state and, on the other hand, the scattering of this partonic state off the target.  The natural theoretical framework for describing such scattering is provided by light-cone perturbation theory (LCPT)~\cite{Bjorken:1970ah} coupled with a Color Glass Condensate (CGC) approach~\cite{Gelis:2010nm} classical color field description of the target. This is the method we also use in this paper. 
At the lowest order in QCD perturbation theory, the relevant partonic state for the virtual photon probe in the deep inelastic scattering (DIS) process is a color-neutral quark-antiquark dipole, hence the term dipole picture. In the dipole picture, the phenomenon of gluon saturation appears simply as the unitarization of the amplitude for a fixed transverse size dipole to scatter off the target. 
One of the major advantages of the dipole picture is that inclusive scattering and diffractive dissociation of the virtual photon are described in terms of this same amplitude.

Both inclusive and diffractive DIS are also experimentally important as probes of the saturation regime of QCD. In particular, the upcoming Electron-Ion Collider (EIC)~\cite{AbdulKhalek:2021gbh} will measure different inclusive and diffractive processes in DIS in the small-$x$ regime, most importantly also for heavier nuclei. Even higher energies, i.e. smaller  values of $x$, could be reached by the LHeC/FCC-he~\cite{LHeC:2020van}. With the prospect of new precise measurements, improved precision is important  also in theoretical calculations. This is due not only to the precision of the  experimental data, but also because higher orders in perturbative calculations in the dipole picture open up  new kinematical regions, and match more smoothly to collinear factorization and DGLAP evolution for high-$Q^2$ physics. Appropriately, there have been many recent advances in taking CGC calculations for DIS to NLO accuracy, such as   for the  high energy BK/JIMWLK evolution~\cite{Balitsky:1995ub,Kovchegov:1999ua,Kovchegov:1999yj,Balitsky:2008zza,Balitsky:2013fea,Kovner:2013ona,Balitsky:2014mca,Beuf:2014uia,Lappi:2015fma,Iancu:2015vea,Iancu:2015joa,Albacete:2015xza,Lappi:2016fmu,Lublinsky:2016meo} and for inclusive DIS cross sections~\cite{Balitsky:2010ze,Balitsky:2012bs,Beuf:2011xd,Beuf:2016wdz,Beuf:2017bpd,Hanninen:2017ddy,Beuf:2020dxl,Lappi:2020srm,Caucal:2021ent,Hanninen:2022gje,Iancu:2022gpw,Taels:2022tza,Beuf:2021qqa,Beuf:2021srj,Beuf:2022ndu,Bergabo:2022tcu,Caucal:2023nci,Caucal:2024cdq,Bergabo:2024ivx}. 

Diffractive DIS is expected to be even more sensitive to gluon saturation than inclusive cross sections~\cite{Golec-Biernat:1999qor,Kowalski:2007rw,Armesto:2019gxy,AbdulKhalek:2021gbh}. There are several recent calculations of diffractive dijet production in high-energy DIS \cite{Boussarie:2014lxa,Boussarie:2016ogo,Iancu:2021rup,Iancu:2022lcw,Hatta:2022lzj,Fucilla:2022wcg}. While these calculations are in many ways close to our present approach, experimentally the necessity to cleanly identify  the two jets tends to push one to higher $x$ and, perhaps more importantly, transverse length scales $\ll 1/\qs$ that are less sensitive to saturation. Indeed, there is a limit in which the NLO dipole picture calculation reduces to collinear factorization in terms of diffractive parton distributions~\cite{Iancu:2021rup,Hatta:2022lzj}. 

Consequently, we will here focus on inclusive diffraction and calculate total diffractive DIS cross sections, conventionally expressed in terms of diffractive structure functions. Our aim in this paper is to provide a full NLO calculation in the dipole picture. A self-contained subset of the contributions considered here (namely the ones with  a $q\bar{q}g$ Fock state crossing both the target gluon field and the cut) were considered earlier in Ref.~\cite{Beuf:2022kyp}. That paper only outlined the missing, calculationally more challenging, contributions but did not calculate them. We will do so here. Ref.~\cite{Beuf:2022kyp} also showed how to reproduce earlier results in the large-$Q^2$ limit (the ``Wüsthoff result''~\cite{Wusthoff:1997fz,Golec-Biernat:1999qor,GolecBiernat:2001gyl}) and in the large-$M_X^2$ limit~\cite{Bartels:1999tn,Kopeliovich:1999am,Kovchegov:2001ni,Munier:2003zb,Golec-Biernat:2005prq} that had already been derived in the literature (see also Refs.~\cite{Kovchegov:1999ji,Kovchegov:2011aa} for a resummation of large-$M_X^2$ logarithms in terms of the Kovchegov--Levin equation and Ref.~\cite{Lappi:2023frf} for a corresponding application to phenomenology). Since the result of Ref.~\cite{Beuf:2022kyp} is fully contained in our present calculation, we will not dwell more on the relation to these earlier results in approximate kinematics in this paper. We believe that together with the total cross section for massless and massive quarks, measurements of diffractive structure functions will provide strong constraints on the structure of the proton and the nucleus in the saturation regime. 

This paper is structured in the following way.  To be concise we will start by presenting our final result, the diffractive structure function at NLO accuracy in the dipole picture for high energy scattering, in Sec.~\ref{sec:final_results}. We then move on to a more detailed exposition of the formalism we use, namely light-cone quantization of the incoming and outgoing states, in Sec.~\ref{sec:computational_setup}. Then, in Sec.~\ref{sec:classification}, we classify the kind of contributions that are needed for calculating the diffractive structure function, according to whether the additional gluon at NLO accuracy in perturbation theory crosses the shockwave (i.e. interacts with the target color field), the cut (i.e. is produced in the final state), both, or neither. Section~\ref{sec:details} then presents the detailed calculation of all the diagrams leading to the full result already written down in Sec.~\ref{sec:final_results}. 
An essential feature of the way the calculation is organized here is that, in order to effectively realize cancellations between radiative and virtual contributions, we combine contributions at the cross section level before integrating over all of the momenta. Thus, for the most intricate parts of the calculation we are working at the level of squared light-cone wave functions.  While it is not always the most convenient thing to do for evaluating the diffractive structure function, it is possible to evaluate the coordinate-space light-cone wave functions separately, before squaring them. This calculation is performed in Appendix~\ref{sec:wave_functions} for future reference. 

\section{Diffractive structure functions: the complete NLO result}
\label{sec:final_results}

In this work we calculate high-energy diffractive cross section for the virtual photon-nucleus scattering at next-to-leading order accuracy. We begin by showing the final result which is itself finite and suitable for numerical evaluations. Details of the NLO calculation are shown in the subsequent sections. The calculated diffractive virtual photon-nucleus cross section $\sigma^{\text{D}}_{\gamma^*_{\lambda} + A }$ is related to the experimentally measured diffractive structure function as 
\begin{equation}
    \xpom F_{\lambda}^{\mathrm{D}(4)}(\xpom, Q^2, M_X^2, t) = \frac{Q^2}{4\pi^2 \aem} \frac{Q^2}{\beta}  \frac{\dd[]{ \sigma^{\text{D}}_{\gamma^*_{\lambda} + A } }}{\dd{t} \dd{M_X^2}}.
\end{equation}
The transverse structure function is defined as the average over the two transverse polarization states
\begin{equation}
   \xpom F_{T}^{\mathrm{D}(4)}(\xpom, Q^2, M_X^2, t) = \half  \sum_{\lambda=\pm}  \xpom F_{\lambda}^{\mathrm{D}(4)}(\xpom, Q^2, M_X^2, t), 
\end{equation}
and the sum of transverse and longitudinal structure functions gives
\begin{equation}
F_{2}^{\mathrm{D}(4)} = F_{T}^{\mathrm{D}(4)} + F_{L}^{\mathrm{D}(4)}
\end{equation}
which is, in the limit of small inelasticity $y$, related to the total diffractive electron-target cross section. 
Here 
\begin{equation}
    \xpom = \frac{M_X^2 + Q^2 - t}{W^2+Q^2-m_N^2}
\end{equation}
is the fraction of the target longitudinal momentum transferred in the process, 
$M_X^2$ is the invariant mass of the diffractively produced system, $Q^2$ is the photon virtuality, $t\approx -\Deltat^2$ is the momentum transfer squared, $\Deltat$ is the transverse momentum transfer, $W$ is the center-of-mass energy of the photon-nucleon system, and $m_N$ is the nucleon mass. The virtual photon polarization is denoted by $\lambda=T,L$.

The dipole picture calculation proceeds by decomposing the incoming virtual photon state in terms of its Fock state components,  the $q\bar{q}$ ``dipole'' and the $q\bar{q}g$ ``tripole'' being the relevant ones at this order. Correspondingly, it is natural to  decompose the diffractive cross section for the virtual photon-nucleus scattering as
\begin{equation}
\label{eq:cross_section_decomposition}
\begin{split}
    \frac{\dd[]{ \sigma^{\text{D}}_{\gamma^*_{\lambda} + A } }}{\dd[2]{\Deltat} \dd{M_X^2}} 
    = \biggl [ \frac{\dd[]{ \sigma^{\text{D}}_{\gamma^*_{\lambda} + A } }}{\dd[2]{\Deltat} \dd{M_X^2}}  \biggr ]_{\text{dip}} 
    + \biggl [ \frac{\dd[]{ \sigma^{\text{D}}_{\gamma^*_{\lambda} + A} }}{\dd[2]{\Deltat} \dd{M_X^2}}  \biggr ]_{\text{trip}} 
    + \biggl [ \frac{\dd[]{ \sigma^{\text{D}}_{\gamma^*_{\lambda} + A } }}{\dd[2]{\Deltat} \dd{M_X^2}}  \biggr ]_{\text{dip-trip}}.
\end{split}
\end{equation}
The individual parts here are as follows:
\begin{itemize}
    \item The ``dip'' term contains the leading-order contribution where only the $|q\bar q\rangle$ Fock state of the virtual photon interacts with the target, including vertex corrections to the $\gamma^*\to q\bar q$ splitting, plus the contributions where a $q\bar q$ system crosses the shockwave and a gluon is exchanged or emitted after the interaction with the target. 
    \item The ``trip'' term contains the contributions where the $q\bar q g$ system interacts with the target in both the amplitude and the conjugate amplitude.  This part of the cross section was already evaluated in Ref.~\cite{Beuf:2022kyp}, but we will repeat the result here for completeness.
    \item The ``dip-trip'' is the interference term: %
    a $q\bar q g$ system interacts with the target in the amplitude and a $q\bar q$ system interacts with the target in the conjugate amplitude, or vice versa. Both real or virtual corrections contribute to the ``dip-trip'' term, whether the gluon crosses the final state cut or not.
\end{itemize}
The contributions with a $|q\bar q\rangle$ interacting with the target in the amplitude, and a $|q\bar q\rangle$ or $|q\bar qg\rangle$ state in the conjugate amplitude, are separately  divergent. However, the ultraviolet and infrared divergences can be canceled between the different contributions. In the  results quoted below these divergences have been canceled by adding and subtracting a subtraction term, so that the three parts of the cross section as presented here are all finite separately. Due to this UV subtraction procedure, discussed in detail in Sec.~\ref{sec:details}, the division of the full NLO cross section to the three parts in Eq.~\eqref{eq:cross_section_decomposition} is not unique but depends on the chosen UV subtraction scheme, similarly as in the case for the inclusive DIS cross section~\cite{Beuf:2016wdz,Beuf:2017bpd,Hanninen:2017ddy}.

In the dipole picture, the interaction with the target is described by Wilson line correlators. 
At leading order, the only correlator that appears is the ``dipole operator''
\begin{equation}
    \dipole_{01} = \frac{1}{\nc} \Tr{V(\xt_0)V^\dagger(\xt_1)},
\end{equation}
where $V(\xt_0)$ is the Wilson line in the fundamental representation describing an eikonal propagation of a quark in the target color field. For the $q\bar{q}g$ state one needs the ``tripole'' correlator, which can be expressed in terms of a product of dipole operators. 
Diffractive cross sections are divided into two categories. Coherent diffraction refers to the events in which the target proton or nucleus remains intact. In the Good--Walker picture~\cite{Good:1960ba} the coherent diffractive cross section is obtained by averaging over the target configurations at the amplitude level, replacing $\dipole_{01}$ by $\left\langle \dipole_{01} \right\rangle$. Incoherent diffraction, on the other hand, refers to the process where the target breaks up, and it can be obtained by calculating the total diffractive cross section and subtracting the coherent contribution. The total diffractive cross section is itself obtained by averaging over the target configurations at the cross section level, in which case the cross section depends both on 2- and higher-point Wilson line correlators such as $\left\langle \dipole_{01} \dipole_{\ov 0 \ov 1}{}^\dag\right\rangle$. The results of this work can be applied to both coherent and total diffraction simply by choosing at which level one averages over the target configurations. 

 The center-of-mass energy dependence of the Wilson lines is given by the JIMWLK evolution equation~\cite{Mueller:2001uk,Balitsky:2013fea,Kovner:2013ona}, from which it is possible to obtain in the large-$\nc$ limit the Balitsky--Kovchegov equation~\cite{Kovchegov:1999yj,Balitsky:1995ub,Balitsky:2008zza} describing the energy dependence of the dipole amplitude $\left\langle \dipole_{01} \right\rangle$.
 The JIMWLK evolution is a resummation of 
 high-energy logarithms, which in our picture correspond to gluons with a small $z_2\to 0 $ fraction of the longitudinal momentum of the probe. Accordingly, the contributions with a $q\bar{q}g$ state with a %
low $z_2$ gluon crossing the target shockwave in the amplitude (conjugate amplitude) will exhibit a large logarithm  arising from this region. This contribution needs to be subtracted from these ``tripole'' terms and absorbed into a renormalization of the dipole operators in the ``dip'' term of the cross section. We will here present our results without this subtraction being done since there are several possible ways to implement it. This subtraction can be done in the same way as for the inclusive cross section~\cite{Ducloue:2017ftk,Beuf:2020dxl} or for exclusive vector meson production~\cite{Mantysaari:2021ryb,Mantysaari:2022bsp,Mantysaari:2022kdm}.

Next we show the complete finite expressions for these different contributions. Let us start with the  ``dip'' term, which can be written as
\begin{equation}
\label{eq:dip}
     \biggl [ \frac{\dd[]{ \sigma^{\text{D}}_{\gamma^*_{\lambda} + A } }}{\dd[2]{\Deltat} \dd{M_X^2}}  \biggr ]_{\text{dip}} =  
     2 \pi\aem  \nc \sum_f e_f^2
     \int [\mathrm{dPS}]_{\text{dip}} \biggl [\mathcal{G}^{\lo}_{\lambda,\text{dip}} + \left (\frac{\as \cf}{2\pi} \right ) \mathcal{G}^{\nlo}_{\lambda,\text{dip}} \biggr ] \left(1-\dipole_{01}\right)
\left(1-\dipole_{\ov 0 \ov 1}\right)^\dag. 
\end{equation}
The hard factors $\mathcal{G}^{\lo}_{\lambda,\text{dip}}$ and $ \mathcal{G}^{\nlo}_{\lambda,\text{dip}}$ are related to the photon wave function and to the perturbative gluon emission from the (color singlet) $q\bar q$ state after the shockwave and are given explicitly below.

We work in the mixed transverse coordinate-longitudinal momentum fraction space which is convenient in the dipole picture as the transverse coordinates of the partons remain fixed when they propagate through the target color field. Throughout this manuscript coordinates with a bar on the index such as $\xt_{\ov 0}$ or on the coordinate such as $\ov \bt$ %
refer to the corresponding coordinates in the complex conjugate amplitude. The 2-particle coordinate-space phase-space integral in \cref{eq:dip} is
\begin{equation}
\label{eq:2partcoordphasesp}
\int [\mathrm{dPS}]_{\text{dip}} = \int \frac{
\dd[2]{\xt_{0}}\dd[2]{\xt_{1}}
\dd[2]{\xt_{\ov 0}}\dd[2]{\xt_{\ov 1}}
    }{(2\pi)^4} \frac{e^{i \Deltat \vdot (\ov \bt -\bt)}}{(2\pi)^2} 
    \int_0^1  \dd[]{z_{0}} \dd[]{z_{1}} \delta(1-z_0 -z_1).  
\end{equation}
Here the fraction of the large photon plus momentum carried by the quark (antiquark) is $z_0$ ($z_1$) %
and $\bt=z_0 \xt_0 + z_1 \xt_1$ is the center-of-mass of the dipole (see also the discussion e.g. in Refs.~\cite{Hatta:2017cte,Beuf:2022kyp}).

The hard factor in \cref{eq:dip} corresponding to the leading-order process is
\begin{equation}
\label{eq:G_LO}
\mathcal{G}^{\lo}_{\lambda,\text{dip}} = 
\mathcal{F}_\lambda(z_0, z_1, z_0, z_1)
J_0\left(\abs{\xt_{01}-\xt_{\ov 0 \ov 1}} \mxbar\right).
\end{equation}
For the leading-order process, the hard factor only depends on two momentum fractions $z_0$ and $z_1$ which are the same in the direct and conjugate amplitudes. However, at NLO this is not always the case because of gluon exchanges between the quark and antiquark. Thus  for future convenience we define here a more general notation
with 
 \begin{equation}
 \label{eq:lo_photon_wf_part}
    \mathcal{F}_\lambda(z_0, z_1, z_{\ov 0}, z_{\ov 1}) =
     \begin{cases}
        4[z_0 z_1 z_{\ov 0} z_{\ov 1}]^{3/2} Q^2 K_0(\abs{\xt_{01}} Q \sqrt{z_0 z_1})K_0(\abs{\xt_{\ov 0 \ov 1}} Q \sqrt{z_{\ov 0} z_{\ov 1}})
        &\lambda = L \\
        z_0 z_1 z_{\ov 0} z_{\ov 1} (z_0 z_{\ov 0} + z_1 z_{\ov 1})
        \frac{\xt_{01} \vdot \xt_{\ov 0 \ov 1}}{\abs{\xt_{01}}\abs{\xt_{\ov 0\ov 1}}}
        Q^2K_1(\abs{\xt_{01}} Q \sqrt{z_0 z_1})K_1(\abs{\xt_{\ov 0 \ov 1}} Q \sqrt{z_{\ov 0} z_{\ov 1}})
        &\lambda = T
     \end{cases}.
 \end{equation}
Here we denote $\xt_{ij}=\xt_i-\xt_j$ and $\mxbar = \sqrt{z_0z_1}M_X$. This form for the coupling of the invariant mass of the $q\bar{q}$ state to the coordinates was obtained in Ref.~\cite{Beuf:2022kyp}\footnote{Note that Eq.~(81) in Ref.~\cite{Beuf:2022kyp} corresponding to the transversely polarized case contains a minus sign typo that has been fixed here.}. 
It is in fact equivalent to the massless-quark limit of the form commonly used in the literature (see e.g. Eqs.~(15) and~(16) in Ref.~\cite{Marquet:2007nf} or Eqs.~(8) and~(9) in Ref.~\cite{Kowalski:2008sa}), but results after performing angular integrals in a different order, as discussed in Ref.~\cite{Beuf:2022kyp}.

The hard factor related to the NLO contributions reads
\begin{equation}
    \mathcal{G}^{\nlo}_{\lambda,\text{dip}} = f^\text{UV}
\mathcal{F}_\lambda(z_0, z_1, z_0, z_1)
    + \int_0^1 \dd{z_{\ov 0}} \dd{z_{\ov 1}} \delta(1-z_{\ov 0}-z_{\ov 1}) f^\text{final state} \mathcal{F}_\lambda(z_0, z_1, z_{\ov 0}, z_{\ov 1}).
\end{equation}
It has been split into two terms that have separate physical origins. The first term includes the sum of all UV divergent diagrams, including the UV limits (subtraction terms) of the diagrams where the $q\bar q g$ system interacts with the target. This combination of  terms is finite and reads
 \begin{equation}
 \label{eq:f_UV}
 \begin{split}
     f^\text{UV} =&
     \left(2\log(z_0 z_1)-3 \right) 
     \times \left[\log(\frac{M_X e^{\gamma_E}  \abs{\xt_{01}}\abs{\xt_{\ov 0\ov 1}} }{2 \abs{\xt_{01}-\xt_{\ov 0\ov 1}} }) 
    J_{0} \left( \mxbar  |\xt_{01}-\xt_{\overline 0 \overline 1}| \right)
    +\frac{\pi}{2} 
    Y_{0} \left( \mxbar |\xt_{01}-\xt_{\overline 0 \overline 1}| \right)
    \right]\\
    &+\left[11 - \pi^2 +\log^2 z_0 + \log^2 z_1 - 4 \log z_0 \log z_1\right]
     J_{0} \left( \mxbar |\xt_{01}-\xt_{\overline 0 \overline 1}| \right).
 \end{split}
 \end{equation}
The second term contains the gluon exchange between the quarks after the interaction with the target: thus it depends in a nontrivial way on the  momentum fractions both in the amplitude and the conjugate amplitude (before or after the gluon exchange) $z_0, z_1$ and $z_{\ov 0}, z_{\ov 1}$ which can be different. Here we write
 \begin{equation}
 \label{eq:f_final_state}
     f^\text{final state} = \Ft+ \Fsub + F^{02} + F_\adiag^{012}+ F_\bdiag^{012} 
 \end{equation}
 where
 \begin{equation}
 \label{eq:Ft}
\begin{split}
    \Ft =&  \int_0^1 \frac{\dd{t}}{t} \frac{1}{\abs{z_0 -z_{\ov 0}}}
    \frac{z_0 z_1+z_{\ov 0} z_{\ov 1}}{\sqrt{z_0 z_1 z_{\ov 0} z_{\ov 1}}} \\
     \times  \Bigg\{ &
    \theta(z_0 -z_{\ov 0})
    \ft\left(t,  
    M_X \sqrt{\frac{z_1}{z_0}}  \abs{ z_{\ov 0} \xt_{\ov 0 \ov 1}-z_0\xt_{01}},
    M_X \abs{\xt_{\ov 0 \ov 1}} \sqrt{(z_0 -z_{\ov 0})\frac{z_{\ov 0}}{z_0}} 
    \right)
     \\ 
    +&\theta(z_0 -z_{\ov 0})
    \ft\left(t,  
    M_X \sqrt{\frac{z_{\ov 0}}{z_{\ov 1}}} \abs{ z_{\ov 1}  \xt_{\ov 0 \ov 1}-z_{ 1}\xt_{01}},
    M_X \abs{\xt_{ 0 1}} \sqrt{(z_0 -z_{\ov 0})\frac{z_{1}}{z_{\ov 1}}}
    \right) \\ 
    +&\theta(z_{\ov 0} - z_{ 0})
    \ft\left(  t,
     M_X\sqrt{\frac{z_0}{z_1}}  \abs{z_{\ov 1} \xt_{\ov 0 \ov 1}- z_1 \xt_{01}}  ,
    M_X \abs{\xt_{\ov 0 \ov 1}} \sqrt{(z_{\ov 0} - z_{ 0})\frac{z_{\ov 1}}{z_1}} 
    \right) \\ 
    +&\theta( z_{\ov 0}-z_0)
    \ft\left(  t,
     M_X \sqrt{\frac{z_{\ov 1}}{z_{\ov 0}}} \abs{ z_{\ov 0} \xt_{\ov 0 \ov 1}-z_{ 0}\xt_{01}}  ,
    M_X \abs{\xt_{ 0 1}} \sqrt{(z_{\ov 0} - z_{ 0})\frac{z_{0}}{z_{\ov 0}}}
    \right) 
    \Bigg\},
\end{split}
\end{equation}
and
\begin{equation}
\label{eq:Fsub}
\begin{split}
    \Fsub =& \int_0^1 \frac{\dd{z_2'}}{z_2'} \frac{1}{2} \frac{z_0 z_1 +z_{\ov 0} z_{\ov 1}}{\sqrt{z_0 z_1 z_{\ov 0} z_{\ov 1}}}\\
    \times\Bigg\{
        4 &\delta(z_0 - z_{\ov 0}) 
        \fsub\left(  
    M_X \sqrt{z_0 z_1} \abs{  \xt_{\ov 0 \ov 1}-\xt_{01}},
    M_X \abs{\xt_{\ov 0 \ov 1}} \sqrt{z_2'} 
    \right)\\
        -&\delta(z_2' - (z_0 -z_{\ov 0})) 
    \fsub\left(  
    M_X \sqrt{\frac{z_1}{z_0}}  \abs{ z_{\ov 0} \xt_{\ov 0 \ov 1}-z_0\xt_{01}},
    M_X \abs{\xt_{\ov 0 \ov 1}} \sqrt{z_2'\frac{z_{\ov 0}}{z_0}} 
    \right)
     \\ 
    -&\delta(z_2' - (z_0 -z_{\ov 0}))
    \fsub\left(  
    M_X \sqrt{\frac{z_{\ov 0}}{z_{\ov 1}}} \abs{ z_{\ov 1}  \xt_{\ov 0 \ov 1}-z_{ 1}\xt_{01}},
    M_X \abs{\xt_{ 0 1}} \sqrt{z_2'\frac{z_{1}}{z_{\ov 1}}}
    \right) \\ 
    -&\delta(z_2' - (z_{\ov 0} - z_{ 0})) 
    \fsub\left(  
     M_X\sqrt{\frac{z_0}{z_1}}  \abs{z_{\ov 1} \xt_{\ov 0 \ov 1}- z_1 \xt_{01}}  ,
    M_X \abs{\xt_{\ov 0 \ov 1}} \sqrt{z_2'\frac{z_{\ov 1}}{z_1}} 
    \right) \\ 
    -&\delta(z_2' - ( z_{\ov 0}-z_0) )
    \fsub\left(  
     M_X \sqrt{\frac{z_{\ov 1}}{z_{\ov 0}}} \abs{ z_{\ov 0} \xt_{\ov 0 \ov 1}-z_{ 0}\xt_{01}}  ,
    M_X \abs{\xt_{ 0 1}} \sqrt{z_2'\frac{z_{0}}{z_{\ov 0}}}
    \right) 
    \Bigg\}.
\end{split}
\end{equation}
Here we have not performed the $z_2'$ integral as we anticipate that in practical numerical implementation it will be more convenient to use the delta function to do the $z_{\ov 0}$ integral. Above we also used the following definitions
\begin{align}
    \ft(t,a,b) &=  \left[ 1 - J_0(tb) \right] J_0\left(a \sqrt{1-t^2}\right), \ \text{and} \\
    \fsub(a,b) &= J_0( a ) \log( \frac{b^2 e^{\gamma_E}}{2a} )
    + \frac{\pi}{2} Y_0(a).
\end{align}
Furthermore we have
\begin{equation}
\begin{split}
    F^{02}
    =& \frac{\pi}{2} \frac{z_0 z_{\ov 0} + z_{1} z_{\ov 1}}{\sqrt{  z_0 z_1 z_{\ov 0} z_{\ov 1}}}
    \times 
    \left[
    J_0 \left( M_X \abs{\xt_{01}} \sqrt{z_0 z_1} \right)
    Y_0 \left( M_X \abs{\xt_{\ov 0\ov 1}} \sqrt{z_{\ov 0 }z_{\ov 1}} \right)
    +
    Y_0 \left( M_X \abs{\xt_{01}} \sqrt{z_0 z_1} \right)
    J_0 \left( M_X \abs{\xt_{\ov 0\ov 1}} \sqrt{z_{\ov 0 }z_{\ov 1}} \right)
    \right],
\end{split}
\end{equation}
\begin{equation}
\label{eq:F_A012_term}
\begin{split}
    F_{\adiag}^{012}
    =& -\frac{1}{4}\int_0^1 \dd{ z_2'} \delta\left(z_2'- (z_0-z_{\ov 0})\right) 
        \int_0^1 \dd{t}\dd{u}\dd{v} \delta(1-t-u-v) \\
        &\times \frac{ z_0 z_1 + z_{\ov 0} z_{\ov 1}}{\left[ z_0 z_{\ov 1}v+z_2' t u \right]^{3/2}} 
        \times M_X
        \sqrt{v \abs{\xt_{01}-\xt_{\ov 0\ov 1}}^2 
        + z_2' \frac{\xt_{01}^2}{z_{\ov 0} z_{\ov 1}} \left(u+z_{\ov 1}v \right) +z_2' \frac{\xt_{\ov 0\ov 1}^2}{z_{ 0} z_{ 1}} \left(t+z_{0}v \right)
        }\\
        &\times  J_1 \left( M_X
        \sqrt{\frac{z_0 z_1 z_{\ov 0}z_{ \ov 1}}{ z_0 z_{\ov 1}v+z_2' t u }}
        \sqrt{v \abs{\xt_{01}-\xt_{\ov 0\ov 1}}^2 
        + z_2' \frac{\xt_{01}^2}{z_{\ov 0} z_{\ov 1}} \left(u+z_{\ov 1}v \right) +  z_2'  \frac{\xt_{\ov 0\ov 1}^2}{z_{ 0} z_{ 1}} \left(t+z_{0}v \right)
        }
        \right)
\end{split}
\end{equation}
and finally
\begin{equation}
\label{eq:F_B012_term}
\begin{split}
    F_{\bdiag}^{012}
    =&
    -\frac{1}{4}\int_0^1 \dd{ z_2'} \delta\left(z_2'- (z_{\ov 0}-z_{ 0})\right) 
        \int_0^1 \dd{t}\dd{u}\dd{v} \delta(1-t-u-v) \\
        &\times \frac{z_0 z_1 + z_{\ov 0} z_{\ov 1}}{\left[ z_1 z_{\ov 0}v+z_2' t u \right]^{3/2}} 
        \times M_X
        \sqrt{v \abs{\xt_{01}-\xt_{\ov 0\ov 1}}^2 
        +z_2' \frac{\xt_{01}^2}{z_{\ov 0} z_{\ov 1}} \left(u+z_{\ov 0}v \right) +z_2' \frac{\xt_{\ov 0\ov 1}^2}{z_{ 0} z_{ 1}} \left(t+z_{1}v \right)
        }\\
        &\times  J_1 \left( M_X
        \sqrt{\frac{z_0 z_1 z_{\ov 0} z_{\ov 1}}{z_1 z_{\ov 0}v+z_2' t u }}
        \sqrt{v \abs{\xt_{01}-\xt_{\ov 0\ov 1}}^2 
        +z_2'  \frac{\xt_{01}^2}{z_{\ov 0} z_{\ov 1}} \left(u+z_{\ov 0}v \right) + z_2'  \frac{\xt_{\ov 0\ov 1}^2}{z_{ 0} z_{ 1}} \left(t+z_{1}v \right)
        }
        \right).
\end{split}
\end{equation}

At the NLO level we can only have one gluon in the process at the cross section level. Thus, if this gluon crosses both shockwaves leading to the ``trip'' Wilson line structure, it must also cross the cut, so that the final state always has three particles. For fixed $M_X$ such contributions are finite by themselves without requiring cancellations with other types of diagrams. This ``trip'' term was already calculated in  Ref.~\cite{Beuf:2022kyp}.  Here we quote again the result, correcting a factor 2 typo in Eq.~(89) of Ref.~\cite{Beuf:2022kyp} for the transverse photon. 
The result reads
\begin{equation}
\begin{split}
      \biggl [ \frac{\dd[]{ \sigma^{\text{D}}_{\gamma^*_{\lambda} + A } }}{\dd[2]{\Deltat} \dd{M_X^2}}  \biggr ]_{\text{trip}} 
      =   2 \pi\aem  \nc \sum_f e_f^2
      \int [\mathrm{dPS}]_{\text{trip}} \left (\frac{\as \cf}{2\pi} \right ) \mathcal{G}^{\nlo}_{\lambda,\text{trip}} \left(1-\tripole_{012}\right)
\left(1-\tripole_{\ov 0 \ov 1 \ov 2}\right)^\dag.
\end{split}
\end{equation}
Here we have three particles (quark, antiquark, and the gluon) in the final state and the phase space integral is more complicated, but on the other hand the momentum fractions $z_0,z_1,z_2$ are always the same in the amplitude and conjugate amplitude:
\begin{equation}
\int [\mathrm{dPS}]_{\text{trip}} = \int \frac{
\dd[2]{\xt_{0}}\dd[2]{\xt_{1}}\dd[2]{\xt_{2}}
\dd[2]{\xt_{\ov 0}}\dd[2]{\xt_{\ov 1}}\dd[2]{\xt_{\ov 2}}
}{(2\pi)^6} \frac{e^{i \Deltat \vdot (\ov \bt -\bt)}}{(2\pi)^2}  
    \int_0^1  \dd[]{z_{0}} \dd[]{z_{1}} \dd[]{z_{2}} \delta(1-z_0 -z_1-z_2).
\end{equation}
The gluon coordinate is denoted by $\xt_2$ (or $\xt_{\ov 2}$) and the plus momentum fraction by $z_2$. The Wilson line operator describing the $q\bar q g$-target interaction can be expressed in terms of the dipole operator as
\begin{equation}
    \tripole_{012} = \frac{\nc}{2 C_F} \left( \dipole_{02} \dipole_{12} - \frac{1}{\nc^2} \dipole_{01}\right).
\end{equation}
The hard factors in the case of longitudinal (L) and transverse (T) photon read
\begin{equation}
\label{eq:G_trip_L}
    \begin{split}
        \mathcal{G}^{\nlo}_{L,\text{trip}}
        =&
        8 z_0 z_1 Q^2 K_0(Q X_{012}) K_0(Q X_{\ov 0 \ov 1 \ov 2})
        \times \frac{M_X}{Y_{012}} J_1(M_X Y_{012})\\
&\times \left\{
z_1^2 \left[ 2z_0 (z_0 +z_2) + z_2^2 \right] \frac{\xt_{20} \vdot \xt_{\ov 2 \ov 0}}{\xt_{20}^2 \xt_{\ov 2 \ov 0}^2}
+
z_0^2 \left[ 2z_1 (z_1 +z_2) + z_2^2 \right] \frac{\xt_{21} \vdot \xt_{\ov 2 \ov 1}}{\xt_{21}^2 \xt_{\ov 2 \ov 1}^2}
\right.\\
&\left.-z_0 z_1 \left[ z_0 (1-z_0) + z_1 (1-z_1) \right]
\left[
\frac{\xt_{20} \vdot \xt_{\ov 2 \ov 1}}{\xt_{20}^2 \xt_{\ov 2 \ov 1}^2 }
+\frac{\xt_{21} \vdot \xt_{\ov 2 \ov 0}}{\xt_{21}^2 \xt_{\ov 2 \ov 0}^2 }
\right]
\right\}
    \end{split}
\end{equation}
and
\begin{equation}
\label{eq:G_trip_T}
    \begin{split}
        \mathcal{G}^{\nlo}_{T,\text{trip}}
        =&
        2 z_0 z_1 Q^2 K_1(Q X_{012}) K_1(Q X_{\ov 0 \ov 1 \ov 2}) \frac{1}{X_{012}} \frac{1}{X_{\ov{012}}}
        \times \frac{M_X}{Y_{012}} J_1(M_X Y_{012})\\
&\times 
\left\{ 
\Upsilonbb + \Upsiloncc + \Upsilond + \Upsilone + \Upsilonbc
\right\}.
\end{split}
\end{equation}
Here the transverse scales are
\begin{align}
\label{eq:Y012}
    Y_{012}^2 &= z_0 z_1 (\xt_{\ov 0\ov 1}- \xt_{01})^2 +  z_2 z_0 (\xt_{\ov 2\ov 0}- \xt_{20})^2 + z_2 z_1 (\xt_{\ov 2\ov 1}- \xt_{21})^2, \\
    \label{eq:X012}
    X_{012}^2 &= z_0 z_1 \xt_{01}^2 + z_0 z_2 \xt_{20}^2 + z_1 z_2 \xt_{21}^2, \ \text{and} \\
    \label{eq:X012ov}
    X_{\ov {012} }^2 &= z_{ 0} z_{ 1} \xt_{ \ov  0  \ov 1}^2 + z_{ 0} z_{ 2} \xt_{\ov 2 \ov  0}^2 + z_{1} z_{2} \xt_{\ov 2 \ov 1}^2.
\end{align}
The coefficient functions read:
\begin{equation}
\begin{split}
    \Upsilonbb =& \frac{z_1^2}{ \xt_{20}^2 \xt_{\ov 2 \ov 0}^2}
    \Big\{ 
    (2z_0^2 +2z_0 z_2 + z_2^2)(1-2z_1(1-z_1))
    \left(\xt_{0+2;1} \vdot \xt_{\ov 0+\ov 2;\ov 1}\right)
    \left( \xt_{20} \vdot \xt_{\ov 2 \ov 0}\right)\\
    &-z_2 (2z_0 + z_2) (2 z_1 -1) 
    \left[ 
    \left(\xt_{20} \vdot \xt_{0+2;1} \right)
    \left(\xt_{\ov 2 \ov 0} \vdot \xt_{\ov 0+\ov 2;\ov 1}\right)
    -
    \left( \xt_{20} \vdot \xt_{\ov 0 +\ov 2; \ov 1}\right)
    \left( \xt_{\ov 2 \ov 0} \vdot \xt_{0 + 2;  1}\right)
    \right]
    \Big\}
\end{split}
\end{equation}
\begin{equation}
\begin{split}
    \Upsiloncc =& \frac{z_0^2}{ \xt_{21}^2 \xt_{\ov 2 \ov 1}^2}
    \Big\{ 
    (2z_1^2 +2z_1 z_2 + z_2^2)(1-2z_0(1-z_0))
    \left(\xt_{0;1+2} \vdot \xt_{\ov 0;\ov 1+\ov 2}\right)
    \left( \xt_{21} \vdot \xt_{\ov 2 \ov 1}\right)\\
    &-z_2 (2z_1 + z_2) (2 z_0 -1) 
    \left[ 
    \left(\xt_{21} \vdot \xt_{0;1+2} \right)
    \left(\xt_{\ov 2 \ov 1} \vdot \xt_{\ov 0;\ov 1+\ov 2}\right)
    -
    \left( \xt_{21} \vdot \xt_{\ov 0;\ov 1 +\ov 2}\right)
    \left( \xt_{\ov 2 \ov 1} \vdot \xt_{0;1 + 2}\right)
    \right]
    \Big\}
\end{split}
\end{equation}
\begin{equation}
    \begin{split}
    \Upsilond =& \frac{z_0^2 z_1^2 z_2^2}{(z_0+z_2)^2}
    -\frac{z_0^2 z_1^3 z_2}{z_0 + z_2} 
    \left( 
    \frac{\xt_{20} \vdot \xt_{0+2;1}}{\xt_{20}^2}
    +\frac{\xt_{\ov 2 \ov 0} \vdot \xt_{\ov 0+\ov 2;\ov 1}}{\xt_{\ov 2 \ov 0}^2}
    \right) \\
    &+ \frac{z_0^2 z_1 z_2 (z_1 +z_2)^2}{z_0 +z_2}
    \left( 
    \frac{\xt_{21} \vdot \xt_{0;1+2}}{\xt_{21}^2}
    +\frac{\xt_{\ov 2 \ov 1} \vdot \xt_{\ov0;\ov 1+\ov 2}}{\xt_{\ov 2 \ov 1}^2}
    \right)
    \end{split}
\end{equation}
\begin{equation}
    \begin{split}
    \Upsilone =& \frac{z_0^2 z_1^2 z_2^2}{(z_1+z_2)^2}
    +\frac{z_0^3 z_1^2 z_2}{z_1 + z_2} 
    \left( 
    \frac{\xt_{21} \vdot \xt_{0;1+2}}{\xt_{21}^2}
    +\frac{\xt_{\ov 2 \ov 1} \vdot \xt_{\ov 0;\ov 1+\ov 2}}{\xt_{\ov 2 \ov 1}^2}
    \right) \\
    &- \frac{z_0 z_1^2 z_2 (z_0 +z_2)^2}{z_1 +z_2}
    \left( 
    \frac{\xt_{20} \vdot \xt_{0+2;1}}{\xt_{20}^2}
    +\frac{\xt_{\ov 2 \ov 0} \vdot \xt_{\ov 0+ \ov 2;\ov 1}}{\xt_{\ov 2 \ov 0}^2}
    \right)
    \end{split}
\end{equation}
\begin{equation}
    \begin{split}
        \Upsilonbc =&
        -z_0 z_1
        (z_0 (1-z_1) + z_1 (1-z_0)) (z_0 (1-z_0) + z_1 (1-z_1)) \\
        &\times \Bigg\{
          \frac{ \left( \xt_{20} \vdot \xt_{\ov 2 \ov 1 } \right)
            \left( \xt_{0+2;1} \vdot \xt_{\ov 0; \ov 1 +\ov 2} \right)}{\xt_{20} \xt_{\ov 2 \ov 1}}
             +
             \left( \xt_{21} \vdot \xt_{\ov 2 \ov 0 } \right)
            \left( \xt_{0;1+2} \vdot \xt_{\ov 0+\ov2; \ov 1 } \right) 
        \Bigg\}\\
        &+z_0 z_1 z_2 (z_0 -z_1)^2
        \Bigg\{
            \frac{
                \left(\xt_{20} \vdot \xt_{0+2;1} \right)
                \left(\xt_{\ov 2\ov 1} \vdot \xt_{\ov 0; \ov 1+\ov 2}\right)
                -
                \left(\xt_{20} \vdot \xt_{\ov 0; \ov 1+\ov 2} \right)
                \left(\xt_{\ov 2\ov 1} \vdot \xt_{0+2;1}\right)
            }{\xt_{20}^2 \xt_{\ov 2 \ov 1}^2}\\
        &+
         \frac{
               \left(\xt_{21} \vdot \xt_{0;1+2} \right)
                \left(\xt_{\ov 2\ov 0} \vdot \xt_{\ov 0 + \ov 2; \ov 1}\right)
                -
                \left(\xt_{21} \vdot \xt_{\ov 0+\ov 2; \ov 1} \right)
                \left(\xt_{\ov 2\ov 0} \vdot \xt_{0;1+2}\right)
            }{\xt_{21}^2 \xt_{\ov 2 \ov 0}^2}
        \Bigg\}.
    \end{split}
\end{equation}
Here we defined
\begin{align}
\label{eq:x021}
    \xt_{0+2;1} &= \xt_{01} + \frac{z_2}{z_0+z_2} \xt_{20}\\
\label{eq:x012}
    \xt_{0;1+2} &= \xt_{01} - \frac{z_2}{z_1+z_2} \xt_{21},
\end{align}
and follow the same notation as in Ref.~\cite{Beuf:2022kyp} where the same contribution (gluon emitted before the target and the $q\bar q g$ system produced in the final state) has been calculated. 

The final ``dip-trip'' term results from the interference  between terms where a gluon crosses the shockwave 
in the amplitude but not in the complex conjugate amplitude, or vice versa. 
We use the symmetry between the amplitude and conjugate amplitude to write this as a factor $2$ times 
the contributions in which the gluon crosses the target only in the amplitude.
These terms  can be written as
\begin{equation}
\label{eq:diptrip}
\begin{split}
      &\biggl [ \frac{\dd[]{ \sigma^{\text{D}}_{\gamma^*_{\lambda} + A} }}{\dd[2]{\Deltat} \dd{M_X^2}}  \biggr ]_{\text{dip-trip}} = 
         2 \pi\aem  \nc \sum_f e_f^2
      \int [\mathrm{dPS}]_{\text{dip-trip}} \frac{\as \cf}{2\pi}\\
      &\times \biggl \{ \mathcal{G}^{\nlo}_{\lambda,\text{dip-trip}}2\Re\left(1-\tripole_{012}\right)
\left(1-\dipole_{\ov 0 \ov 1 }\right)^\dag  
- \mathcal{G}^{\nlo}_{\lambda,\text{dip sub}}2\Re\left(1-\dipole_{01}\right)
 \left(1-\dipole_{\ov 0 \ov 1 }\right)^\dag  \biggr \}.
\end{split}
\end{equation}
The phase space integral in this case also has a gluon with momentum fraction $z_2$ in the amplitude at coordinate $\xt_{2}$, but not in the leading-order conjugate amplitude:
\begin{equation}
\begin{split}
\int [\mathrm{dPS}]_{\text{dip-trip}} = & \int \frac{
\dd[2]{\xt_{0}}\dd[2]{\xt_{1}}\dd[2]{\xt_{2}}
\dd[2]{\xt_{\ov 0}}\dd[2]{\xt_{\ov 1}}
}{(2\pi)^5}  \frac{e^{i \Deltat \vdot (\ov \bt -\bt)}}{(2\pi)^2}
\int_0^1  \dd[]{z_{0}} \dd[]{z_{1}} \dd[]{z_{2}} \delta(1-z_0 -z_1-z_2). 
\end{split}
\end{equation}
We have used the momentum conservation so that term by term the momentum fractions in the conjugate amplitude ($z_{\ov 0}+z_{\ov 1}=1$) are 
either $z_{\ov 0}=z_0$ and~$z_{\ov 1}=z_1+z_2$ or $z_{\ov 0}=z_0+z_2$  and~$z_{\ov 1}=z_1$ in the individual terms. 
The two terms in \cref{eq:diptrip} correspond to the genuine interference contribution (first term) and a UV subtraction term, which has been subtracted here and added to the ``dip'' term presented above to render it finite. 
It is convenient to use the following short-hand notations when writing the results for the inference terms:
\begin{equation}
\label{eq:shorthand_notations}
\begin{aligned}
    \mxbar[0]^2 &= z_1 (1-z_1) M_X^2, & \mxbar[1]^2 &= z_0 (1-z_0) M_X^2, \\
    \Qbar[0]^2  &= z_{1}(1-z_{1}) Q^2,  &  \Qbar[1]^2 &= z_{0}(1-z_{0}) Q^2, \\
    \omega_{0} &= \frac{z_{0} z_{2}}{z_{1}(z_{0}+z_{2})^2}, & \omega_{1} &= \frac{z_{1} z_{2}}{z_{0} (z_{1}+z_{2})^2} .
\end{aligned}
\end{equation}
The results for longitudinal and transverse photons are then given by
\begin{equation}
\label{eq:G_diptrip_L}
\begin{split}
    \mathcal{G}^{\nlo}_{L,\text{dip-trip}} =&
   \frac{8}{z_2} Q^2 K_0(X_{012} Q) \\
&\times     \Bigg[
     z_1^2 (1-z_1) K_0\left( \abs{\xt_{\ov 0 \ov 1}} \Qbar[0] \right)
     \times 
    J_0\left( \mxbar[0] \sqrt{
    \left( \xt_{0+2;1} -\xt_{\ov 0 \ov 1}\right)^2
    + \omega_0\xt_{20}^2}
    \right) \\
     &\times\left( z_1 \left(2 z_0 (z_0+z_2) + z_2^2 \right) \frac{1}{\xt_{20}^2}
     - z_0 \left(z_0(1-z_0)+z_1(1-z_1) \right) \frac{\xt_{20} \vdot \xt_{21}}{\xt_{20}^2 \xt_{21}^2}
     \right)
     \\
    &+ 
     z_0^2 (1-z_0) K_0\left( \abs{\xt_{\ov 0 \ov 1}}\Qbar[1] \right)
     \times
    J_0\left(\mxbar[1] \sqrt{
    \left( \xt_{\ov 0 \ov 1}-\xt_{0;1+2} \right)^2
    + \omega_1 \xt_{21}^2}
    \right)\\
    &\times\left( z_0 \left( 2z_1 (z_1+z_2) + z_2^2 \right) \frac{1}{\xt_{21}^2}
     - z_1 \left( z_0(1-z_0)+z_1(1-z_1) \right) \frac{\xt_{20} \vdot \xt_{21}}{\xt_{20}^2 \xt_{21}^2}
     \right)
    \Bigg]
\end{split}
\end{equation}
and 
\begin{equation}
\label{eq:G_diptrip_T}
\begin{split}
    \mathcal{G}^{\nlo}_{T,\text{dip-trip}} =&
   \frac{2}{z_2} \frac{Q}{X_{012}\abs{\xt_{\ov{01}}}}K_1(X_{012} Q) \\
&\times     \Bigg[
     z_1 \Qbar[0] K_1\left( \abs{\xt_{\ov 0 \ov 1}} \Qbar[0]\right)
     \times
    J_0\left( \mxbar[0] \sqrt{
    \left( \xt_{0+2;1} -\xt_{\ov 0 \ov 1}\right)^2
    +\omega_0 \xt_{20}^2}
    \right) 
     \\&\times\left(
     \Upsilonuone + \Upsilonutwo  + \Upsilonuthree + \Upsilonufour
     \right)
     \\
    &+ 
     z_0 \Qbar[1] K_1\left( \abs{\xt_{\ov 0 \ov 1}} \Qbar[1] \right)
     \times
    J_0\left( \mxbar[1] \sqrt{
    \left(\xt_{\ov 0 \ov 1}-\xt_{0;1+2} \right)^2
    + \omega_1 \xt_{21}^2}
    \right)
     \\&\times\left(
     \Upsilonvone + \Upsilonvtwo  + \Upsilonvthree + \Upsilonvfour
     \right)
    \Bigg].
\end{split}
\end{equation}
Here we have defined:
\begin{equation}
\label{eq:Upsilonuone}
\begin{split}
    \Upsilonuone =& \frac{z_1}{\xt_{20}^2}
    \left[ 2z_0^2+2z_0z_2+z_2^2 \right] \left[ 1-2z_1(1-z_1) \right]
    \left( \xt_{0+2;1} \vdot \xt_{\ov{01}}\right)
\end{split}
\end{equation}
\begin{equation}
\begin{split}
    \Upsilonutwo =& \frac{z_0}{\xt_{20}^2 \xt_{21}^2} \bigg\{ 
    \left[ 2z_0 z_1 -(z_0+z_1) \right] \left[ z_0 (1-z_0) + z_1(1-z_1) \right]
    \left(\xt_{20} \vdot \xt_{21}\right)\left( \xt_{0;1+2} \vdot \xt_{\ov{01}}\right)
     \\
    & +z_2 \left[z_0 - z_1 \right]^2
    \left[\left(\xt_{21} \vdot \xt_{0;1+2}\right)\left( \xt_{20} \vdot \xt_{\ov{01}}\right)
    -
    \left(\xt_{21} \vdot \xt_{\ov{01}}\right)\left( \xt_{20} \vdot \xt_{0;1+2}\right)
      \right]
    \bigg\}
\end{split}
\end{equation}
\begin{equation}
\begin{split}
    \Upsilonuthree =& - \frac{z_0^2 z_1^2 z_2}{z_0 + z_2}
    \frac{\xt_{20} \vdot \xt_{\ov{01}}}{\xt_{20}^2 } 
\end{split}
\end{equation}
\begin{equation}
\begin{split}
    \Upsilonufour =& - \frac{z_0 z_1 z_2 (1-z_1)^2}{z_1 + z_2}
    \frac{\xt_{20} \vdot \xt_{\ov{01}}}{\xt_{20}^2 } 
\end{split}
\end{equation}
\begin{equation}
\begin{split}
    \Upsilonvone =& \frac{z_0}{\xt_{21}^2}
    \left[ 2z_1^2+2z_1z_2+z_2^2 \right] \left[ 1-2z_0(1-z_0) \right]
    \left( \xt_{0;1+2} \vdot \xt_{\ov{01}}\right)
\end{split}
\end{equation}
\begin{equation}
\begin{split}
    \Upsilonvtwo =& \frac{z_1}{\xt_{20}^2 \xt_{21}^2} \bigg\{ 
    \left[ 2z_0 z_1 -(z_0+z_1) \right] \left[ z_0 (1-z_0) + z_1(1-z_1) \right]
    \left(\xt_{20} \vdot \xt_{21}\right)\left( \xt_{0+2;1} \vdot \xt_{\ov{01}}\right)
     \\
    & +z_2 \left[z_0 - z_1 \right]^2
    \left[\left(\xt_{20} \vdot \xt_{0+2;1}\right)\left( \xt_{21} \vdot \xt_{\ov{01}}\right)
    -
    \left(\xt_{20} \vdot \xt_{\ov{01}}\right)\left( \xt_{21} \vdot \xt_{0+2;1}\right)
      \right]
    \bigg\}
\end{split}
\end{equation}
\begin{equation}
\begin{split}
    \Upsilonvthree =&  \frac{z_0^2 z_1^2 z_2}{z_1 + z_2}
    \frac{\xt_{21} \vdot \xt_{\ov{01}}}{\xt_{21}^2} 
\end{split}
\end{equation}
\begin{equation}
\label{eq:Upsilonvfour}
\begin{split}
    \Upsilonvfour =&  \frac{z_0 z_1 z_2 (1-z_0)^2}{z_0 + z_2}
    \frac{\xt_{21} \vdot \xt_{\ov{01}}}{\xt_{21}^2} .
\end{split}
\end{equation}

The second term in \cref{eq:diptrip}, originating from the UV subtraction part, reads (in the subtraction scheme applied in this work)
\begin{equation}
\label{eq:G_dipsub}
\begin{split}
    \mathcal{G}^{\nlo}_{\lambda,\text{dip sub}} =&
    \frac{2}{ z_2 (1-z_1)^2} \left[z_0^2 +(1-z_1)^2 \right] \frac{1}{\xt_{20}^2} \exp{-\frac{\xt_{20}^2}{\xt_{01}^2 e^{\gamma_E}}} \\
    &\times \mathcal{F}_\lambda(1-z_1, z_1, 1-z_1, z_1)J_0\left(\abs{\xt_{01}-\xt_{\ov 0 \ov 1}} \mxbar[0]\right) \\
    +&
    \frac{2}{ z_2 (1-z_0)^2} \left[z_1^2 +(1-z_0)^2 \right] \frac{1}{\xt_{21}^2} \exp{-\frac{\xt_{21}^2}{\xt_{01}^2 e^{\gamma_E}}} \\
    &\times \mathcal{F}_\lambda(z_0, 1-z_0, z_0, 1-z_0) J_0\left(\abs{\xt_{01}-\xt_{\ov 0 \ov 1}} \mxbar[1] \right). 
\end{split}
\end{equation}
This subtraction term renders Eq.~\eqref{eq:diptrip} UV-finite, canceling the UV divergences in the integration over $\xt_2$ in the limits $\xt_{20}^2\to 0$, $\xt_{21}^2 \to 0 $.

Detailed derivations of all these expressions are given in the following sections. However, we emphasize again that all terms quoted in this section are finite and can in principle be implemented numerically. The multi-dimensional phase-space integral especially in the ``trip'' term is computationally challenging. In practice, this can be simplified by considering the $t$-integrated cross sections (the experimental measurements are also typically integrated over $t$), and by employing some common approximations for the dipole-target scattering amplitude. An integration over $t$ results in a delta function setting $\ov \bt \to \bt$. Some additional integrals can be performed analytically by assuming that the dipole amplitude does not depend on the azimuthal angles and that the impact parameter dependence factorizes from $1-S_{01}$, see Ref.~\cite{Beuf:2022kyp} for details. In this work we show the most general form for the NLO cross sections and leave practical numerical implementation with potential simplifying assumptions and phenomenological applications for the future.

\section{Scattering off an external field in light-cone perturbation theory}
\label{sec:computational_setup}
\subsection{In- and out-states}

We follow the standard light-cone perturbation theory approach for high-energy scattering off an external classical field~\cite{Bjorken:1970ah}. However, because we are here not dealing with a fully inclusive observable, we have to consider  the possibility of final-state interactions between multiparticle final states. This makes the situation more complicated and warrants a slightly more detailed discussion. In particular, the final-state interactions mean that we must be slightly more careful with the ``$i\varepsilon$-prescription'' for regularizing light-cone energy denominators; we will henceforth denote the infinitesimal variable used here by $i\delta$ to differentiate it from the $\varepsilon$ of dimensional regularization. 

The incoming and outgoing states in the scattering process are what we call ``dressed states''. The dressed states are interaction-picture states evaluated at $x^+=0$, obtained with the interaction-picture time-evolution operator $U_{\text{I}}(x^+,x^+_0)$ from free states at $x^+\to \mp \infty$:
\begin{eqnarray}
\ket{\phi}_\iin &=& U_{\text{I}}(0,-\infty) \ket{\phi}
\\
\ket{\phi}_\oout &=& U_{\text{I}}(0,\infty) \ket{\phi} = U^\dag_{\text{I}}(\infty,0) \ket{\phi}.
\end{eqnarray}
The free-theory eigenstates $\ket{\phi}$ are time-independent, since for the free theory the interaction picture is equivalent to the Heisenberg picture where states do not depend on time. The interaction-picture states are time dependent, but they are needed at the time $x^+=0$ where we have placed the shockwave corresponding to the target. 

Since the time-evolution operator is unitary, both the free states $\ket{\phi}$ and the dressed states $\ket{\phi}_\iin$, $\ket{\phi}_\oout$ form an orthonormal basis in terms of Fock states. A Fock state is specified by the number of particles of different species, their quantum numbers, and momenta, which we here denote by a collective notation $a,n,m,\dots$. For a theory with a (light cone) Hamiltonian $\hat{H} = \hat{H}_0+\hat{V}$ the relation between the in-, out- and free Fock states can be obtained by the well-known (Dyson) perturbative series
\begin{eqnarray}
\label{eq:dysonin}
 \ket{a}_\iin & = &U_{\text{I}}(0,-\infty) \ket{a} = 
 \ket{a} + \sum_n\ket{n} \frac{\bra{n}\hat{V} \ket{a}}{E_a-E_n+i\delta}
 + \sum_{m,n}\ket{n} 
 \frac{ \bra{n}\hat{V} \ket{m}}{E_a-E_n+i\delta}
\frac{ \bra{m}\hat{V} \ket{a}}{E_a-E_m+i\delta}
+ \dots
\\
\label{eq:dysonout}
 \ket{a}_\oout & = &U_{\text{I}}(0,+\infty) \ket{a} = 
 \ket{a} + \sum_n\ket{n} \frac{\bra{n}\hat{V}\ket{a}}{E_a-E_n-i\delta}
 + \sum_{m,n}\ket{n} 
 \frac{\bra{n}\hat{V}\ket{m}}{E_a-E_n-i\delta}
\frac{\bra{m}\hat{V}\ket{a}}{E_a-E_m-i\delta}
+ \dots \ ,
\end{eqnarray}
where the energies $E_a,E_n, E_m, \dots$ are the eigenvalues of the free part of the Hamiltonian for the free-particle states: $\hat{H}_0\ket{a}= E_a\ket{a}, \dots$.
The only difference between the perturbative series for the in- and out-states is the different sign of $i\delta$, which is dictated  by the fact that they must reduce to the free states in the past/future. This means that the evolution from the asymptotic state to the interaction at $x^+=0$ is forward/backward in time, i.e. with time-evolution operators that are Hermitian conjugates of each other \cite{Bjorken:1970ah}. 

Now it is of course possible that in the perturbative series \eqref{eq:dysonin}, \eqref{eq:dysonout} the real parts of some energy denominators $E_a-E_n$ become zero. While it is possible to just let this be regulated by the $i\delta$~\cite{Chen:1995pa}, it is more convenient to define renormalization coefficients for the states.
In the classic treatment~\cite{Bjorken:1970ah} one is only considering loop diagrams for single particle outgoing states. Here, we also have loop contributions for multiparticle states and must be a bit more specific, and separate two different cases of the energy denominator going to zero. 

The first case is terms  where the intermediate state $\ket{n}$ has the same particle content as the asymptotic state $\ket{a}$, with every particle having exactly the same  momentum in both states, i.e. identically $\ket{n}= \ket{a}$. By ``identically''  we  mean here that the matrix elements in Eqs.~\eqref{eq:dysonin} and~\eqref{eq:dysonout} leave, after integrating over the other intermediate states $\ket{m}, \dots$, exactly the same amount of independent momentum-conserving delta functions as there are particles in the state $\ket{n}= \ket{a}$. Thus, one is not merely integrating over the state $n$,  in a way that might include the state $\ket{n}=\ket{a}$ in a set of measure zero. Instead, for identically same states this integration includes a delta function causing even this set %
to give a finite contribution. This is always the case if $\ket{a}$ and $\ket{n}$ are single-particle states with the same quantum numbers because for a single-particle state the conservation of total momentum implies the conservation of the momentum of the single particle. But this is not necessarily the case for multiparticle states: since the overall conserved total momentum can be split in different ways between the particles, the momentum of each individual particle is individually conserved only for some diagrams, not all. 
In terms of a diagrammatic notation, the contributions with identically the same state correspond to diagrams that are 1-particle reducible by cutting one or several of the lines leading to the asymptotic state. For this first case we employ the usual procedure and remove these contributions from the perturbative series \eqref{eq:dysonin}, \eqref{eq:dysonout}. Instead, their contributions are included by adding a \emph{single-particle} renormalization 
coefficient\footnote{
It might be tempting to try to introduce a ``multi-particle renormalization coefficient'' for  any state $\ket{a}$, and determine it from a normalization condition. However, scattering amplitudes and consequently also LCWFs of multiparticle states have both  real and imaginary parts, and such a normalization condition cannot give both. 
Moreover, the LCWF of a multiparticle state is a non-trivial function of the kinematics of that state. But the normalization condition can determine only a number, not a function.
}
for each of the $N_a$ particles in the state $\ket{a}$, and writing the series as
\begin{equation}
\label{eq:dysonZ}
 \ket{a}_{\iin,\oout}  = \left[\prod_{i=1}^{N_a}\sqrt{Z_{i}} \right]
\left[
\ket{a} + {\sum_n}' \ket{n} \frac{\bra{n}\hat{V} \ket{a}}{E_a-E_n \pm i\delta}
 + {\sum_{m,n}}'\ket{n} 
 \frac{ \bra{n}\hat{V} \ket{m}}{E_a-E_n\pm i\delta}
\frac{ \bra{m}\hat{V} \ket{a}}{E_a-E_m \pm  i\delta}
+ \dots
\right],
\end{equation}
where $\sum'$ denotes the fact that the states where \emph{identically} $\ket{n}=\ket{a}, \ket{m}=\ket{a}, \dots$ are left out from the sum and the upper (lower) signs apply for the in (out) states.
The values of the renormalization coefficients are determined from the normalizations of the single-particle states: $\prescript{}{\iin,\oout}{
\bra{i}\ket{i}}_{\iin,\oout}=1$ for the single-particle state
\begin{equation}
\ket{i}_{\iin,\oout}=\sqrt{Z_i}\left[\ket{i} + {\sum_n}'\ket{n} \frac{{}\bra{n}\hat{V} \ket{i}}{E_a-E_n \pm i\delta} + \dots\right].
\end{equation}
Because the renormalization coefficient $Z_i$ is determined from a normalization condition, it can be chosen to be real, and will be the same for the in- and out-states.

The second case for the energy denominator approaching zero is the cases where the states are not identically the same: either $\ket{n}\neq\ket{a}$ or $\ket{n}=\ket{a}$ only in a set of measure zero. In fact, this can happen even for single-particle asymptotic states when one has a collinear divergence associated with the emission of a massless particle. In such a case the energy denominator between a single-particle state, e.g. $\ket{a}=\ket{q}$ and a two-particle state $\ket{n}= \ket{qg}$ reaches zero at the $\ktt^2\to 0$ limit of the phase space integration, but is not \emph{identically} zero (there is no $\delta^{(2)}(\kt)$ from the $q\to qg$ matrix element). Such collinear divergences are not cured by wave function renormalization but must be regulated by other means. In our present calculation of diffractive DIS, we also have such nonidentically zero energy denominators from soft gluon exchanges between the outgoing quarks. In this case we have an asymptotic state $\ket{a}=\ket{q\bar{q}}$ and intermediate states $\ket{n}=\ket{q\bar{q}}$. Now again the integration over the momenta in the state $\ket{n}$ can reach a value such that $E_a=E_n$, but $\ket{a}$ and $\ket{n}$ are not identically the same. In other words,  there is no delta function setting separately the momentum of every particle in $\ket{n}$ to be the same as the corresponding particle in $\ket{a}$, which would cause this set of zero measure to have a finite contribution. It is for such configurations that the correct sign of the $i\delta$ in Eqs.~\eqref{eq:dysonin}, \eqref{eq:dysonout} matters, because the integration over the intermediate states $n$ is regularized by the $i\delta$.  This pole contribution gives rise to an imaginary part in the amplitude, and the sign of $i\delta $  determines the sign of this  imaginary part.  This contribution to a loop amplitude from the pole corresponds to the imaginary part of the amplitude resulting from a cut in covariant theory.

Because the time evolution is unitary and the free states orthonormal, the multiparticle states given by Eq.~\eqref{eq:dysonZ} satisfy orthonormality conditions. To be more specific,
the in-states are orthonormal to each other,  and the out-states likewise.  For example for a quark-antiquark state (suppressing color and helicity indices)
\begin{eqnarray}
\prescript{}{\iin}{
\bra{q(\kpvec)\bar{q}(\qpvec)} \ket{q(\kvec)\bar{q}(\qvec)}_\iin 
}
&=& 
2k^+(2\pi)^3\delta^{(3)}(\kvec-\kpvec)\ 2q^+(2\pi)^3\delta^{(3)}(\qvec-\qpvec)
\\
\prescript{}{\oout}{ \bra{q(\kpvec)\bar{q}(\qpvec)}\ket{q(\kvec)\bar{q}(\qvec)}_\oout } &=& 
2k^+(2\pi)^3\delta^{(3)}(\kvec-\kpvec)\ 2q^+(2\pi)^3\delta^{(3)}(\qvec-\qpvec),
\end{eqnarray}
where the 3-dimensional delta function is defined as $\delta^{(3)}(\kvec - \kpvec) = \delta(k^+ - k^{\prime+}) \delta^{(2)}(\kt -\kt')$.
However, in- and out-states are not orthogonal to each other: e.g. the matrix element $\prescript{}{\oout}{\bra{q(\kpvec)\bar{q}(\qpvec)}\ket{q(\kvec)\bar{q}(\qvec)}_\iin} $ has both a disconnected part $\sim 2k^+(2\pi)^3\delta^{(3)}(\kvec-\kpvec)\ 2q^+(2\pi)^3\delta^{(3)}(\qvec-\qpvec)$ and a connected part $\sim 2(k^++q^+)(2\pi)^3\delta^{(3)}(\kvec+\qvec -\kpvec-\qpvec)$. The latter gives the scattering amplitude for the physical process of $q\bar{q} \to q\bar{q}$ scattering, which can happen even in the absence of an external field. The correct signs of the $i\delta$'s are crucial to get this matrix element correctly: due to the $i\delta$'s the energy denominators combine to produce the energy-conserving delta function $\delta(k^-+q^- -\smash{k'}^- - \smash{q'}^-)$ that this matrix element must, again in the absence of an external field, have.

\subsection{Fock states for diffractive DIS}

Let us now specialize the general strategy for the case of interest, namely diffractive DIS. Here our incoming Fock state is a virtual photon, and the outgoing asymptotic states (at this order) are $q\bar{q}$ and $q\bar{q} g$ states. Thus we are dealing with the following system of states: 
\begin{align}
\label{eq:fock1}
\begin{split}
    \ket{\gamma(\qvec)}_\iin =&{} \sqrt{Z_\gamma}\left[
    \ket{\gamma(\qvec)}+
    \sum_{q\bar{q}~ \text{F. states}} 
    \Psi^{\gamma^*\to q\bar{q}}_\iin
    \ket{q(\kvec_0)\bar{q}(\kvec_1) } \right. \\
&\left.+
    \sum_{q\bar{q}g~ \text{F. states}} 
        \Psi^{\gamma^*\to q\bar{q}g}_\iin
    \ket{q(\kvec_0)\bar{q}(\kvec_1) g(\kvec_2)}
+ \dots
    \right] 
\end{split}\\
\label{eq:fock2}
    \begin{split}
    \ket{q(\pvec_0)\bar{q}(\pvec_1) }_\oout =&{}
\sqrt{Z_q Z_{\bar{q}}}\left[
    \ket{q(\pvec_0)\bar{q}(\pvec_1) }
+
{\sum_{q \bar q~ \text{F. states}}}'\;     \Psi^{ q\bar{q} \to q \bar q}_\oout \ket{q(\kvec_0)\bar{q}(\kvec_1)} \right.\\
&\left.+
    \sum_{q\bar{q}g~ \text{F. states}}   
\Psi^{ q\bar{q} \to q\bar{q}g }_\oout 
    \ket{q(\kvec_0)\bar{q}(\kvec_1) g(\kvec_2) }
+ \dots
\right]
    \end{split}
\\
\label{eq:fock3}
    \ket{q(\pvec_0)\bar{q}(\pvec_1) g(\pvec_2)}_\oout =&{}
\sqrt{Z_q Z_{\bar{q}} Z_g }\left[
   \ket{q(\pvec_0)\bar{q}(\pvec_1) g(\pvec_2)}
 +
    \sum_{q\bar{q}~ \text{F. states}}   
\Psi^{ q\bar{q}g \to q\bar{q} }_\oout 
    \ket{q(\kvec_0)\bar{q}(\kvec_1)  }
+ \dots
\right],
\end{align}
where we have defined a shorthand notation for the momentum space integrations for the different states, together with the overall momentum-conserving delta functions:
\begin{equation}
\label{eq:qq_ps_momentumspace}
\begin{split}
\sum_{q\bar{q}~ \text{F. states}} & \! \! = \!
\sum_{\substack{h_0,h_1\\\alpha_0,\alpha_1}}
\left (\prod_{i=0}^{1}\int_{0}^{\infty} \frac{\ud k^+_i \theta(k^+_i)}{2k^+_i(2\pi)}\right )
2q^+2\pi\delta\left(q^+ - \sum_{j=0}^{1}k^+_j\right)
\left (\prod_{i=0}^{1}\int\!\!\frac{\ud^{2-2\varepsilon}\kt_i}{(2\pi)^{2-2\varepsilon}}\right )(2\pi)^{2-2\varepsilon}\delta^{(2-2\varepsilon)}\!\left(\qt-\sum_{j=0}^{1}\kt_j\right),
\\
\sum_{q\bar{q}g~ \text{F. states}} & \! \!\!\! = \!
\sum_{\substack{h_0,h_1\\ \sigma\\\alpha_0,\alpha_1,a}}
\left (\prod_{i=0}^{2}\int_{0}^{\infty} \frac{\ud k^+_i \theta(k^+_i)}{2k^+_i(2\pi)}\right )
2q^+2\pi\delta\left(q^+ - \sum_{j=0}^{2}k^+_j\right)
\left (\prod_{i=0}^{2}\int\!\!\frac{\ud^{2-2\varepsilon}\kt_i}{(2\pi)^{2-2\varepsilon}}\right )(2\pi)^{2-2\varepsilon}\delta^{(2-2\varepsilon)}\!\left(\qt - \sum_{j=0}^{2} \kt_j\right).
\end{split}
\end{equation}
The Fock state  decompositions of the outgoing $q\bar{q}$ and $q\bar{q}g$ states \eqref{eq:fock2}, \eqref{eq:fock3} also include an outgoing photon state among the terms denoted by $\dots$. We have not written this term explicitly, because this is not a final state that is needed in our present calculation. The phase space integration for these $\Psi^{ q\bar{q} \to \gamma^* }_\oout $ and $\Psi^{ q\bar{q}g \to \gamma^* }_\oout $ terms 
 would be given by 
\begin{equation}
\sum_{\gamma^*~ \text{F. state}}  = 
\sum_{\lambda}
\int \frac{\ud q^+ \theta(q^+)}{2q^+(2\pi)}
2q^+2\pi\delta\left(q^+ - \sum_{j}p^+_j\right)
(2\pi)^{2-2\varepsilon}\delta^{(2-2\varepsilon)}\!\left(\qt-\sum_{j}\pt_j\right),
\end{equation}
with  the summation  over $j$  going from $j=0$ to $j=1$ for the $q\bar{q}$ state and to  $j=2$ for the $q\bar{q}g$ state. Equations~\eqref{eq:fock1}--\eqref{eq:fock3} define what is meant by the \emph{light-cone wave functions} (LCWF) $\Psi^{a \to b}_{\iin/\oout}$. Comparing the Fock state expansions in terms of the LCWFs~\eqref{eq:fock1}--\eqref{eq:fock3} to the general perturbative expansion of the asymptotic states
 \eqref{eq:dysonZ} one can derive the expressions (or diagrammatical rules) for calculating the LCWFs in terms of matrix elements of the interaction part of the Hamiltonian. 
 Our calculation is leading order in the electromagnetic coupling constant, and thus $Z_\gamma=1$ at this order~\cite{Beuf:2016wdz}. Since only the QCD NLO corrections include a gluon, the gluon wave function renormalization coefficient $Z_g$ would be needed only at the following order in QCD (NNLO). The quark wave function renormalization coefficients $Z_{q},Z_{\bar{q}}$, on the other hand, are needed to one loop order. 

 The only difference between the in- and out wave functions is the sign of the $i\delta$ used to regularize the energy denominators. The orthogonality of the in- and out-states can be used to derive constraint (unitarity) relations between the different wave functions. It is, for example, easy to see that at the lowest order in perturbation theory (where the $q\bar{q}g$ state does not contibute), we have $\Psi^{q\bar{q}\to \gamma^*}_{\lo, \iin/\oout} = - \left(\Psi^{\gamma^*\to q\bar{q}}_{\lo, \iin/\oout}\right)^*$. Diagrammatically this can be interpreted in the following way.  
 Firstly, switching the asymptotic and bare states from $\gamma \to q\bar{q}$ to $  q\bar{q} \to \gamma $ turns the matrix element $\bar{u}\slashed{\varepsilon}v$ into its complex conjugate $\bar{v} \slashed{\varepsilon}^* u$. Swapping the asymptotic and bare states also exchanges the sign of the real part of the energy denominator. Extending the complex conjugation from the numerator to the whole wave function also changes the sign of the $i \delta$, so that an overall minus sign can be pulled out. 
 
The diffractive DIS cross section for a nucleus $A$ (including the proton $A=1$) is defined as
the cross section $\gamma^*+A \to X+A$,
where the diffractive system $X$ is any color-singlet hadronic state with an invariant mass $M_X$ and a total transverse momentum $\Deltat$:
\begin{equation}
    \frac{\dd[]{ \sigma^{\text{D}}_{\gamma^*_\lambda + A } }}{\dd[2]{\Deltat} \dd{M_X^2}} 
    = \sum_{\text{color-singlet }n} \int \dd{ [\mathcal{PS}]_n } \frac{\dd{ \sigma_{\gamma^*_\lambda + A \to n + A}  }}{\dd{[\mathcal{PS}]_n}} \delta(M_X^2- M_n^2)
    \delta^{(2)}\left(\Deltat - \left(\sum_i \pt_i-\qt\right)\right).
\end{equation}
At the NLO accuracy in the dipole picture we need to  sum over the $n=q\bar{q}, q\bar{q}g$ systems.
The invariant phase space measure is an integral over the single-particle  phase space elements 
\begin{equation} 
\widetilde{\ud p_i} \equiv \frac{\ud^ 2 \pt_i \ud p_i^+}{2p_i^ + (2\pi)^3}, 
\end{equation}
\begin{equation}
    \dd{[\mathcal{PS}]_n} = 2 q^+ (2\pi) \delta\left( q^+- \sum_{i \in \text{F.s.} n} p_i^+\right)  \prod_{i \in \text{F.s.} n} \widetilde{ \dd{p_i}} .
\end{equation}
The phase space-differential  cross section is given in terms of the invariant scattering amplitude as 
\begin{equation}
    \frac{\dd{  \sigma_{\gamma^*_\lambda + A \to n + A}  }}{\dd{[\mathcal{PS}]_n}}
    =  \abs{\mathcal{M}_{\gamma_\lambda^* \to n}}^2,
\end{equation}
where the amplitude $\mathcal{M}_{\gamma_\lambda^* \to n}$ corresponds to the matrix element of the asymptotic states with the target field scattering operator $\hat{S}$, stripped of the longitudinal momentum-conserving delta function:
\begin{equation}
\label{eq:amplitude_from_S-matrix}
    \prescript{}{\oout}{\bra{ \text{F.s.} n }} \hat{S} -1 \ket{ \gamma^*_\lambda}_\iin
    = 2q^+ (2\pi) \delta\left( q^+- \sum_{i \in \text{F.s.} n} p_i^+\right) i\mathcal{M}_{\gamma^*_\lambda \to n}.
\end{equation}
These conventions follow the formalism laid out in Ref.~\cite{Bjorken:1970ah}.

The diffractive cross section corresponds to color-singlet outgoing states, and thus we take the color structure of the outgoing states $\ket{ \text{F.s.} n }_\oout= \ket{q(0)\bar{q}(1)}_\oout, \ \ket{q(0)\bar{q}(1)g(2)}_\oout$ to be a normalized color-singlet state:
\begin{eqnarray}
    \ket{ q_0 \bar q_1 }^\text{singlet} &=& \frac{\delta_{\alpha_0}^{\alpha_1}}{\sqrt{\nc}} \ket{ q_0(\alpha_0) \bar q_1(\alpha_1)},
\\
    \ket{ q_0 \bar q_1 g_2 }^\text{singlet} &=& \frac{t_{\alpha_0 \alpha_1}^a}{\sqrt{\cf \nc}} \ket{ q_0(\alpha_0) \bar q_1(\alpha_1) g_2(a)}.
\end{eqnarray}
In our notations the subscript $0$ always refers to the quark, the subscript $1$ to the antiquark and the subscript $2$ to the gluon, both for momentum and color indices. For brevity of notation we will mostly not write out the color structure explicitly.

The interaction with the target color field is most naturally expressed in terms of transverse coordinates. For this purpose, it is convenient to define the mixed space (longitudinal momentum and transverse coordinate, sometimes loosely referred to simply as ``coordinate space'') Fock states  as transverse Fourier transforms
\begin{equation}
|q(k^+_i, \xt_i, h_i,\alpha_i) \rangle = \int \frac{\ud^{2-2\varepsilon}\kt_i}{(2\pi)^{2-2\varepsilon}} e^{-i\kt_i \cdot \xt_i} |q(k^+_i, \kt_i, h_i,\alpha_i \rangle.
\label{FT_Fock}
\end{equation}
The corresponding phase space sums are:  
\begin{equation}
\label{eq:defphasesp}
\begin{split}
\widetilde{\sum_{q\bar{q}~ \text{F. states}}} & = 
\left (\prod_{i=0}^{1}  \int_{0}^{\infty} \frac{\ud k^+_i \theta(k^+_i)}{2k^+_i(2\pi)}\right )
2q^+ 2\pi\delta\left(q^+ - \sum_{j=0}^{1}k^+_j\right) \left (\prod_{k=0}^{1}\int \ud^{2-2\varepsilon}\xt_k\right ),
\\
\widetilde{\sum_{q\bar{q}g~ \text{F. states}}} & = 
\left (\prod_{i=0}^{2}\int_{0}^{\infty} \frac{\ud k^+_i \theta(k^+_i)}{2k^+_i(2\pi)}\right )
2q^+2\pi\delta\left(q^+ - \sum_{j=0}^{2}k^+_j\right)\left (\prod_{k=0}^{2}\int \ud^{2-2\varepsilon}\xt_k\right ).
\end{split}
\end{equation}
Note that, unlike in Ref.~\cite{Hanninen:2017ddy} but similarly to Refs.~\cite{Beuf:2016wdz,Beuf:2022kyp} we use a notation where a factor $1/(2q^+)$ is included in the wave function, so that the phase space integration measure has a factor $2q^+$. This has the effect that a delta function $\delta(q^+-\dots)$ always appears in the combination $2q^+\delta(q^+-\dots)$.
The motivation for this convention is that the light-cone wave functions are invariant under longitudinal boosts.

For loop diagrams, one encounters divergences that need to be regularized. Similarly to Refs.~\cite{Hanninen:2017ddy,Beuf:2021qqa,Beuf:2021srj} we regularize transverse momentum integrals in both the CDR (conventional dimensional regularization) and FDH (four-dimensional helicity scheme) variants of dimensional regularization (see Ref.~\cite{Beuf:2021qqa} for a discussion). 
Here, in addition to the dimension of space for the integration variable momenta, $D=4-2\varepsilon$, one needs the ``spin dimension'' $\Ds$ of the space where the helicities live. 
For this spin dimension we use the notation $\Ds = 4-2\deltas \varepsilon$, where the value of $\deltas$ determines the variant of dimensional regularization as
\begin{equation}
    \deltas = 
    \begin{cases}
        1  & \text{ for CDR} \\
        0 & \text{ for FDH}.
    \end{cases}
\end{equation}
The dependence on the parameter $\deltas$ must cancel in the final result, which it does. 
The reason for this is that the only nonperturbative objects in this calculation are related to the Wilson lines which do not contain transverse divergences whose regularization would depend on $\deltas$. The situation  would be different for a scattering process involving  nonperturbative quantities whose renormalization absorbs large transverse momentum logarithms in a potentially scheme-dependent way, such as parton distribution functions or fragmentation functions. The cancellation of $\deltas$ provides an additional check for our calculation. The divergences of 
 the transverse integrals regularized in such  a way by dimensional regularization are  either UV or collinear divergences. 

 Moreover, we also include a lower cut-off $\alpha q^+$ on integrals over light-cone plus momenta. Indeed, various types of divergences can occur in that  regime in individual diagrams: soft divergences, rapidity divergences, or spurious divergences due to the use of LCPT.

Now we have defined all of the observables to be computed and our normalization conventions. In the next section we proceed to a classification of the different kinds of contributions to the full result.

\section{Light-cone wave functions and scattering amplitudes for diffractive DIS}
\label{sec:classification}

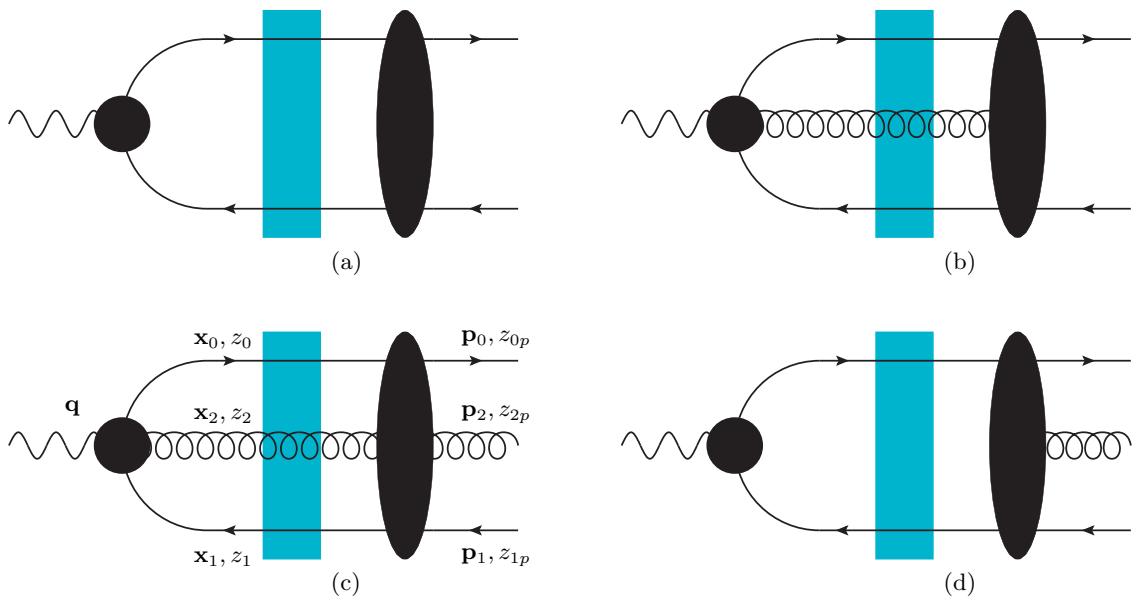
\begin{figure*}[tbp!]
\centering
\includegraphics[width=0.4\textwidth]{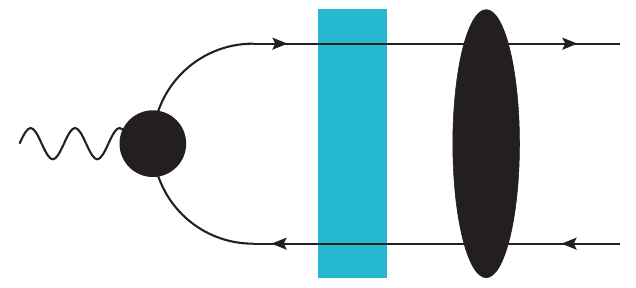}
\begin{tikzpicture}[overlay]
\node[anchor=south west] at (-3cm,-0.5cm) {\namediag{diag:gamma_qq_qq}};
\end{tikzpicture}
\rule{2em}{0pt}
\includegraphics[width=0.4\textwidth]{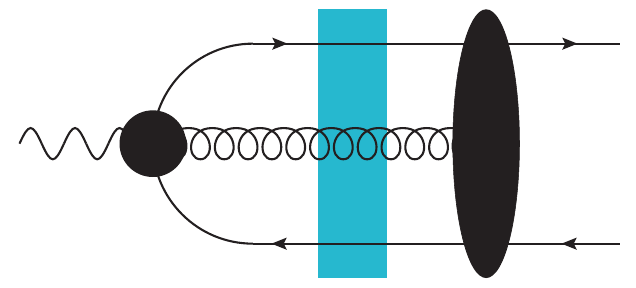}
\begin{tikzpicture}[overlay]
\node[anchor=south west] at (-3cm,-0.5cm) {\namediag{diag:gamma_qqg_qq}};
\end{tikzpicture}
\\
\vspace{1cm}
\includegraphics[width=0.4\textwidth]{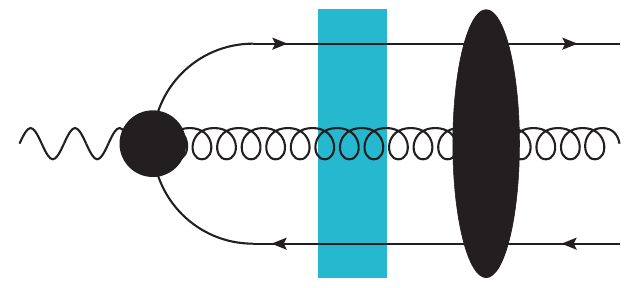}
\begin{tikzpicture}[overlay]
\node[anchor=south east] at (-6cm,1.9cm) {$\qt$};
\node[anchor=south east] at (-3.7cm,2.8cm) {$\xt_0, z_0$};
\node[anchor=south east] at (-3.7cm,-0.1cm) {$\xt_1, z_1$};
\node[anchor=south east] at (-3.7cm,1.8cm) {$\xt_2, z_2$};
\node[anchor=south east] at (0cm,2.8cm) {$\pt_0, z_{0p}$};
\node[anchor=south east] at (0cm,-0.1cm) {$\pt_1, z_{1p}$};
\node[anchor=south east] at (0cm,1.8cm) {$\pt_2, z_{2p}$};
\node[anchor=south west] at (-3cm,-0.5cm) {\namediag{diag:gamma_qqg_qqg}};
\end{tikzpicture}
\rule{2em}{0pt}
\includegraphics[width=0.4\textwidth]{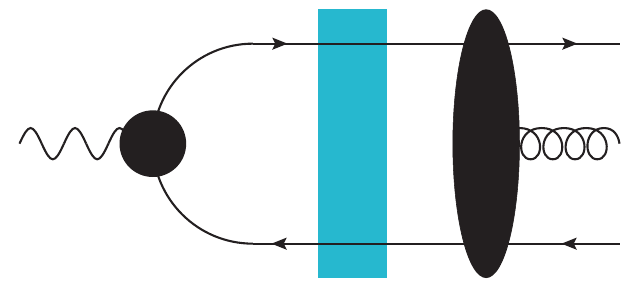}
\begin{tikzpicture}[overlay]
\node[anchor=south west] at (-3cm,-0.5cm) {\namediag{diag:gamma_qq_qqg}};
\end{tikzpicture}
\caption{
Different contributions to the invariant amplitudes. 
}\label{fig:invariant_amplitudes}
\end{figure*}

The diffractive cross section can be calculated from the invariant amplitudes $\mathcal{M}_{\gamma_\lambda^* \to n}$ for different color-singlet final states $n$. These invariant amplitudes can be further separated into cases with different Fock states interacting with the target, as depicted in Fig.~\ref{fig:invariant_amplitudes}.
These are the four different contributions to the invariant amplitudes that appear at NLO, corresponding to a $q\bar{q}$ state passing through both the target and the cut \ref{diag:gamma_qq_qq}, an additional gluon passing through the target but not the cut \ref{diag:gamma_qqg_qq}, the gluon passing through both the target and the cut \ref{diag:gamma_qqg_qqg} and the gluon passing through the cut but not the target \ref{diag:gamma_qq_qqg}. Amplitudes with the same final state are added at the amplitude level and can interfere, whereas different final states are only added at the cross section level. Thus we group these four contributions into the two amplitudes
\begin{align}
    \mathcal{M}_{q \bar q} =& \mathcal{M}_{\ref{diag:gamma_qq_qq}} + 
    \mathcal{M}_{\ref{diag:gamma_qqg_qq}}, \\
    \mathcal{M}_{q \bar qg} =& \mathcal{M}_{\ref{diag:gamma_qqg_qqg}} + 
    \mathcal{M}_{\ref{diag:gamma_qq_qqg}} .
\end{align}
The vertex $\gamma \to q \bar q$ drawn as a black blob corresponds to a general wave function $\Psi^{\gamma^* \to q \bar q}$ which includes all higher-order corrections. Similarly, the black blobs after the target depict interactions between the particles to all orders. They also include the non-interacting case. It should be noted that at the NLO accuracy, the black blob in Diagram~\ref{diag:gamma_qqg_qqg} is only needed at the non-interacting level and is shown only for consistency with the general framework.

This classification of the invariant amplitudes is very natural when thinking in terms of light-cone wave functions. All of the invariant amplitudes can be written as 
\begin{equation}
    i\mathcal{M}_n \sim \widetilde \Psi_\iin^{\gamma_\lambda^* \to m} \left(\widetilde \Psi_\oout^{n \to m} \right)^\ast
    \left(1- \hat S^{(m)}\right),
\end{equation}
where $n$ is the final state and $m$ is the intermediate state interacting with the target. The wave functions $\widetilde \Psi_\iin^{\gamma_\lambda^* \to m}$ and $\widetilde \Psi_\oout^{n \to m} $ are written in terms of momenta for the initial and final states $\gamma_\lambda^*$ and $n$, and for the intermediate state $m$ they are expressed in the mixed space of coordinates $(\xt_i,z_i)$ that is useful calculating the eikonal interaction with the target. It is also useful to factor the dependence on the overall momentum and the color factors out of the wave functions, and write the amplitudes in terms of the so-called reduced wave functions $\widetilde \psi$. The required reduced virtual photon wave functions are:
\begin{equation}
    \widetilde \Psi^{\gamma^*(q) \to q(\xt_0, z_0) \bar q(\xt_1, z_1)} = e^{i \qt \vdot \bt} \delta_{\alpha_0}^{\alpha_1} \widetilde \psi^{\gamma^* \to q \bar q}(\xt_{01}, z_i),
\end{equation}
\begin{equation}
    \widetilde \Psi^{\gamma^*(q) \to q(\xt_0, z_0) \bar q(\xt_1, z_1) g(\xt_2, z_2)} = e^{i \qt \vdot \bt} t_{\alpha_0 \alpha_1}^a \widetilde \psi^{\gamma^* \to q \bar q g}(\xt_{ij}, z_i).
\end{equation}
Similarly, we need the reduced wave functions describing the $q\bar q \to q\bar q$, $q\bar q \to q\bar q g$ and $q\bar q g \to q\bar q$ transitions:
\begin{equation}
    \widetilde \Psi^{q(p_0) \bar q(p_1) \to q(\xt_0, z_0) \bar q(\xt_1, z_1)} = e^{i \left(\sum \pt_i\right) \vdot \bt} \frac{1}{\nc} 
    C_{S,i}^{q\bar q} C_{S,f}^{q\bar q} 
    \widetilde \psi^{q \bar q \to q \bar q}(\Pt_{01}, z_{ip};\xt_{01}, z_i)
    + \octet
\end{equation}
\begin{equation}
    \widetilde \Psi^{q(p_0) \bar q(p_1) \to q(\xt_0, z_0) \bar q(\xt_1, z_1) g(\xt_2, z_2)} = e^{i \left(\sum \pt_i\right) \vdot \bt} \frac{1}{\nc}  
    C_{S,i}^{q\bar q} C_{S,f}^{q\bar q g} 
    \widetilde \psi^{q \bar q \to q \bar q g}(\Pt_{01}, z_{ip};\xt_{ij}, z_i)
    + \octet
\end{equation}
\begin{equation}
    \widetilde \Psi^{q(p_0) \bar q(p_1)g(p_2) \to q(\xt_0, z_0) \bar q(\xt_1, z_1)} = e^{i \left(\sum \pt_i\right) \vdot \bt} \frac{1}{\nc}
    C_{S,i}^{q\bar q g} C_{S,f}^{q\bar q}
   \widetilde  \psi^{q \bar q g \to q \bar q}(\Pt_{ij}, z_{ip};\xt_{01}, z_i)
    + \octet
\end{equation}
Here  the notation $\octet$ refers to the color-octet part which is not needed in our calculation, since diffractive scattering requires a color-neutral interaction with the target.
The singlet color factors are 
\begin{equation}
    \begin{aligned}
    C_{S}^{q\bar q} =& \delta_{\alpha_0}^{\alpha_1}, &
    C_{S}^{q\bar q g} =& t^a_{\alpha_0 \alpha_1}  .
    \end{aligned}
\end{equation}
Furthermore, the subscripts $i, f$ refer to whether the color indices correspond to the initial or final state. We use the following notations for the momenta and the coordinates:
\begin{equation}
    \begin{aligned}
        \Pt_{ij} &= z_{jp} \pt_i - z_{ip} \pt_j ,
        &
        \Deltat &= \sum_i \pt_i - \qt, \\
        \xt_{ij} &= \xt_i -\xt_j ,
        &
        \bt &= \sum_i z_i \xt_i, \\
        \xt_{\ov{ij}} &= \xt_{\ov i} -\xt_{\ov j} ,
        &
        \ov \bt &= \sum_i z_{\ov i} \xt_{\ov i}.
    \end{aligned}
\end{equation}
Here the barred quantities correspond to particles in the complex conjugate amplitude that appears in the cross section.
Note that the center-of-mass $\bt$  of the partonic system  is the Fourier conjugate to 
the total transverse momentum of the system (see also discussion in Ref.~\cite{Beuf:2022kyp}). 
It should be noted that the wave functions also depend on the helicities of the particles. We use a simplified notation where for each $z_i$ the corresponding helicity $h_i$ is not written out explicitly; the eikonal interaction with the target conserves the helicity as it does  $z_i$. The emitted (or absorbed) gluon always carries a polarization $\sigma$.
Sums over helicities and gluon polarizations are left implicit and can be deduced from the momentum fractions $z_i$.
The momentum fractions are defined as $z_{ip} = p_i^+ / q^+$ and $z_i \equiv z_{ik} = k_i^+ /p^+$ where $k_i$ refers to the momentum of the particle before transforming to the mixed space. The transverse coordinates $\xt_i$ are correspondingly obtained by Fourier transforming the transverse momenta $\kt_i$.

In these notations, the invariant amplitudes in Fig.~\ref{fig:invariant_amplitudes} can be written as:
\begin{equation}
\begin{split}
    i \mathcal{M}_{\ref{diag:gamma_qq_qq}} 
=&\sqrt{\nc }\int \dd[2-2\varepsilon]{\xt_{0}} \dd[2-2\varepsilon]{\xt_1} \int \frac{\dd{z_{0}} \dd{z_{1}}}{(4\pi)^2 z_{0} z_{1}} (4\pi) \delta(1-z_{0}-z_{1})
e^{i \left( \qt - \sum_i \pt_i \right) \vdot \bt } 
\left(1-\dipole_{01}\right)\\
&\times \widetilde \psi^{\gamma^*_\lambda \to q \bar q}_\iin(\xt_{01},z_i)
\left[ 
(4\pi) z_0 z_1 \delta(z_0 - z_{0p})\delta_{h_0}^{h_{0p}} \delta_{h_1}^{h_{1p}} e^{i \Pt_{01} \vdot \xt_{01}}
+\widetilde \psi_\oout^{q \bar q \to q \bar q}(\Pt_{01},z_{ip}, ; \xt_{01}, z_i) 
\right]^\ast,
\end{split}
\label{eq:amplia}
\end{equation}

\begin{equation}
\begin{split}
    i \mathcal{M}_{\ref{diag:gamma_qqg_qq}} 
=&\cf\sqrt{\nc} \int \dd[2-2\varepsilon]{\xt_0} \dd[2-2\varepsilon]{\xt_{1}} \dd[2-2\varepsilon]{\xt_{2}} \int \frac{\dd{ z_{0}} \dd{ z_{1}} \dd{ z_{2}}}{(4\pi)^3 z_{0} z_{1} z_{2}} (4\pi) \delta(1-z_{0}-z_{1} -z_{2})
 \\
&\times  \widetilde \psi_\iin^{\gamma^*_\lambda \to q \bar q g}(\xt_{ij},z_i)
\left[  \widetilde \psi_\oout^{q \bar q \to q \bar q g}(\Pt_{01}, z_{i p};  \xt_{ij}, z_i) \right]^\ast
\left(1-\tripole_{012}\right) 
e^{i \left( \qt - \sum_i \pt_i \right) \vdot \bt },
\end{split}
\label{eq:amplib}
\end{equation}

\begin{equation}
\begin{split}
    i \mathcal{M}_{\ref{diag:gamma_qqg_qqg}} 
=&\sqrt{\nc \cf} \int \dd[2-2\varepsilon]{\xt_{0}} \dd[2-2\varepsilon]{\xt_{1}} \dd[2-2\varepsilon]{\xt_2}
 e^{i\left( \qt - \sum_i \pt_i \right) \vdot \bt} \\
&\times \widetilde \psi_\iin^{\gamma^*_\lambda \to q \bar q g}( \xt_{ij},z_{ip})
e^{-i \Pt_{01} \vdot \xt_{01}  -i \Pt_{20} \vdot \xt_{20} -i \Pt_{21} \vdot \xt_{21}} 
\left(1-\tripole_{012}\right),
\end{split}
\label{eq:amplic}
\end{equation}

\begin{equation}
\begin{split}
    i \mathcal{M}_{\ref{diag:gamma_qq_qqg}}   
=&\sqrt{\nc \cf} \int \dd[2-2\varepsilon]{\xt_{0}} \dd[2-2\varepsilon]{\xt_1} \int \frac{\dd{ z_{0}} \dd{ z_{1}}}{(4\pi)^2 z_{0} z_{1}} (4\pi) \delta(1-z_{0}-z_{1})
e^{i\left( \qt - \sum_i \pt_i \right) \vdot \bt } \\
&\times \widetilde \psi_\iin^{\gamma^*_\lambda \to q \bar q}( \xt_{01},z_i)
\left[  \widetilde \psi_\oout^{q \bar q g \to q \bar q}(\Pt_{ij}, z_{ip};  \xt_{01},z_{i}) \right]^\ast
\left(1-\dipole_{01} \right).
\end{split}
\label{eq:amplid}
\end{equation}
It should be noted that in these equations we have already used the fact that the renormalization coefficients satisfy $Z_q=Z_{\bar q}=Z_g=1$ at this order in dimensional regularization with the $\varepsilon$ used for both IR and UV divergences. This feature leads to what effectively looks like cancellation between UV and IR divergences  (e.g. UV from the one loop $\gamma^*\to q\bar{q}$ vertex and IR from final state gluon radiation). However, what is actually happening physically is that the UV divergences from other diagrams cancel against the UV divergences in $Z_q,Z_{\bar q}$, and IR divergences from final state radiation against IR divergences in $Z_q,Z_{\bar q}$.

We show explicit expressions for the wave functions in Eqs.~\eqref{eq:amplia}--\eqref{eq:amplid} at the order needed for this calculation Appendix~\ref{sec:wave_functions}. However, rather than using the explicit expressions for the wave functions as a starting point, it is more convenient to work at the cross section level where many divergences cancel. The cross section can be divided into different contributions from these invariant amplitudes as
\begin{equation}
\begin{split}
    \frac{\ud \sigma^\textrm{D}_{\gamma^*_\lambda + A}}{\dd[2]{\Deltat} \dd{M_X^2} } 
    =&
    \left[ \frac{\ud \sigma^\textrm{D}_{\gamma^*_\lambda + A}}{\dd[2]{\Deltat} \dd{M_X^2} } \right]_{\abs{\ref{diag:gamma_qq_qq}}^2}
    +
    \left[ \frac{\ud \sigma^\textrm{D}_{\gamma^*_\lambda + A}}{\dd[2]{\Deltat} \dd{M_X^2} } \right]_{\abs{\ref{diag:gamma_qqg_qq}}^2}
    +
    \left[ \frac{\ud \sigma^\textrm{D}_{\gamma^*_\lambda + A}}{\dd[2]{\Deltat} \dd{M_X^2} } \right]_{\abs{\ref{diag:gamma_qqg_qqg}}^2}
    +
    \left[ \frac{\ud \sigma^\textrm{D}_{\gamma^*_\lambda + A}}{\dd[2]{\Deltat} \dd{M_X^2} } \right]_{\abs{\ref{diag:gamma_qq_qqg}}^2} \\
    &+2 \Re
    \left[ \frac{\ud \sigma^\textrm{D}_{\gamma^*_\lambda + A}}{\dd[2]{\Deltat} \dd{M_X^2} } \right]_{\ref{diag:gamma_qqg_qq}\times \ref{diag:gamma_qq_qq}^*}
    + 2 \Re
    \left[ \frac{\ud \sigma^\textrm{D}_{\gamma^*_\lambda + A}}{\dd[2]{\Deltat} \dd{M_X^2} } \right]_{\ref{diag:gamma_qqg_qqg}\times \ref{diag:gamma_qq_qqg}^*}
\end{split}
\end{equation}
Let us now consider these terms separately. First of all, the contribution $\abs{\ref{diag:gamma_qqg_qq}}^2$ is not needed at NLO as the amplitude $i \mathcal{M}_{\ref{diag:gamma_qqg_qq}}$ is already at the order $\mathcal{O}(\as)$. Second, the contribution $\abs{\ref{diag:gamma_qqg_qqg}}^2$ has already been calculated in Ref.~\cite{Beuf:2022kyp}, and it gives the term
\begin{equation}
    \left[ \frac{\dd[]{ \sigma^{\text{D}}_{\gamma^*_{\lambda} + A } }}{\dd[2]{\Deltat} \dd{M_X^2}}  \right]_{\text{trip}} 
    = 
    \left[ \frac{\dd[]{ \sigma^{\text{D}}_{\gamma^*_{\lambda} + A } }}{\dd[2]{\Deltat} \dd{M_X^2}}  \right]_{\abs{\ref{diag:gamma_qqg_qqg}}^2} 
\end{equation}
in the final result Eq.~\eqref{eq:cross_section_decomposition}.
Third, the term $\abs{\ref{diag:gamma_qq_qq}}^2$ can be separated even further: it contains the leading-order part, the NLO correction for the wave function $\Psi^{\gamma^* \to q \bar q}$, and a correction from a gluon exchange in the final state. This gluon exchange correction is related to the rest of the final-state corrections in the term $\abs{\ref{diag:gamma_qq_qqg}}^2$, and thus it is convenient to consider them together.
Similarly, the terms ${\ref{diag:gamma_qqg_qq}\times \ref{diag:gamma_qq_qq}^*}$ and ${\ref{diag:gamma_qqg_qqg}\times \ref{diag:gamma_qq_qqg}^*}$ are also related to each other such that there are cancellations between them. The term ${\ref{diag:gamma_qqg_qq}\times \ref{diag:gamma_qq_qq}^*}$ contains a UV divergence that needs to be subtracted out, and after the subtraction these cross terms are related to the ``dip-trip'' term in the final result shown in Eq.~\eqref{eq:cross_section_decomposition} by
\begin{equation}
    \left[ \frac{\dd[]{ \sigma^{\text{D}}_{\gamma^*_{\lambda} + A } }}{\dd[2]{\Deltat} \dd{M_X^2}}  \right]_{\text{dip-trip}} 
    = 
    2\Re
    \left[ \frac{\dd[]{ \sigma^{\text{D}}_{\gamma^*_{\lambda} + A } }}{\dd[2]{\Deltat} \dd{M_X^2}}  \right]_{\ref{diag:gamma_qqg_qq}\times \ref{diag:gamma_qq_qq}^*}
    +
    2\Re
    \left[ \frac{\dd[]{ \sigma^{\text{D}}_{\gamma^*_{\lambda} + A } }}{\dd[2]{\Deltat} \dd{M_X^2}}  \right]_{\ref{diag:gamma_qqg_qqg}\times \ref{diag:gamma_qq_qqg}^*}
    - \left[ \frac{\dd[]{ \sigma^{\text{D}}_{\gamma^*_{\lambda} + A } }}{\dd[2]{\Deltat} \dd{M_X^2}}  \right]_\UV.
\end{equation}
The UV subtraction should then be combined with $\abs{\ref{diag:gamma_qq_qq}}^2$ and $\abs{\ref{diag:gamma_qq_qqg}}^2$, which gives the final ``dip'' term in Eq.~\eqref{eq:cross_section_decomposition}:
\begin{equation}
    \left[ \frac{\dd[]{ \sigma^{\text{D}}_{\gamma^*_{\lambda} + A } }}{\dd[2]{\Deltat} \dd{M_X^2}}  \right]_{\text{dip}} 
    = 
    \left[ \frac{\dd[]{ \sigma^{\text{D}}_{\gamma^*_{\lambda} + A } }}{\dd[2]{\Deltat} \dd{M_X^2}}  \right]_{\abs{\ref{diag:gamma_qq_qq}}^2}
    +
    \left[ \frac{\dd[]{ \sigma^{\text{D}}_{\gamma^*_{\lambda} + A } }}{\dd[2]{\Deltat} \dd{M_X^2}}  \right]_{\abs{\ref{diag:gamma_qq_qqg}}^2}
    + \left[ \frac{\dd[]{ \sigma^{\text{D}}_{\gamma^*_{\lambda} + A } }}{\dd[2]{\Deltat} \dd{M_X^2}}  \right]_\UV.
\end{equation}
These observations are made explicit in Sec.~\ref{sec:details}.

\section{Details of the calculation}
\label{sec:details}

\subsection{Contribution ${\abs{\ref{diag:gamma_qq_qq}}^2}$: Leading order }

At leading order, we only need the invariant amplitude
\begin{equation}
\begin{split}
    &i \mathcal{M}_{\ref{diag:gamma_qq_qq}}^\lo 
    =\sqrt{\nc }\int \dd[2-2\varepsilon]{\xt_{0}} \dd[2-2\varepsilon]{\xt_1}
e^{i \left( \qt - \sum_i \pt_i \right) \vdot \bt - i \Pt_{01} \vdot \xt_{01} } 
\left(1-\dipole_{01}\right)\widetilde \psi^{\gamma^*_\lambda \to q \bar q}_{\iin,\lo}(\xt_{01},z_{ip})
\end{split}
\end{equation}
which allows us to write the leading-order cross section as
\begin{multline}
\label{eq:LO_cross_section}
     \left[\frac{\dd[]{\sigma^\textrm{D}_{\gamma^*_\lambda +A}}}{\dd[2]{\Deltat}\dd{M_X^2}} \right]^\lo_{\abs{\ref{diag:gamma_qq_qq}}^2} =
   \sum_f \frac{\nc}{(4\pi)^2}
    \int 
    \dd[2-2\varepsilon]{\xt_{0}}
    \dd[2-2\varepsilon]{\xt_{1}}
    \dd[2-2\varepsilon]{\xt_{\ov 0}} 
    \dd[2-2\varepsilon]{\xt_{\ov 1}} 
    \int_0^1  \dd[]{z_{0}} \dd[]{z_{1}} \delta(1-z_0 -z_1)
    \\
    \times \widetilde \psi_{\iin,\lo}^{\gamma^*_\lambda \to q \bar q }( \xt_{01},z_i) \left[\widetilde \psi_{\iin,\lo}^{\gamma^*_\lambda \to q \bar q }( \xt_{\overline 0 \overline 1}, z_i) \right]^*
    J_{-\varepsilon} \left( \mxbar |\xt_{01}-\xt_{\overline 0 \overline 1}| \right) \left( \frac{\mxbar}{2\pi |\xt_{01}-\xt_{\overline 0 \overline 1}|}\right)^{-\varepsilon}    
    \\
     \times
\left(1-\dipole_{01}\right)
\left(1-\dipole_{\ov 0 \ov 1}\right)^\dag 
\frac{1}{(2\pi)^{2-2\varepsilon}}e^{i \Deltat \vdot (\ov \bt -\bt)}.
\end{multline}
Note that we also sum over the helicities of the quark and antiquark which are the same in the wave functions $\widetilde \psi_{\iin,\lo}^{\gamma^*_\lambda \to q \bar q }( \xt_{01},z_i)$ and $ \left[\widetilde \psi_{\iin,\lo}^{\gamma^*_\lambda \to q \bar q }( \xt_{\overline 0 \overline 1}, z_i) \right]^*$. This results in 
\begin{equation}
\label{eq:wave_function_sum}
    \begin{split}
        &\sum_{h_0 h_1 }\widetilde \psi_{\iin,\lo}^{\gamma_\lambda^{\ast} \to q \bar q} ( \xt_{01}, z_i)
\left[\widetilde \psi_{\iin,\lo}^{\gamma_\lambda^{\ast} \to q \bar q}( \xt_{\ov 0 \ov 1},z_{ i})\right]^\ast 
= 2 \frac{\aem}{\pi} Q^2  e_f^2 \, \mathcal{F}_\lambda(z_0, z_1, z_0, z_1),
    \end{split}
\end{equation}
where $\mathcal{F}_\lambda$ is defined in Eq.~\eqref{eq:lo_photon_wf_part}.
We can then take $\varepsilon \to 0$ and read the LO result for the full cross section, Eq.~\eqref{eq:G_LO}.

\subsection{Contribution ${\abs{\ref{diag:gamma_qq_qq}}^2}$: NLO corrections to the wave function $\Psi^{\gamma^*_\lambda \to q \bar q}$}

Including the NLO corrections to the wave function $\Psi^{\gamma^*_\lambda \to q \bar q}$ we get the same result as the LO case, Eq.~\eqref{eq:LO_cross_section}, except now with the full NLO photon wave function. It turns out that the NLO corrections to the wave function $\Psi^{\gamma^*_\lambda \to q \bar q}$ factorize into the form
\begin{equation}
    \widetilde \psi_{\iin,\nlo}^{\gamma_\lambda^{\ast} \to q \bar q} ( \xt_{01}, z_i)
    = \frac{\as \cf}{2\pi }K^{\gamma^*}(\xt_{01}, z_i) \times 
    \widetilde \psi_{\iin,\lo}^{\gamma_\lambda^{\ast} \to q \bar q} ( \xt_{01}, z_i)
\end{equation}
where $K^{\gamma^*}$ has been defined in Eq.~\eqref{eq:K}.
The contribution from these corrections to the cross section can be written as
\begin{equation}
\label{eq:cross_section_photon_wf_nlo}
\begin{split}
     \left[\frac{\dd[]{\sigma^\textrm{D}_{\gamma^*_\lambda +A}}}{\dd[2]{\Deltat}\dd{M_X^2}} \right]_{\abs{\ref{diag:gamma_qq_qq}}^2}^{\gamma^* \to q\bar q  \text{ at NLO}} =
     &
    \sum_f\frac{\nc}{(4\pi)^2}
    \frac{\as \cf}{2\pi}
    \int 
    \dd[2-2\varepsilon]{\xt_{0}}
    \dd[2-2\varepsilon]{\xt_{1}}
    \dd[2-2\varepsilon]{\xt_{\ov 0}} 
    \dd[2-2\varepsilon]{\xt_{\ov 1}} 
    \int_0^1  \dd[]{z_{0}} \dd[]{z_{1}} \delta(1-z_0 -z_1)
    \\
    &\times
    \widetilde \psi_{\iin,\lo}^{\gamma^*_\lambda \to q \bar q }( \xt_{01},z_i) \left[\widetilde \psi_{\iin,\lo}^{\gamma^*_\lambda \to q \bar q }( \xt_{\overline 0 \overline 1}, z_i) \right]^*
    \left(  K^{\gamma^*}(\xt_{01}, z_i)+ K^{\gamma^*}(\xt_{\ov 0\ov 1}, z_i)  \right)
    \\
    & \times
    J_{-\varepsilon} \left(\mxbar |\xt_{01}-\xt_{\overline 0 \overline 1}| \right) \left( \frac{\mxbar}{2\pi |\xt_{01}-\xt_{\overline 0 \overline 1}|}\right)^{-\varepsilon}
 \left(1-\dipole_{01}\right)
\left(1-\dipole_{\ov 0 \ov 1}\right)^\dag
\times\frac{1}{(2\pi)^{2-2\varepsilon}}e^{i \Deltat \vdot (\ov \bt -\bt)}.
\end{split}
\end{equation}
Note that this contribution contains both UV divergences in $\varepsilon$ and divergences of the form $\log \alpha$ (where $\alpha$ regulates the gluon plus momentum) which need to cancel with the rest of the calculation.

\subsection{Contribution $\ref{diag:gamma_qqg_qq}\times \ref{diag:gamma_qq_qq}^*$}

The part of the cross section corresponding to the interference between the $\gamma^*\to q\bar{q}$ vertex correction $\ref{diag:gamma_qqg_qq}$ and the gluon crossing the shockwave $\ref{diag:gamma_qq_qq}$
can be written as
\begin{equation}
\label{eq:bxastart}
\begin{split}
 \left[ \frac{\ud \sigma^\textrm{D}_{\gamma^*_\lambda +A}}{\dd[2]{\Deltat} \dd{M_X^2} } \right]_{\ref{diag:gamma_qqg_qq} \times \ref{diag:gamma_qq_qq}^\ast }
=& \sum_f
\cf \nc \int \frac{\ud z_{ 0} \ud z_{ 1} \dd{z_{ 2}}}{(4\pi)^3 z_{ 0} z_{ 1} z_{ 2} } 4\pi \delta(1-z_{ 0}-z_{ 1}-z_{ 2}) 
\int \frac{\ud z_{\ov 0} \ud z_{\ov 1}}{(4\pi)^2 z_{\ov 0} z_{\ov 1}} 4\pi \delta(1-z_{\ov 0}-z_{\ov 1}) \\
&\times 
\int  \frac{\ud^{2-2\varepsilon} \Pt_{01} }{(2\pi)^{2-2\varepsilon}} 
 \delta \left(M_X^2 - \frac{\Pt_{01}^2}{z_{\ov 0} z_{\ov 1}} \right) 
 \int 
 \dd[2-2\varepsilon]{\xt_0}
 \dd[2-2\varepsilon]{\xt_1}
 \dd[2-2\varepsilon]{\xt_2}
 \dd[2-2\varepsilon]{\xt_{\ov 0} }
 \dd[2-2\varepsilon]{\xt_{\ov 1}}
\\
&\times 
\widetilde \psi_\iin^{\gamma_\lambda^* \to q \bar q g}( \xt_{ij},z_i) 
\left[\widetilde \psi_\oout^{q \bar q \to q \bar q g}(\Pt_{01}, z_{\ov i };  \xt_{ i  j},z_i)\right]^\ast
e^{i \Pt_{01} \vdot \xt_{\ov 0 \ov 1}}
\left[\widetilde \psi_\iin^{\gamma_\lambda^{\ast} \to q \bar q } (z_{\ov i}, \xt_{\ov 0 \ov 1})\right]^\ast\\
&\times 
\left(1-\tripole_{012}\right)
\left(1-\dipole_{\ov 0 \ov 1 }\right)^\dag 
\times \frac{1}{(2\pi)^{2-2\varepsilon}} e^{i \Deltat \vdot (\ov \bt -\bt)}
\end{split}
\end{equation}
When the gluon crossing the shockwave is emitted and absorbed by the same particle, there are UV divergences in $\varepsilon$ that are related to the gluon having the same transverse coordinate as the emitting particle (i.e. $\xt_2 \to \xt_0$ or $\xt_2 \to \xt_1$).
Although this means that we strictly speaking cannot evaluate Eq.~\eqref{eq:bxastart} in $D=4$ dimensions, we may first subtract the UV-divergent part out, after which the resulting subtracted expression is UV finite and we can take $\varepsilon \to 0$.
Anticipating the UV subtraction, we show here Eq.~\eqref{eq:bxastart} in $D=4$ dimensions with the momentum integrals done.
Substituting the wave function $\widetilde \psi_\oout^{q \bar q \to q \bar q g}$ from Eq.~\eqref{eq:wf_qq_to_qqg_red}, this is given by:
\begin{equation}
\label{eq:ab}
\begin{split}
 \left[ \frac{\ud \sigma^\textrm{D}_{\gamma^*_\lambda +A}}{\dd[2]{\Deltat} \dd{M_X^2} } \right]_{\ref{diag:gamma_qqg_qq} \times \ref{diag:gamma_qq_qq}^\ast } 
=& \sum_f
\cf \nc \int \frac{\ud z_{ 0} \ud z_{ 1} \dd{z_{ 2}}}{(4\pi)^3 z_{ 0} z_{ 1} z_{ 2} } (4\pi) \delta(1-z_{ 0}-z_{ 1}-z_{ 2}) \int \frac{\ud z_{\ov 0} \ud z_{\ov 1}}{(4\pi)^2 z_{\ov 0} z_{\ov 1}} 4\pi \delta(1-z_{\ov 0}-z_{\ov 1}) \\
 &\times
 \int 
 \dd[2]{\xt_0}
 \dd[2]{\xt_1}
 \dd[2]{\xt_{2}}
 \dd[2]{\xt_{\ov  0} }
 \dd[2]{\xt_{ \ov 1}}
\times 
\left(1-\tripole_{012}\right)
\left(1-\dipole_{\ov 0 \ov 1 }\right)^\dag
\times \frac{1}{(2\pi)^2} e^{i \Deltat \vdot (\ov \bt -\bt)}
\\
&\times 
\widetilde \psi_\iin^{\gamma_\lambda^* \to q \bar q g}( \xt_{ij},z_i) 
\left[\widetilde \psi_\iin^{\gamma_\lambda^{\ast} \to q \bar q } ( \xt_{\ov 0 \ov 1},z_{\ov i})\right]^\ast
\times g \varepsilon^{j}_{\sigma}
\times \frac{-i}{2\pi} 
   \\
    &\times \Bigg[ 
     z_{1}^2 \delta(z_{\ov 1}-z_{ 1}) \delta_{h_{1}}^{h_{\ov 1}}
       \frac{\xt_{2 0}^i}{\xt_{ 2  0}^2}
     \times J_0\left( \mxbar[0] 
     \abs{\xt_{0+2;1} - \xt_{\ov 0 \ov 1} }\right)
     S^{ij}_0(z_{\ov 0},  z_{0})\\
    &\hspace{0.5em}- z_{0}^2 \delta(z_{\ov 0}-z_0) \delta_{h_{0}}^{h_{\ov 0}}
      \frac{\xt_{21}^i}{\xt_{21}^2}
     \times J_0\left( \mxbar[1] 
     \abs{ (\xt_{\ov 0 \ov 1} -\xt_{0;1+2} }\right) 
     S^{ij}_1(z_1,  z_{\ov 1})
    \Bigg]
\end{split}
\end{equation}
where $\mxbar[0]$ and $\mxbar[1]$ have been defined in Eq.~\eqref{eq:shorthand_notations},
and the spinor terms are given by
\begin{equation}
\label{eq:spinor0}
    S^{ij}_0(z_L,  z_R) =  \frac{1}{2q^+}\bar u(k_L^+) \gamma^+\left\{ 
    (z_L + z_R) \delta^{ij}   +\frac{1}{2} (z_L -z_R) \left[ \gamma^i, \gamma^j \right] 
    \right\}u(k_R^+) ,
\end{equation}
\begin{equation}
\label{eq:spinor1}
    S^{ij}_1(z_L,  z_R) = \frac{1}{2q^+} \bar v(k_L) \gamma^+ \left\{ 
    (z_L + z_R) \delta^{ij}  +\frac{1}{2} (z_L -z_R ) \left[ \gamma^i, \gamma^j \right] 
    \right\} v(k_R^+).
\end{equation}
Note that $S_0$ and $S_1$ also depend on the helicities of the spinors, but not on the transverse momenta.
They are also boost invariant, meaning that they do not depend on the plus momenta $k_i^+$ but only on the momentum fractions $z_i = k_i^+ /  q^+$ instead.

Equation~\eqref{eq:ab} is divergent when integrated over $\xt_2$ and hence must be accompanied by the UV subtraction term at the integrand level.
The UV divergences come from the same particle emitting and absorbing the gluon, and hence
a suitable UV subtraction term can be figured out by considering an emission from the quark and the antiquark separately. 
For a gluon emission from the quark (antiquark), we can take the UV limit $\xt_2 \to \xt_0$ ($\xt_2 \to \xt_1$) in the photon wave function $\widetilde \psi_\iin^{\gamma^* \to q \bar q g}$ by using Eq.~\eqref{eq:wf_UV_quark} (Eq.~\eqref{eq:wf_UV_antiquark}) as described in Appendix~\ref{sec:UV_divergence}.
Similarly, 
we replace 
$\tripole_{012} \to \dipole_{01} $, and 
for the wave function $\widetilde \psi_\oout^{q \bar q \to q \bar q g}$ we take $\xt_{0+2;1} \to \xt_{01}$ and $\xt_{0;1+2} \to \xt_{01}$ in the exponent in Eq.~\eqref{eq:wf_qq_to_qqg_red}, which also corresponds to the UV limit  $\xt_2 \to \xt_0$ or $\xt_2 \to \xt_1$.
At the amplitude level, the UV-divergent part  for a gluon emission from the quark then reads

\begin{equation}
\begin{split}
    i \mathcal{M}^{q\to q g }_{\ref{diag:gamma_qqg_qq}_\text{UV}}  
=&\cf\sqrt{\nc}\int \dd[2-2\varepsilon]{\xt_{01}}\dd[2-2\varepsilon]{\xt_{20} }\dd[2-2\varepsilon]{\bt} 
\int \frac{\dd{ z_{0}} \dd{ z_{1}} \dd{ z_{2}}}{(4\pi)^3 z_{0} z_{1} z_{2}} (4\pi) \delta(1-z_{0}-z_{1} -z_{2})
e^{-i \left( \qt - \sum_i \pt_i \right) \vdot \bt } \\
&\times \widetilde \psi^{\gamma_\lambda^* \to q \bar q g}_{\iin,\ref{diag:X_q_qg},\text{UV}}(\xt_{ij},z_i)
\left[  \widetilde \psi^{q \bar q \to q \bar q g}_{\oout,\ref{diag:qqbar_qqbarg_a},\text{UV}}( \Pt_{01},z_{ip}; \xt_i,z_i) \right]^\ast
\left(1-\dipole_{01}\right)
\exp(-\frac{\xt_{20}^2}{\xt_{01}^2 e^{\gamma_E}}) \\
=&\frac{\as \cf}{2\pi}\sqrt{\nc}\int \dd[2-2\varepsilon]{\xt_{01}}\dd[2-2\varepsilon]{\xt_{20} }\dd[2-2\varepsilon]{\bt} 
\int\dd{ z_{0}} \dd{ z_{1}} \dd{ z_{2}} \delta(1-z_{0}-z_{1} -z_{2})
e^{-i \left( \qt - \sum_i \pt_i \right) \vdot \bt } \\
&\times 
\widetilde \psi^{\gamma_\lambda^* \to q \bar q}_{\iin}(\xt_{ij},z_{ip})
e^{-i \Pt_{01} \vdot \xt_{01}}
\left(1-\dipole_{01}\right)
 \\
&\times  \frac{1}{2\pi} \delta(z_1-z_{1p}) \frac{2}{z_2 z_{0p}^2} \left[ z_{0}^2+z_{0p}^2-\varepsilon \deltas z_2^2\right]\times \Gamma(1-\varepsilon)^2 \times 
 \frac{1}{\xt_{20}^2} ( \mu \pi \xt_{20}^2)^{2\varepsilon} \times \exp(-\frac{\xt_{20}^2}{\xt_{01}^2 e^{\gamma_E}}) \\
=&\frac{\as \cf}{2\pi}\sqrt{\nc}\int \dd[2-2\varepsilon]{\xt_{01}}\dd[2-2\varepsilon]{\bt} 
e^{-i \left( \qt - \sum_i \pt_i \right) \vdot \bt }
\widetilde \psi^{\gamma_\lambda^* \to q \bar q}_{\iin}(\xt_{ij},z_{ip})
e^{-i \Pt_{01} \vdot \xt_{01}}
\left(1-\dipole_{01}\right)\\
&\times \Gamma(\varepsilon) \Gamma(1-\varepsilon) (\pi \mu^2 e^{\gamma_E} \xt_{01}^2)^{\varepsilon} \left[ -\frac{3}{2} +2 \log( \frac{z_{0p}}{\alpha})- \frac{1}{2} \varepsilon \deltas \right].
\end{split}
\end{equation}
 Here we have introduced the factor $\exp(-\xt_{20}^2/(\xt_{01}^2 e^{\gamma_E}))$ to regulate the $\xt_{20}$-integral in the IR region and prevent the UV-subtraction from introducing a separate IR divergence. There is no unique choice for the precise functional form chosen here, as long as the UV limit $\xt_{20}\to 0$ is the correct one. Since this term is subtracted from the $\ref{diag:gamma_qqg_qq}\times \ref{diag:gamma_qq_qq}^*$ contribution and then added to the other terms, the choice of scheme does not affect the final result.   This particular choice leads to simple expressions as it also subtracts all of the $\alpha$-divergences except for the divergence that will be related to the JIMWLK equation. It should be noted that this scheme for the UV subtraction is the same as the one used in Ref.~\cite{Hanninen:2017ddy}, and is discussed in more detail in Appendix E of Ref.~\cite{Hanninen:2017ddy}.
The contribution with a gluon emission from the antiquark is, correspondingly,
\begin{equation}
\begin{split}
    i \mathcal{M}^{\bar q \to \bar  q g }_{\ref{diag:gamma_qqg_qq}_\text{UV}}  
=&\cf\sqrt{\nc}\int \dd[2-2\varepsilon]{\xt_{01}}\dd[2-2\varepsilon]{\xt_{20} }\dd[2-2\varepsilon]{\bt} \int \frac{\dd{ z_{0}} \dd{ z_{1}} \dd{ z_{2}}}{(4\pi)^3 z_{0} z_{1} z_{2}} (4\pi) \delta(1-z_{0}-z_{1} -z_{2})
e^{-i \left( \qt - \sum_i \pt_i \right) \vdot \bt } \\
&\times \widetilde \psi^{\gamma_\lambda^* \to q \bar q g}_{\iin,\ref{diag:X_qbar_qbarg},\text{UV}}(\xt_{ij},z_i)
\left[  \widetilde \psi^{q \bar q \to q \bar q g}_{\oout,\ref{diag:qqbar_qqbarg_b},\text{UV}}(\Pt_{01},z_{ip}; \xt_i,z_i) \right]^\ast
\left(1-\dipole_{01}\right)
\exp(-\frac{\xt_{21}^2}{\xt_{01}^2 e^{\gamma_E}}) \\
=&\frac{\as \cf}{2\pi}\sqrt{\nc}\int \dd[2-2\varepsilon]{\xt_{01}}\dd[2-2\varepsilon]{\xt_{20} }\dd[2-2\varepsilon]{\bt} 
\int\dd{ z_{0}} \dd{ z_{1}} \dd{ z_{2}} \delta(1-z_{0}-z_{1} -z_{2})
e^{-i \left( \qt - \sum_i \pt_i \right) \vdot \bt } \\
&\times 
\widetilde \psi^{\gamma_\lambda^* \to q \bar q}_{\iin}(\xt_{ij},z_{ip})
e^{-i \Pt_{01} \vdot \xt_{01}}
\left(1-\dipole_{01}\right)
 \\
&\times  \frac{1}{2\pi} \delta(z_0-z_{0p}) \frac{2}{z_2 z_{1p}^2} \left[ z_{1}^2+z_{1p}^2-\varepsilon \deltas z_2^2\right]\times \Gamma(1-\varepsilon)^2 \times 
 \frac{1}{\xt_{21}^2} ( \mu \pi \xt_{21}^2)^{2\varepsilon} \times \exp(-\frac{\xt_{21}^2}{\xt_{01}^2 e^{\gamma_E}}) \\
=&\frac{\as \cf}{2\pi}\sqrt{\nc}\int \dd[2-2\varepsilon]{\xt_{01}}\dd[2-2\varepsilon]{\bt} 
e^{-i \left( \qt - \sum_i \pt_i \right) \vdot \bt }
\widetilde \psi^{\gamma_\lambda^* \to q \bar q}_{\iin}(\xt_{ij},z_{ip})
e^{-i \Pt_{01} \vdot \xt_{01}}
\left(1-\dipole_{01}\right)\\
&\times \Gamma(\varepsilon) \Gamma(1-\varepsilon) (\pi \mu^2 e^{\gamma_E} \xt_{01}^2)^{\varepsilon} \left[ -\frac{3}{2} +2 \log( \frac{z_{1p}}{\alpha})- \frac{1}{2} \varepsilon \deltas \right].
\end{split}
\end{equation}
Note that in this UV-limit expression the amplitude does not depend on the Wilson lines at the transverse coordinate of the gluon $\xt_2$, which can then be integrated over analytically. 
At the cross section level, the UV-divergent part integrated over the gluon coordinate is given by
\begin{equation}
\label{eq:UV_divergence_cross_section}
\begin{split}
    \left[\frac{\dd[]{\sigma^\textrm{D}_{\gamma^*_\lambda +A}}}{\dd[2]{\Deltat}\dd{M_X^2}} \right]_{\ref{diag:gamma_qqg_qq}_\text{UV} \times \ref{diag:gamma_qq_qq}^* + 
    \ref{diag:gamma_qq_qq} \times \ref{diag:gamma_qqg_qq}_\text{UV}^*
    }
    =& \sum_f
    \frac{\nc}{(4\pi)^2} \frac{\as \cf}{2\pi} \int 
    \dd[2-2\varepsilon]{\xt_{0}}
    \dd[2-2\varepsilon]{\xt_{1}}
    \dd[2-2\varepsilon]{\xt_{\ov 0}} 
    \dd[2-2\varepsilon]{\xt_{\ov 1}} 
    \int_0^1  \dd[]{z_{0}} \dd[]{z_{1}} \delta(1-z_0 -z_1)\\
    &\times \widetilde \psi_\iin^{\gamma^*_\lambda \to q \bar q }( \xt_{01},z_i) \left[\widetilde \psi_\iin^{\gamma^*_\lambda \to q \bar q }( \xt_{\overline 0 \overline 1}, z_i) \right]^* 
    J_{-\varepsilon} \left( \mxbar |\xt_{01}-\xt_{\overline 0 \overline 1}| \right) \left( \frac{\mxbar}{2\pi |\xt_{01}-\xt_{\overline 0 \overline 1}|}\right)^{-\varepsilon}\\
    &\times 
\left(1-\dipole_{01}\right)
\left(1-\dipole_{\ov 0 \ov 1}\right)^\dag 
\times\frac{1}{(2\pi)^{2-2\varepsilon}}e^{i \Deltat \vdot (\ov \bt -\bt)} 
    \\
    &\times  \Gamma(\varepsilon) \Gamma(1-\varepsilon) (\pi \mu^2 e^{\gamma_E} )^{\varepsilon} \left[ -3 +2 \log  \frac{z_0 z_{1}}{\alpha^2} - \varepsilon \deltas \right]
    \times \left[ (\xt_{01}^2)^\varepsilon + (\xt_{\overline 0 \overline 1}^2)^\varepsilon \right].
\end{split}
\end{equation}
Note that this has the structure of the LO part, Eq.~\eqref{eq:LO_cross_section}, times a divergent NLO part, in particular containing only dipole Wilson line operators. Thus it is in a form similar to the $| \ref{diag:gamma_qq_qq}|^2$ term and can be 
combined with the rest of the calculation.

In  order to subtract the UV divergent contribution from the full equation for the term ${\ref{diag:gamma_qqg_qq} \times \ref{diag:gamma_qq_qq}^\ast }$, Eq.~\eqref{eq:ab}, we need it in a form that is not integrated over the gluon coordinate, in contrast to the integrated form~\eqref{eq:UV_divergence_cross_section}. 
Here we will use the fact that 
the UV-subtracted equation
is UV finite and  can  be calculated it in $D=4$ dimensions,
making the resulting integrals easier.
The UV-subtraction terms in $D=4$ dimensions,  with the gluon coordinate not integrated over,
read
\begin{equation}
\begin{split}
 \left[ \frac{\ud \sigma^\textrm{D}_{\gamma^*_\lambda +A}}{\dd[2]{\Deltat} \dd{M_X^2} } \right]_{\ref{diag:gamma_qqg_qq}_\text{UV} \times \ref{diag:gamma_qq_qq}^\ast }^{q \to q g} 
=& \sum_f
\frac{\nc}{(4\pi)^2} \frac{\as \cf}{2\pi} \int \ud z_{ 0} \ud z_{ 1} \dd{z_{ 2}}  \delta(1-z_{ 0}-z_{ 1}-z_{ 2}) \int \ud z_{\ov 0} \ud z_{\ov 1}\delta(1-z_{\ov 0}-z_{\ov 1}) \\
&\times 
 \int 
 \dd[2]{\xt_0}
 \dd[2]{\xt_1}
 \dd[2]{\xt_{2}}
 \dd[2]{\xt_{\ov 0} }
 \dd[2]{\xt_{\ov 1}}
\times 
\left(1-\dipole_{01}\right)
\left(1-\dipole_{\ov 0 \ov 1 }\right)^\dag
\times \frac{1}{(2\pi)^2} e^{i \Deltat \vdot (\ov \bt -\bt)}
\\
&\times 
\widetilde \psi_\iin^{\gamma_\lambda^* \to q \bar q}( \xt_{01},z_{\ov i}) 
\left[\widetilde \psi_\iin^{\gamma_\lambda^{\ast} \to q \bar q } ( \xt_{\ov 0 \ov 1},z_{\ov i})\right]^\ast
\times \delta(z_1-z_{\ov 1}) \frac{1}{\pi z_2 z_{\ov 0}^2} \left[ z_{0}^2+z_{\ov 0}^2\right] \times 
 \frac{1}{\xt_{20}^2} \times \exp(-\frac{\xt_{20}^2}{\xt_{01}^2 e^{\gamma_E}}) 
\end{split}
\end{equation}
and
\begin{equation}
\begin{split}
 \left[ \frac{\ud \sigma^\textrm{D}_{\gamma^*_\lambda +A}}{\dd[2]{\Deltat} \dd{M_X^2} } \right]_{\ref{diag:gamma_qqg_qq}_\text{UV} \times \ref{diag:gamma_qq_qq}^\ast }^{\bar q \to \bar q g} 
=& \sum_f
\frac{\nc}{(4\pi)^2} \frac{\as \cf}{2\pi} \int \ud z_{ 0} \ud z_{ 1} \dd{z_{ 2}}  \delta(1-z_{ 0}-z_{ 1}-z_{ 2}) \int \ud z_{\ov 0} \ud z_{\ov 1}\delta(1-z_{\ov 0}-z_{\ov 1}) \\
&\times 
 \int 
 \dd[2]{\xt_0}
 \dd[2]{\xt_1}
 \dd[2]{\xt_{2}}
 \dd[2]{\xt_{\ov  0} }
 \dd[2]{\xt_{ \ov 1}}
\times 
\left(1-\dipole_{01}\right)
\left(1-\dipole_{\ov 0 \ov 1 }\right)^\dag
\times \frac{1}{(2\pi)^2} e^{i \Deltat \vdot (\ov \bt -\bt)}
\\
&\times 
\widetilde \psi_\iin^{\gamma_\lambda^* \to q \bar q}( \xt_{01},z_{\ov i}) 
\left[\widetilde \psi_\iin^{\gamma_\lambda^{\ast} \to q \bar q } ( \xt_{\ov 0 \ov 1},z_{\ov i})\right]^\ast
\times \delta(z_0-z_{\ov 0}) \frac{1}{\pi z_2 z_{\ov 1}^2} \left[ z_{1}^2+z_{\ov 1}^2\right] \times 
 \frac{1}{\xt_{21}^2} \times \exp(-\frac{\xt_{21}^2}{\xt_{01}^2 e^{\gamma_E}}).
\end{split}
\end{equation}
These  give us the subtraction term of Eq.~\eqref{eq:G_dipsub} in the final result after integrating over $z_{\ov 0}$ and $z_{\ov 1}$. 
Note that in both these expressions  and  the unsubtracted result~\eqref{eq:ab} the  integrals over the transverse dipole sizes $\xt_{2i}$ are  divergent. Thus strictly speaking they cannot be taken in $D=4$ dimensions separately, but only when combined with each other  in the full finite result~\eqref{eq:diptrip}.

We could open the expressions for the wave functions $\widetilde \psi_\iin^{\gamma_\lambda^* \to q \bar q g}$ and $\left[\widetilde \psi_\iin^{\gamma_\lambda^{\ast} \to q \bar q } ( \xt_{\ov 0 \ov 1},z_{\ov i})\right]^\ast $  appearing in Eq.~\eqref{eq:ab} at this point, but as it turns out that this can combined with the contribution ${\ref{diag:gamma_qqg_qqg} \times \ref{diag:gamma_qq_qqg}^\ast } $ we will leave this for Sec.~\ref{sec:cd}. Also, it should be noted that even after combining this with the UV subtraction the resulting expression still contains a $\log \alpha$ divergence. This is related to the JIMWLK equation for the Wilson lines and will be discussed in Sec.~\ref{sec:JIMWLK}.

\subsection{ Contribution $ \ref{diag:gamma_qqg_qqg} \times \ref{diag:gamma_qq_qqg}^\ast$}
\label{sec:cd}

The cross section for the contribution $ \ref{diag:gamma_qqg_qqg} \times \ref{diag:gamma_qq_qqg}^\ast$, with a gluon crossing the cut and being emitted  before the shockwave in the amplitude and after it in the conjugate, reads
\begin{equation}
\label{eq:c_times_d}
\begin{split}
 \left[ \frac{\dd{ \sigma^\textrm{D}_{\gamma^*_\lambda +A}}}{\dd[2]{\Deltat} \dd{M_X^2} } \right]_{ \ref{diag:gamma_qqg_qqg} \times \ref{diag:gamma_qq_qqg}^\ast}  
=&\sum_f \nc \cf
\int \frac{\dd{z_{0}}\dd{z_{1}}\dd{z_{2}}}{(4\pi)^3 z_{0} z_{1} z_{2}}  (4\pi) \delta(1-z_{0}-z_{1}-z_{2})
\int \frac{\dd{z_{\ov 0}}\dd{z_{\ov 1}}}{(4\pi)^2 z_{\ov 0} z_{\ov 1} }  (4\pi) \delta(1-z_{\ov 0}-z_{\ov 1})\\
&\times \int 
\dd[2-2\varepsilon]{\xt_{0}} 
\dd[2-2\varepsilon]{\xt_{ 1} } 
\dd[2-2\varepsilon]{\xt_{2}} 
\dd[2-2\varepsilon]{\xt_{\ov 0} }
\dd[2-2\varepsilon]{\xt_{\ov 1} }
\times  
\left(1-\tripole_{012}\right)
\left(1-\dipole_{\ov 0 \ov 1}\right)^\dag 
\times e^{i \Deltat \vdot (\ov \bt - \bt)}
 \\
& \times 
\int
\prod_{i=0,1,2}
\left[
\frac{\dd[2-2\varepsilon] \pt_i}{(2\pi)^{2-2\varepsilon}}  \right] \delta^{(2-2\varepsilon)}\left(\Deltat - \left( \sum_i \pt_i - \qt \right) \right)
\delta\left(M_X^2 -  \left[ 
\sum_i \frac{\pt_i^2}{z_i} - \left( \sum_i \pt_i \right)^2
\right] \right) \\
&\times e^{
-i\Pt_{01} \vdot  \xt_{01}
-i\Pt_{20} \vdot  \xt_{20}
-i\Pt_{21} \vdot  \xt_{21}}
\times \widetilde \psi_\iin^{\gamma_\lambda^{\ast} \to q \bar q g} ( \xt_{ij},z_i)
\left[\widetilde \psi_\iin^{\gamma_\lambda^{\ast} \to q \bar q}( \xt_{\ov 0 \ov 1},z_{\ov i})\right]^\ast
\widetilde \psi_\oout^{q \bar q g  \to q \bar q }( \Pt_{ij},z_{i};  \xt_{\ov 0 \ov 1},z_{\ov i}).
\end{split}
\end{equation}
Here the momentum integral is written in terms of the momenta $\pt_i$ in the final state.
It is useful to change the momentum variables in the integral such that one is chosen as the total momentum $\sum_i \pt_i$, and the remaining variables are relative momenta such as $\Pt_{ij}$.
\begin{figure*}[tbp!]
\newdiag{diag:cd}
\centerline{\includegraphics[width=4cm]{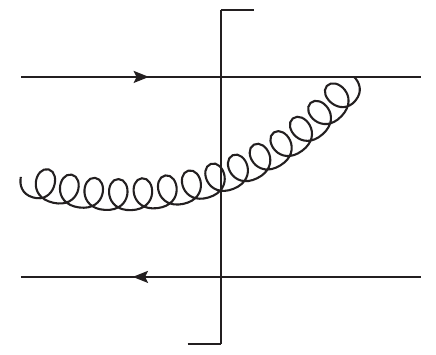}
\begin{tikzpicture}[overlay]
\node[anchor=south west] at (-3cm,-0.5cm) {\namediag{diag:cd_quark}};
            \draw [to-to](-0.4cm,0.8cm) -- (-0.4cm,2.4cm);
            \draw [to-to](-3.5cm,1.7cm) -- (-3.5cm,2.4cm);
         \node[anchor=south east] at (-3.5cm,1.8cm) {$\Pt_{20}$};
         \node[anchor=south east] at (-0.3cm,1.3cm) {$\Kt_{01}$};
         \node[anchor=south east] at (-2.0cm,2.5cm) {$p_{0}$};
         \node[anchor=south east] at (-2.0cm,0.7cm) {$p_{1}$};
         \node[anchor=south east] at (-2.0cm,1.7cm) {$p_{2}$};
\end{tikzpicture}
\rule{2em}{0pt}
\includegraphics[width=4cm]{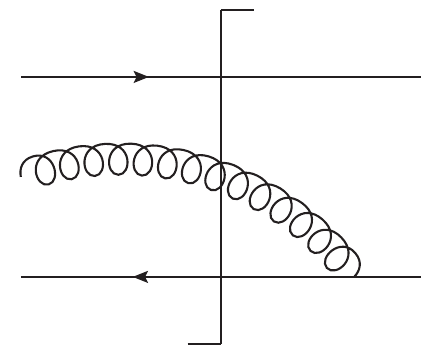}
\begin{tikzpicture}[overlay]
\node[anchor=south west] at (-3cm,-0.5cm) {\namediag{diag:cd_antiquark}};
            \draw [to-to](-0.4cm,0.8cm) -- (-0.4cm,2.4cm);
            \draw [to-to](-3.5cm,0.8cm) -- (-3.5cm,1.5cm);
         \node[anchor=south east] at (-3.5cm,0.9cm) {$\Pt_{21}$};
         \node[anchor=south east] at (-0.3cm,1.3cm) {$\Kt_{01}$};
         \node[anchor=south west] at (-2.0cm,2.5cm) {$p_{0}$};
         \node[anchor=south west] at (-2.0cm,0.7cm) {$p_{1}$};
         \node[anchor=south west] at (-2.0cm,1.7cm) {$p_{2}$};
\end{tikzpicture}
}
\caption{
Diagrams for the final-state contributions in the term $ \ref{diag:gamma_qqg_qqg} \times \ref{diag:gamma_qq_qqg}^\ast$ at next-to-leading order. The cut represents the final state.
}\label{fig:cd}
\end{figure*}
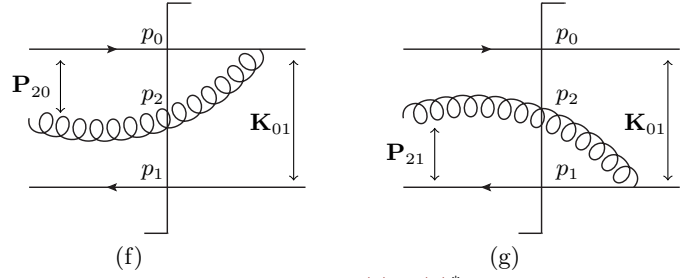
The best choice for the relative momenta depends on the exact form of the integrand.
It turns out that the integral is easiest to do by writing the expression in terms of the relative momentum of the $q \bar q$ state before the gluon emission, $\Kt_{01}$, and the relative momentum of the $q g$ or $\bar q g$ pair depending on which particle emits the gluon, see Fig.~\ref{fig:cd}.
If the gluon is emitted from the quark (Diagram~\ref{diag:cd_quark}),
we have $\Kt_{01} = z_{1p} (\pt_0 + \pt_2) - (z_{0p} + z_{2p}) \pt_1$ 
and the momentum integral can be written as
\begin{multline}
   \int
\prod_{i=0,1,2}
\left[
\frac{\dd[2] \pt_i}{(2\pi)^{2}}  \right] \delta^{(2)}\left(\Deltat - \left( \sum_i \pt_i - \qt \right) \right)
\delta\left(M_X^2 -  \left[ 
\sum_i \frac{\pt_i^2}{z_i} - \left( \sum_i \pt_i \right)^2
\right] \right) \\
= \frac{1}{(2\pi)^2 (1-z_1)^2}
\int \frac{\dd[2]{\Pt_{20}} \dd[2]{\Kt_{01}}}{(2\pi)^{4}} 
    \delta\left(M_X^2 -  \left[ 
    \frac{\Kt_{01}^2}{z_1 (1-z_1)} + \frac{\Pt_{20}^2}{z_0 z_2 (1-z_1)}
    \right] \right)
.
\end{multline}
Note that these are final state emissions, thus the momenta are the same in the amplitude and the conjugate and do not have a bar, while the coordinates in Eq.~\eqref{eq:c_times_d} are separate in the amplitude and the conjugate.
For the antiquark emitting the gluon (Diagram~\ref{diag:cd_antiquark}) we have 
$\Kt_{01} = (z_{1p} + z_{2p}) \pt_0 - z_{0p} (\pt_1 + \pt_2)$, and the integral becomes
\begin{multline}
    \int
\prod_{i=0,1,2}
\left[
\frac{\dd[2] \pt_i}{(2\pi)^{2}}  \right] \delta^{(2)}\left(\Deltat - \left( \sum_i \pt_i - \qt \right) \right)
\delta\left(M_X^2 -  \left[ 
\sum_i \frac{\pt_i^2}{z_i} - \left( \sum_i \pt_i \right)^2
\right] \right) \\
= \frac{1}{(2\pi)^2 (1-z_0)^2}
\int \frac{\dd[2]{\Pt_{21}} \dd[2]{\Kt_{01}}}{(2\pi)^{4}} 
    \delta\left(M_X^2 -  \left[ 
    \frac{\Kt_{01}^2}{z_0 (1-z_0)} + \frac{\Pt_{21}^2}{z_1 z_2 (1-z_0)}
    \right] \right)
.
\end{multline}
We note that we are allowed to perform these transverse momentum integrals in 2 dimensions as
the contribution in Eq.~\eqref{eq:c_times_d} is finite in the limit $\varepsilon \to 0$. 
Substituting the wave function $\widetilde \psi_\oout^{q \bar q g  \to q \bar q }( \Pt_{ij},z_{i};  \xt_{\ov 0 \ov 1},z_{\ov i})$ from Eq.~\eqref{eq:wf_qqg_to_qq_red} we get a contribution from both of these cases, which leads to the momentum integrals
\begin{multline}
    \int \frac{\dd[2]{\Pt_{20}} \dd[2]{\Kt_{01}}}{(2\pi)^{4}} 
    e^{i \Kt_{01} \vdot (\xt_{\ov{01}}-\xt_{0+2;1})- \frac{1}{z_0 +z_2} i \Pt_{20} \vdot \xt_{20}}
    \delta\left(M_X^2 -  \left[ 
    \frac{\Kt_{01}^2}{z_1 (1-z_1)} + \frac{\Pt_{20}^2}{z_0 z_2 (1-z_1)}
    \right] \right)
    \frac{\Pt_{02}^i}{\Pt_{02}^2} \\
    =  z_1 (1-z_1)^2  \frac{i}{8\pi^2} \frac{\xt_{20}^i}{\xt_{20}^2}
    \times 
    \left[
    J_0\left( \mxbar[0]
    \abs{\xt_{0+2;1}-\xt_{\ov 0 \ov 1}}  \right)
    -
    J_0\left( \mxbar[0] \sqrt{
    \left(\xt_{0+2;1}-\xt_{\ov 0 \ov 1}\right)^2
    + \omega_0 \xt_{20}^2}
    \right)
    \right]
\end{multline}
and
\begin{multline}
\int \frac{\dd[2]{\Pt_{21}} \dd[2]{\Kt_{01}}}{(2\pi)^{4}} 
    e^{i \Kt_{01} \vdot (-\xt_{\ov{01}}+\xt_{0;1+2})- \frac{1}{z_1 +z_2} i \Pt_{21} \vdot \xt_{21}}
    \delta\left(M_X^2 -  \left[ 
    \frac{\Kt_{01}^2}{z_0 (1-z_0)} + \frac{\Pt_{21}^2}{z_1 z_2 (1-z_0)}
    \right] \right)
    \frac{\Pt_{12}^i}{\Pt_{12}^2} \\
    =  z_0 (1-z_0)^2  \frac{i}{8\pi^2} \frac{\xt_{21}^i}{\xt_{21}^2}
    \times 
    \left[
    J_0\left( \mxbar[1]
    \abs{\xt_{\ov 0 \ov 1}-\xt_{0;1+2}}  \right)
    -
    J_0\left( \mxbar[1] \sqrt{
    \left(\xt_{\ov 0 \ov 1}-\xt_{0;1+2}\right)^2
    + \omega_1 \xt_{21}^2}
    \right)
    \right].
\end{multline}
As we will see, the fact that the results of the final state momentum integrals with a fixed invariant mass can be written as differences of two Bessel $J_0$ terms, is a wonderful simplification, because the first term is in fact like the one resulting from the two-particle phase space.
After these integrals we are left with
\begin{equation}
\label{eq:cd_after_integration}
\begin{split}
 &\left[ \frac{\dd{ \sigma^\textrm{D}_{\gamma^*_\lambda +A}}}{\dd[2]{\Deltat} \dd{M_X^2} } \right]_{ \ref{diag:gamma_qqg_qqg} \times \ref{diag:gamma_qq_qqg}^\ast} \\
=&\sum_f \nc \cf
\int \frac{\dd{z_{0}}\dd{z_{1}}\dd{z_{2}}}{(4\pi)^3 z_{0} z_{1} z_{2}}  (4\pi) \delta(1-z_{0}-z_{1}-z_{2})
\int \frac{\dd{z_{\ov 0}}\dd{z_{\ov 1}}}{(4\pi)^2 z_{\ov 0} z_{\ov 1} }  (4\pi) \delta(1-z_{\ov 0}-z_{\ov 1})\\
&\times \int 
\dd[2]{\xt_{0}} 
\dd[2]{\xt_{1} } 
\dd[2]{\xt_{2}} 
\dd[2]{\xt_{\ov 0} }
\dd[2]{\xt_{\ov 1} }
\times  
\left(1-\tripole_{012}\right)
\left(1-\dipole_{\ov 0 \ov 1}\right)^\dag 
\times\frac{1}{(2\pi)^{2}}e^{i \Deltat \vdot (\ov \bt - \bt)} 
 \\
&\times \widetilde \psi_\iin^{\gamma_\lambda^{\ast} \to q \bar q g} ( \xt_{ij},z_i)
\left[\widetilde \psi_\iin^{\gamma_\lambda^{\ast} \to q \bar q}( \xt_{\ov 0 \ov 1},z_{\ov i})\right]^\ast
\times 
   g  \varepsilon_\sigma^j \times \frac{i}{2\pi}
\\
& 
     \times \Bigg\{ z_{1}^2 \delta( z_1 - z_{\ov 1} ) \delta_{h_{\ov 1}}^{h_{1}} \times
   \frac{\xt_{20}^i}{\xt_{20}^2}
   \times S_0^{ij}(z_{\ov 0}, z_0)
   \times 
    \left[
    J_0\left( \mxbar[0]
    \abs{\xt_{0+2;1}-\xt_{\ov 0 \ov 1}}  \right)
    -
    J_0\left( \mxbar[0] \sqrt{
    \left(\xt_{0+2;1}-\xt_{\ov 0 \ov 1}\right)^2
    + \omega_0 \xt_{20}^2}
    \right)
    \right]\\
    &-  z_{0}^2 \delta( z_0 - z_{\ov 0} ) \delta_{h_{\ov 0}}^{h_{0}}
    \times 
    \frac{\xt_{21}^i}{\xt_{21}^2}
    \times S_1^{ij}(z_1, z_{\ov 1})
    \times 
    \left[
    J_0\left( \mxbar[1]
    \abs{\xt_{\ov 0 \ov 1}-\xt_{0;1+2}}  \right)
    -
    J_0\left( \mxbar[1] \sqrt{
    \left(\xt_{\ov 0 \ov 1}-\xt_{0;1+2} \right)^2
    + \omega_1 \xt_{21}^2}
    \right)
    \right]
    \Bigg\}
\end{split}
\end{equation}
where $S_0$ and $S_1$ are defined in Eqs.~\eqref{eq:spinor0} and \eqref{eq:spinor1}.
We note that this expression is exactly the same as the one for the $\ref{diag:gamma_qqg_qq} \times \ref{diag:gamma_qq_qq}^\ast$ contribution where the gluon does not cross the cut, in  Eq.~\eqref{eq:ab}, except for the sign and the additional Bessel function term.
We can then combine these two terms which leaves only one of the Bessel functions remaining, giving us:
\begin{equation}
\label{eq:ab+cd}
\begin{split}
 \left[ \frac{\dd{ \sigma^\textrm{D}_{\gamma^*_\lambda +A}}}{\dd[2]{\Deltat} \dd{M_X^2} } \right]_{ \ref{diag:gamma_qqg_qqg} \times \ref{diag:gamma_qq_qqg}^\ast}
 +  &\left[ \frac{\dd{ \sigma^\textrm{D}_{\gamma^*_\lambda +A}}}{\dd[2]{\Deltat} \dd{M_X^2} } \right]_{ \ref{diag:gamma_qqg_qq} \times \ref{diag:gamma_qq_qq}^\ast}
\\
=&\sum_f \nc \cf
\int \frac{\dd{z_{0}}\dd{z_{1}}\dd{z_{2}}}{(4\pi)^3 z_{0} z_{1} z_{2}}  (4\pi) \delta(1-z_{0}-z_{1}-z_{2})
\int \frac{\dd{z_{\ov 0}}\dd{z_{\ov 1}}}{(4\pi)^2 z_{\ov 0} z_{\ov 1} }  (4\pi) \delta(1-z_{\ov 0}-z_{\ov 1})\\
&\times \int 
\dd[2]{\xt_{0}} 
\dd[2]{\xt_{1} } 
\dd[2]{\xt_{2}} 
\dd[2]{\xt_{\ov 0} }
\dd[2]{\xt_{\ov 1} }
\times  
\left(1-\tripole_{012}\right)
\left(1-\dipole_{\ov 0 \ov 1}\right)^\dag
\times\frac{1}{(2\pi)^{2}}e^{i \Deltat \vdot (\ov \bt - \bt)} 
 \\
&\times \widetilde \psi_\iin^{\gamma_\lambda^{\ast} \to q \bar q g} ( \xt_{ij},z_i)
\left[\widetilde \psi_\iin^{\gamma_\lambda^{\ast} \to q \bar q}( \xt_{\ov 0 \ov 1},z_{\ov i})\right]^\ast
\times 
   g  \varepsilon_\sigma^j  \times \frac{-i}{2\pi} \\
&\times     \Bigg\{ z_{1}^2 \delta( z_1 - z_{\ov 1} ) \delta_{h_{\ov 1}}^{h_{1}} \times
     \frac{\xt_{20}^i}{\xt_{20}^2}
    \times S_0^{ij}(z_{\ov 0}, z_0)
    \times
    J_0\left( \mxbar[0] \sqrt{
    \left(\xt_{0+2;1}-\xt_{\ov 0 \ov 1}\right)^2
    + \omega_0\xt_{20}^2}
    \right)\\ 
    &- z_{0}^2 \delta( z_0 - z_{\ov 0} ) \delta_{h_{\ov 0}}^{h_{0}}
    \times 
      \frac{\xt_{21}^i}{\xt_{21}^2}
    \times S_1^{ij}(z_1, z_{\ov 1})
    \times
    J_0\left( \mxbar[1] \sqrt{
    \left(\xt_{\ov 0 \ov 1}-\xt_{0;1+2} \right)^2
    + \omega_1 \xt_{21}^2}
    \right)
    \Bigg\}.
\end{split}
\end{equation}
Note that while Eq.~\eqref{eq:cd_after_integration} is finite in the limit $\alpha \to 0$, the combination   in Eq.~\eqref{eq:ab} contains a divergence of the form $\log \alpha$ inherited from Eq.~\eqref{eq:ab+cd}. This is related to the JIMWLK evolution of the Wilson lines as discussed in Sec.~\ref{sec:JIMWLK}.

At this point it is convenient to use the explicit expressions the photon wave functions and perform the helicity sums. 
For longitudinal photons we get
\begin{equation}
\begin{split}
 \left[ \frac{\dd{ \sigma^\textrm{D}_{\gamma^*_L +A}}}{\dd[2]{\Deltat} \dd{M_X^2} } \right]_{ \ref{diag:gamma_qqg_qqg} \times \ref{diag:gamma_qq_qqg}^\ast}
 +& \left[ \frac{\dd{ \sigma^\textrm{D}_{\gamma^*_L +A}}}{\dd[2]{\Deltat} \dd{M_X^2} } \right]_{ \ref{diag:gamma_qqg_qq} \times \ref{diag:gamma_qq_qq}^\ast}
\\
=&8 \frac{\nc}{(2\pi)^4} \aem \frac{\as C_F}{2\pi} Q^2 \sum_f  e_f^2
\int \dd{z_{0}}\dd{z_{1}}\dd{z_{2}}   \delta(1-z_{0}-z_{1}-z_{2})
\\
&\times \int 
\dd[2]{\xt_{0}} 
\dd[2]{\xt_{1} } 
\dd[2]{\xt_{2}} 
\dd[2]{\xt_{\ov 0} }
\dd[2]{\xt_{\ov 1} }
\times  
\left(1-\tripole_{012}\right)
\left(1-\dipole_{\ov 0 \ov 1}\right)^\dag \times\frac{1}{(2\pi)^{2}}e^{i \Deltat \vdot (\ov \bt - \bt)} \times
   \frac{1}{z_2} K_0(X_{012} Q) \\
&\times     \Bigg[
     z_1^2 (1-z_1) K_0\left( \abs{\xt_{\ov 0 \ov 1}} \Qbar[0] \right)
     J_0\left( \mxbar[0] \sqrt{
    \left( \xt_{0+2;1}-\xt_{\ov 0 \ov 1}\right)^2
    + \omega_0 \xt_{20}^2}
    \right)
     \\&\times\left( z_1 \left(2 z_0 (z_0+z_2) + z_2^2 \right) \frac{1}{\xt_{20}^2}
     - z_0 \left(z_0(1-z_0)+z_1(1-z_1) \right) \frac{\xt_{20} \vdot \xt_{21}}{\xt_{20}^2 \xt_{21}^2}
     \right)
     \\
    &+ 
     z_0^2 (1-z_0) K_0\left( \abs{\xt_{\ov 0 \ov 1}} \Qbar[1] \right)
     J_0\left( \mxbar[1] \sqrt{
    \left(\xt_{\ov 0 \ov 1}-\xt_{0;1+2} \right)^2
    + \omega_1 \xt_{21}^2}
    \right)
     \\
     &\times\left( z_0 \left( 2z_1 (z_1+z_2) + z_2^2 \right) \frac{1}{\xt_{21}^2}
     - z_1 \left( z_0(1-z_0)+z_1(1-z_1) \right) \frac{\xt_{20} \vdot \xt_{21}}{\xt_{20}^2 \xt_{21}^2}
     \right)
    \Bigg],
\end{split}
\end{equation}
and for transverse ones
\begin{equation}
\begin{split}
 \left[ \frac{\dd{ \sigma^\textrm{D}_{\gamma^*_T +A}}}{\dd[2]{\Deltat} \dd{M_X^2} } \right]_{ \ref{diag:gamma_qqg_qqg} \times \ref{diag:gamma_qq_qqg}^\ast}
 + & \left[ \frac{\dd{ \sigma^\textrm{D}_{\gamma^*_T +A}}}{\dd[2]{\Deltat} \dd{M_X^2} } \right]_{ \ref{diag:gamma_qqg_qq} \times \ref{diag:gamma_qq_qq}^\ast}
\\
=&2 \frac{\nc}{(2\pi)^4} \aem \frac{\as C_F}{2\pi}\sum_f e_f^2
\int \dd{z_{0}}\dd{z_{1}}\dd{z_{2}}   \delta(1-z_{0}-z_{1}-z_{2})
\\
&\times \int 
\dd[2]{\xt_{0}} 
\dd[2]{\xt_{1} } 
\dd[2]{\xt_{2}} 
\dd[2]{\xt_{\ov 0} }
\dd[2]{\xt_{\ov 1} }
\times  
\left(1-\tripole_{012}\right)
\left(1-\dipole_{\ov 0 \ov 1}\right)^\dag
\times\frac{1}{(2\pi)^{2}}e^{i \Deltat \vdot (\ov \bt - \bt)} \times
   \frac{1}{z_2}  \frac{Q}{X_{012}\abs{\xt_{\ov{01}}}} K_1(X_{012} Q) \\
&\times     \Bigg[
     z_1 \Qbar[0] K_1\left( \abs{\xt_{\ov 0 \ov 1}} \Qbar[0]\right)
     \times
    J_0\left( \mxbar[0] \sqrt{
    \left(\xt_{0+2;1}-\xt_{\ov 0 \ov 1}\right)^2
    + \omega_0 \xt_{20}^2}
    \right)
      \times\left(
     \Upsilonuone + \Upsilonutwo  + \Upsilonuthree + \Upsilonufour
     \right)\\
    &+ 
     z_0 \Qbar[1] K_1\left( \abs{\xt_{\ov 0 \ov 1}} \Qbar[1] \right)
     \times
    J_0\left( \mxbar[1] \sqrt{
    \left(\xt_{\ov 0 \ov 1}-\xt_{0;1+2} \right)^2
    + \omega_1 \xt_{21}^2}
    \right)
     \times\left(
     \Upsilonvone + \Upsilonvtwo  + \Upsilonvthree + \Upsilonvfour
     \right)
    \Bigg].
\end{split}
\end{equation}
Here the
$\Upsilon$ terms have been defined in Sec.~\ref{sec:final_results}, Eqs.~\eqref{eq:Upsilonuone}--\eqref{eq:Upsilonvfour}.
This  gives us the expressions in Eqs.~\eqref{eq:G_diptrip_L} and \eqref{eq:G_diptrip_T}.

\subsection{Contribution $\abs{\ref{diag:gamma_qqg_qqg}}^2$}

The contribution from the $\abs{\ref{diag:gamma_qqg_qqg}}^2$ term involves gluon emissions before the shockwave both in the amplitude and the complex conjugate. This expression has been already calculated in Ref.~\cite{Beuf:2022kyp}, but we briefly go through the calculation as a cross-check of the results presented there. 
We can write this contribution to the cross section as
\begin{equation}
\begin{split}
 \left[ \frac{\ud \sigma^\textrm{D}_{\gamma^*_\lambda +A}}{ \dd[2]{\Deltat}\dd{M_X^2} } \right]_{\abs{\ref{diag:gamma_qqg_qqg}}^2 }
&=\sum_f\nc \cf
\int \frac{\dd{z_{0}}\dd{z_{1}}\dd{z_{2}}}{(4\pi)^3 z_{0} z_{1} z_{2}}  (4\pi) \delta(1-z_{0}-z_{1}-z_{2})\\
&\times \int 
\dd[2-2\varepsilon]{\xt_{0}}
\dd[2-2\varepsilon]{\xt_{1}} 
\dd[2-2\varepsilon]{\xt_{2}}
\dd[2-2\varepsilon]{\xt_{\ov 0}}
\dd[2-2\varepsilon]{\xt_{\ov 1} }
\dd[2-2\varepsilon]{\xt_{\ov 2} }
 \\
 & \times \int \prod_{i=0,1,2} \left[ \frac{\dd[2-2\varepsilon]{\pt_i}}{(2\pi)^{2-2\varepsilon}} \right]
\delta^{(2-2\varepsilon)}\left(\Deltat - \left(\sum_i \pt_i -\qt \right)\right)
\delta\left( M_X^2 - \left[\sum_i \frac{\pt_i^2}{z_i} - \left( \sum_i \pt_i \right)^2 \right] \right) \\
&\times \widetilde \psi_\iin^{\gamma_\lambda^{\ast} \to q \bar q g} ( \xt_{ij},z_i)
\left[\widetilde \psi_\iin^{\gamma_\lambda^{\ast} \to q \bar qg}( \xt_{\ov i \ov j},z_{ i})\right]^\ast 
\left(1-\tripole_{012}\right)
\left(1-\tripole_{\ov 0 \ov 1 \ov 2}\right)^\dag
e^{i \Deltat \vdot (\ov \bt -\bt)}
.
\end{split}
\end{equation}
As this contribution is finite, we can set $\varepsilon = 0$ and do the transverse momentum integrals analytically~\cite{Beuf:2022ndu}. 
This leads to
\begin{equation}
\begin{split}
 \left[ \frac{\ud \sigma^\textrm{D}_{\gamma^*_\lambda +A}}{\dd[2]{\Deltat} \dd{M_X^2} } \right]_{\abs{\ref{diag:gamma_qqg_qqg}}^2 }
=&\sum_f \frac{2}{(4\pi)^4}\nc \cf
\int {\dd{z_{0}}\dd{z_{1}}\dd{z_{2}}} \delta(1-z_{0}-z_{1}-z_{2})\\
& \times
\int 
\dd[2]{\xt_{0}}
\dd[2]{\xt_{1} }
\dd[2]{\xt_{2}}
\dd[2]{\xt_{\ov 0}}
\dd[2]{\xt_{\ov 1} }
\dd[2]{\xt_{\ov 2} }
\times\frac{1}{(2\pi)^{2}}e^{i \Deltat \vdot (\ov \bt - \bt)}
 \\
 &\times  \frac{M_X}{Y_{012}} J_1 (M_X Y_{012})
 \widetilde \psi_\iin^{\gamma_\lambda^{\ast} \to q \bar q g} ( \xt_{ij},z_i)
\left[\widetilde \psi_\iin^{\gamma_\lambda^{\ast} \to q \bar qg}( \xt_{\ov i \ov j},z_{ i})\right]^\ast 
\left(1-\tripole_{012}\right)
\left(1-\tripole_{\ov 0 \ov 1 \ov 2}\right)^\dag
\end{split}
\end{equation}
where $Y_{012}$ has been defined in Eq.~\eqref{eq:Y012}.
The photon wave functions $\widetilde \psi_\iin^{\gamma_\lambda^* \to q \bar q g}$ can now be substituted into this expression.
For longitudinal photons, this yields
\begin{equation}
\begin{split}
 \left[ \frac{\ud \sigma^\textrm{D}_{\gamma^*_L +A}}{\dd[2]{\Deltat} \dd{M_X^2} } \right]_{\abs{\ref{diag:gamma_qqg_qqg}}^2 }
=&\frac{8}{(2\pi)^6}\nc \cf \aem \as Q^2 \sum_f e_f^2
\int {\dd{z_{0}}\dd{z_{1}}\dd{z_{2}}} \delta(1-z_{0}-z_{1}-z_{2}) \\
&\times \int 
\dd[2]{\xt_{0}}
\dd[2]{\xt_{1} }
\dd[2]{\xt_{2}}
\dd[2]{\xt_{\ov 0}}
\dd[2]{\xt_{\ov 1} }
\dd[2]{\xt_{\ov 2}}
\times \left(1-\tripole_{012}\right)
\left(1-\tripole_{\ov 0 \ov 1 \ov 2}\right)^\dag 
\times\frac{1}{(2\pi)^{2}}e^{i \Deltat \vdot (\ov \bt - \bt)}\\
&\times \frac{M_X}{Y_{012}} J_1 (M_X Y_{012}) 
\times z_0 z_1 K_0(Q X_{012}) K_0(Q X_{\ov 0 \ov 1 \ov 2}) \\
&\times 
\left\{
z_1^2 \left[ 2z_0 (z_0 +z_2) + z_2^2 \right] \frac{\xt_{20} \vdot \xt_{\ov 2 \ov 0}}{\xt_{20}^2 \xt_{\ov 2 \ov 0}^2}
+
z_0^2 \left[ 2z_1 (z_1 +z_2) + z_2^2 \right] \frac{\xt_{21} \vdot \xt_{\ov 2 \ov 1}}{\xt_{21}^2 \xt_{\ov 2 \ov 1}^2}
\right.\\
&\left.-z_0 z_1 \left[ 2z_0z_1 + z_2(z_0+z_1) \right]
\left[
\frac{\xt_{20} \vdot \xt_{\ov 2 \ov 1}}{\xt_{20}^2 \xt_{\ov 2 \ov 1}^2 }
+\frac{\xt_{21} \vdot \xt_{\ov 2 \ov 0}}{\xt_{21}^2 \xt_{\ov 2 \ov 0}^2 }
\right]
\right\}
\end{split}
\end{equation}
in agreement with Ref.~\cite{Beuf:2022kyp}. Here $X_{012}$ and $X_{\ov{012}}$ have been defined in Eqs.~\eqref{eq:X012} and \eqref{eq:X012ov}.
For transversely polarized photons, we get
\begin{equation}
\begin{split}
 \left[ \frac{\ud \sigma^\textrm{D}_{\gamma^*_T +A}}{\dd[2]{\Deltat} \dd{M_X^2} } \right]_{\abs{\ref{diag:gamma_qqg_qqg}}^2 }
=&\frac{2}{(2\pi)^6}\nc \cf \alpha_\text{em} \as Q^2 \sum_f e^2_f
\int {\dd{z_{0}}\dd{z_{1}}\dd{z_{2}}} \delta(1-z_{0}-z_{1}-z_{2})\\
& \times \int 
\dd[2]{\xt_{0}}
\dd[2]{\xt_{1} }
\dd[2]{\xt_{2}}
\dd[2]{\xt_{\ov 0}}
\dd[2]{\xt_{\ov 1} }
\dd[2]{\xt_{\ov 2} }
\times \left(1-\tripole_{012}\right)
\left(1-\tripole_{\ov 0 \ov 1 \ov 2}\right)^\dag
\times\frac{1}{(2\pi)^{2}}e^{i \Deltat \vdot (\ov \bt - \bt)} 
 \\
 &\times  \frac{M_X}{Y_{012}} J_1 (M_X Y_{012}) 
 \times z_0 z_1 K_1(Q X_{012}) K_1(Q X_{\ov 0 \ov 1 \ov 2})  \frac{1}{X_{012} X_{\ov 0 \ov 1 \ov 2}} \\
& \times 
\left\{ 
\Upsilonbb + \Upsiloncc + \Upsilond + \Upsilone + \Upsilonbc
\right\}
\end{split}
\end{equation}
where the $\Upsilon$ terms follow the terminology from Ref.~\cite{Beuf:2022kyp} and are shown in Sec.~\ref{sec:final_results}.
We note that we have corrected here a factor of 2 that was missing in Ref.~\cite{Beuf:2022kyp}.
From these expressions we can read Eqs.~\eqref{eq:G_trip_L} and \eqref{eq:G_trip_T} in the final result.

\subsection{Contributions $\abs{\ref{diag:gamma_qq_qq}}^2$ and $\abs{\ref{diag:gamma_qq_qqg}}^2$: final-state corrections}
\label{sec:final-state_corrections}

The final contributions to calculate are related to gluon emission and absorption in the final state.
The NLO corrections corresponding to ``virtual'' gluon exchanges between the shockwave and the cut are defined as part of the contribution ${\abs{\ref{diag:gamma_qq_qq}}^2}$, and they can be written as
\begin{equation}
\label{eq:a2_final-state}
\begin{split}
 \left[ \frac{\ud \sigma^\textrm{D}_{\gamma^*_\lambda +A}}{\dd[2]{\Deltat} \dd{M_X^2} } \right]_{\abs{\ref{diag:gamma_qq_qq}}^2}^\text{final-st. corr.}
=&\sum_f \nc  
\int \frac{\ud z_{0} \ud z_{1}}{(4\pi)^2 z_{0} z_{1}} (4\pi) \delta(1-z_{0}-z_{1})
\int \frac{\ud z_{\ov 0} \ud z_{\ov 1}}{(4\pi)^2 z_{\ov 0} z_{\ov 1}} (4\pi) \delta(1-z_{\ov 0}-z_{\ov 1})\\
&\times 
\int \frac{\ud z_{0p} \ud z_{1p}}{(4\pi)^2 z_{0p} z_{1p}} 4\pi \delta(1-z_{0p}-z_{1p}) 
\int  \frac{\ud^{2-2\varepsilon} \Pt_{01} }{(2\pi)^{2-2\varepsilon}}  
\delta \left(M_X^2 - \frac{\Pt_{01}^2}{z_{0p} z_{1p}} \right)  \\
& \times 
\int 
\dd[2-2\varepsilon]{\xt_0}
\dd[2-2\varepsilon]{\xt_1}
\dd[2-2\varepsilon]{\xt_{\ov 0}}
\dd[2-2\varepsilon]{\xt_{ \ov 1}}
\times 
\left(1-\dipole_{01}\right)
\left(1-\dipole_{\ov 0 \ov 1}\right)^\dag 
\times \frac{1}{(2\pi)^{2-2\varepsilon}}e^{i \Deltat \vdot (\ov \bt -\bt)} \\
&\times 
\widetilde \psi_\iin^{\gamma_\lambda^{\ast} \to q \bar q} ( \xt_{01}, z_i)
\left[\widetilde \psi_\iin^{\gamma_\lambda^{\ast} \to q \bar q}( \xt_{\ov 0 \ov 1},z_{\ov i})\right]^\ast \\
&\times 
\left\{
(4\pi) z_{\ov 0} z_{\ov 1} \delta(z_{\ov 0} - z_{0p}) \delta_{h_{\ov 0}}^{h_{0p}} \delta_{h_{\ov 1}}^{h_{1p}} e^{i \Pt_{01} \vdot \xt_{\ov 0 \ov 1}}
\left[\widetilde \psi_\oout^{q \bar q \to q \bar q}(\Pt_{01},z_{ip}, ; \xt_{01}, z_i) 
\right]^\ast \right.\\
&\hspace{1.3ex}+\left.
(4\pi) z_0 z_1 \delta(z_0 - z_{0p})  \delta_{h_0}^{h_{0p}} \delta_{h_1}^{h_{1p}} e^{-i \Pt_{01} \vdot \xt_{01}} \widetilde \psi_\oout^{q \bar q \to q \bar q}(\Pt_{01},z_{ip}, ; \xt_{\ov 0 \ov 1}, z_{\ov i}) 
\right\}.
\end{split}
\end{equation}
The ``real'' gluon emission terms, where the gluon is emitted after the shockwave part of the final state, correspond to the term ${\abs{\ref{diag:gamma_qq_qqg}}^2}$. The corresponding cross section can be written as
\begin{equation}
\label{eq:d2}
\begin{split}
 \left[ \frac{\ud \sigma^\textrm{D}_{\gamma^*_\lambda +A}}{\dd[2]{\Deltat} \dd{M_X^2} } \right]_{\abs{\ref{diag:gamma_qq_qqg}}^2}
=&\sum_f \nc \cf
 \int \frac{\ud z_{0} \ud z_{1} }{(4\pi)^2  z_{0} z_{1} } (4\pi) \delta(1- z_{0}-z_{1})
  \int \frac{ \ud z_{\ov 0} \ud z_{\ov 1}}{(4\pi)^2  z_{\ov 0}  z_{\ov 1}} (4\pi) \delta(1- z_{\ov 0}-z_{\ov 1})
 \\
&\times 
\int \frac{\dd{z_{0p}}\dd{z_{1p}}\dd{z_{2p}}}{(4\pi)^3 z_{0p} z_{1p} z_{2p}}  (4\pi) \delta(1-z_{0p}-z_{1p}-z_{2p})
\\
& \times \int \prod_{i=0,1,2} \left[ \frac{\dd[2-2\varepsilon]{\pt_i}}{(2\pi)^{2-2\varepsilon}} \right]
\delta^{(2-2\varepsilon)}\left(\Deltat - \left(\sum_i \pt_i-\qt\right)\right)
\delta\left( M_X^2 - \left[\sum_i \frac{\pt_i^2}{z_{ip}} - 
\left( \sum_i \pt_i \right)^2 \right] \right) \\
& \times 
\int
\dd[2-2\varepsilon]{\xt_{0}}
\dd[2-2\varepsilon]{\xt_{1}}
\dd[2-2\varepsilon]{\xt_{\ov 0} }
\dd[2-2\varepsilon]{\xt_{\ov 1} }
\times
\left(1-\dipole_{01}\right)
\left(1-\dipole_{\ov 0 \ov 1}\right)^\dag 
\times e^{i \Deltat \vdot (\ov \bt -\bt)}
\\
&\times  \widetilde \psi_\iin^{\gamma_\lambda^{\ast} \to q \bar q} ( \xt_{01},z_i)
\left[\widetilde \psi_\iin^{\gamma_\lambda^{\ast} \to q \bar q}( \xt_{\ov 0 \ov 1},z_{\ov i})\right]^\ast 
\left[  \widetilde \psi_\oout^{q \bar q g \to q \bar q}( \Pt_{ij},z_{ip};  \xt_{01},z_i) \right]^\ast
  \widetilde \psi_\oout^{q \bar q g \to q \bar q}( \Pt_{ij},z_{ip};  \xt_{\ov 0 \ov 1},z_{\ov i}).
\end{split}
\end{equation}

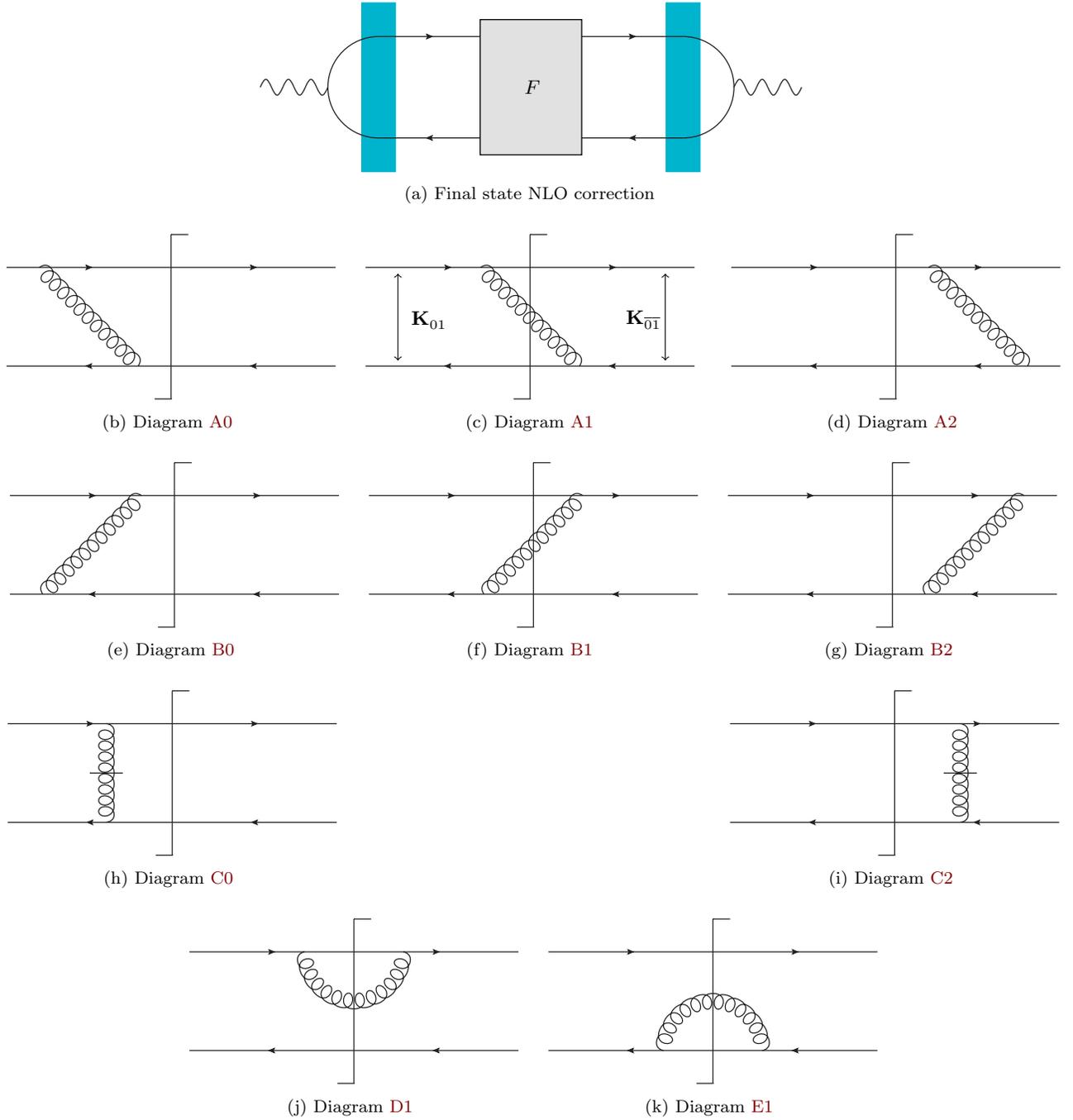
\begin{figure}
     \centering
    \subfloat[Final state NLO correction \label{fig:F} ]{
         \centering
         \includegraphics[width=0.5\columnwidth]{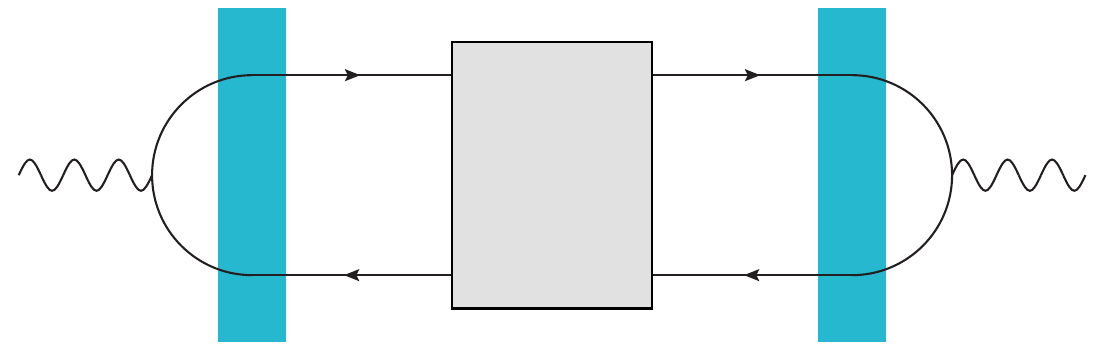}
          \begin{tikzpicture}[overlay]
        \node[anchor=south east] at (-4.3cm,1.2cm) {$F$};
        \end{tikzpicture}
         }\\
    \subfloat[Diagram $\azero$ \label{fig:A0}]{
         \centering
         \includegraphics[width=0.31\columnwidth]{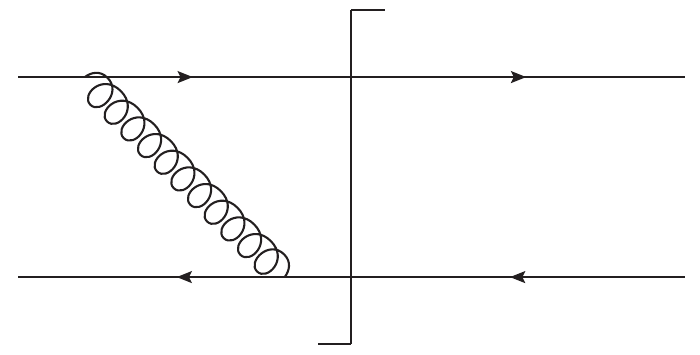}
         }
    \subfloat[Diagram $\aone$ \label{fig:A1}]{
         \centering
         \includegraphics[width=0.31\columnwidth]{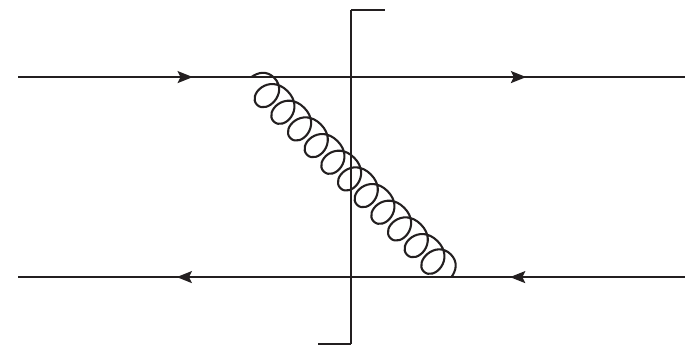}
          \begin{tikzpicture}[overlay]
            \draw [to-to](-0.7cm,0.7cm) -- (-0.7cm,2.1cm);
            \draw [to-to](-5cm,0.7cm) -- (-5cm,2.1cm);
         \node[anchor=south east] at (-4.1cm,1.1cm) {$\Kt_{01}$};
         \node[anchor=south east] at (-0.65cm,1.1cm) {$\Kt_{\ov 0 \ov 1}$};
        \end{tikzpicture}
         } 
    \subfloat[Diagram $\atwo$ \label{fig:A2}]{
         \centering
         \includegraphics[width=0.31\columnwidth]{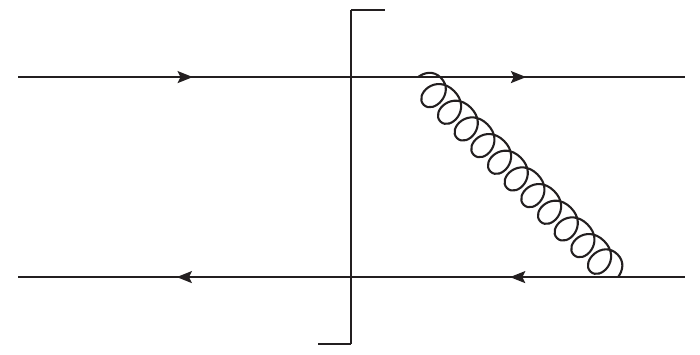}
         } \\
    \subfloat[Diagram $\bzero$ \label{fig:B0}]{
         \centering
         \includegraphics[width=0.31\columnwidth]{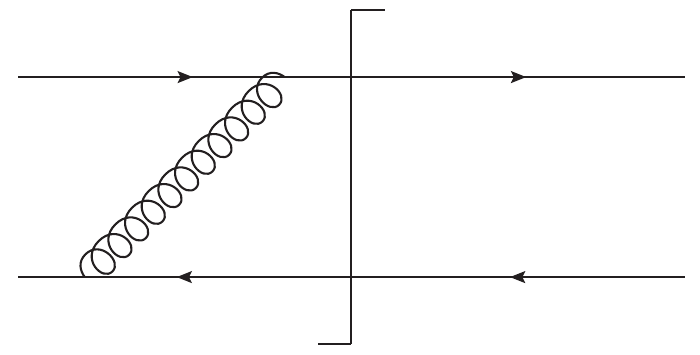}
         }
    \subfloat[Diagram $\bone$ \label{fig:B1}]{
         \centering
         \includegraphics[width=0.31\columnwidth]{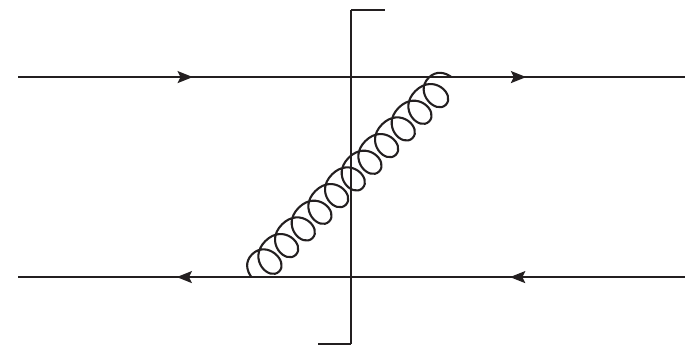}
         }
    \subfloat[Diagram $\btwo$ \label{fig:B2}]{
         \centering
         \includegraphics[width=0.31\columnwidth]{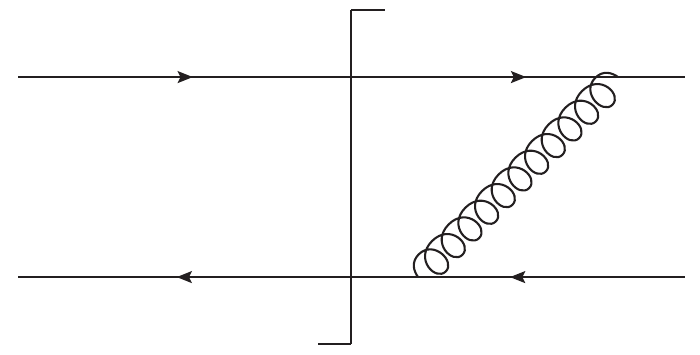}
         }\\
    \subfloat[Diagram $\czero$ \label{fig:C0}]{
         \centering
         \includegraphics[width=0.31\columnwidth]{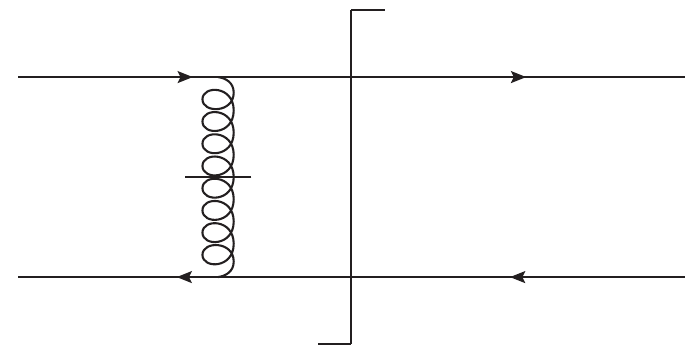}
         } 
         \hspace{0.32\textwidth}
    \subfloat[Diagram $\ctwo$ \label{fig:C2}]{
         \centering
         \includegraphics[width=0.31\columnwidth]{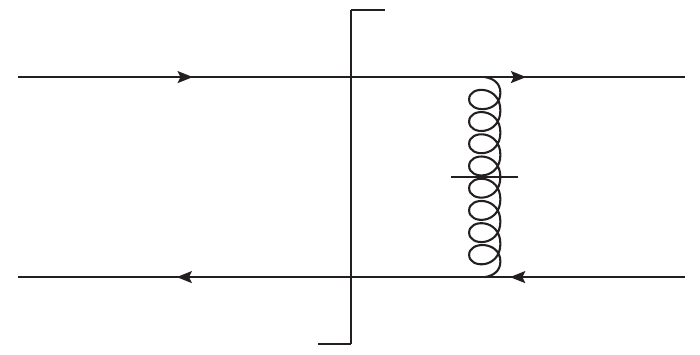}
         } \\
    \subfloat[Diagram $\done$ \label{fig:D1}]{
         \centering
         \includegraphics[width=0.31\columnwidth]{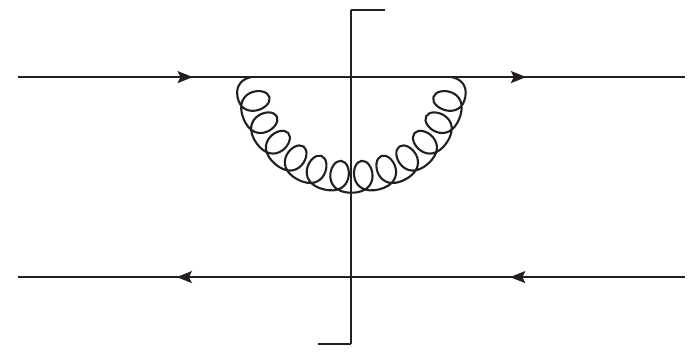}
         }
    \subfloat[Diagram $\eone$ \label{fig:E1}]{
         \centering
         \includegraphics[width=0.31\columnwidth]{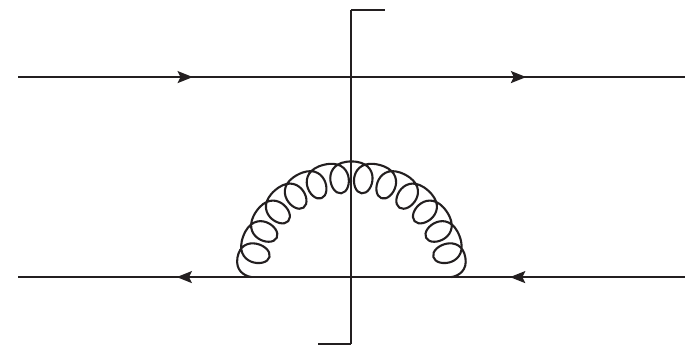}
         }
     \caption{Feynman diagrams contributing to final state corrections $F$ at next-to-leading order. The cut represents the final state.
     }
\label{fig:final_state_NLO}
\end{figure}

Evaluating these terms can be simplified by noting that the final-state contributions essentially factorize out of the rest of the diagram. 
This is shown in Fig.~\ref{fig:final_state_NLO}, where the sum of the final-state contributions is defined as $F$, and the individual diagrams contributing to the final-state corrections are also shown.
We can then write the corresponding cross section as
\begin{equation}
\begin{split}
 \left[ \frac{\dd{ \sigma^\textrm{D}_{\gamma^*_\lambda +A}}}{\dd[2]{\Deltat}\dd{M_X^2}} \right]^{\text{final-st. corr.}}_{\abs{\ref{diag:gamma_qq_qq}}^2+\abs{\ref{diag:gamma_qq_qqg}}^2}
=&\sum_f \frac{\nc}{(4\pi)^2} \frac{\as \cf}{2\pi} \int_0^1 \dd{ z_0} \dd{ z_1} \delta(1-z_0-z_1) \int_0^1 \dd{ z_{\overline 0}} \dd{ z_{\overline 1}} \delta(1-z_{\ov 0} -z_{\ov 1})\\
& \times \int 
\dd[2-2\varepsilon]{\xt_0}
\dd[2-2\varepsilon]{\xt_1}
\dd[2-2\varepsilon]{\xt_{\ov 0}}
\dd[2-2\varepsilon]{\xt_{\ov 1}}
\times
\left(1-\dipole_{01}\right)
\left(1-\dipole_{\ov 0 \ov 1}\right)^\dag 
\times 
\frac{1}{(2\pi)^{2-2\varepsilon}} e^{i \Deltat \vdot (\ov \bt -\bt)}
\\
&\times 
\delta_{h_0}^{h_{\ov 0}} \delta_{h_1}^{h_{\ov 1}}
\times
\widetilde \psi_\iin^{\gamma_\lambda^{\ast} \to q \bar q} (\xt_{01},z_{i})
\left[\widetilde \psi_\iin^{\gamma_\lambda^{\ast} \to q \bar q}( \xt_{\overline 0\overline 1},z_{\overline i})\right]^\ast 
\times
F(z_0, z_{\overline 0}, \xt_{01} ,\xt_{\overline 0 \overline 1})
\end{split}
\end{equation}
where $F(z_0, z_{\overline 0}, \xt_{01} ,\xt_{\overline 0 \overline 1}) = \sum_{i \in \text{diagrams}} F_i$ contains contributions from diagrams corresponding to the process $q \bar q \to q \bar q$ with the cut representing the final state. 
We have also written the helicity-conserving Kronecker deltas $\delta_{h_0}^{h_{\ov 0 }} \delta_{h_1}^{h_{\ov 1 }}$ out from the factor $F$ for simplicity.
The helicity conservation is a consequence of neglecting the quark mass.
Combining these with the photon wave functions,
the sums over the helicities can be written as
\begin{equation}
        \sum_{h_0 h_1 h_{\ov 0} h_{\ov 1} }\delta_{h_{\ov 0}}^{h_{0}} \delta_{h_{ \ov 1}}^{h_{1}}   \times \widetilde \psi_\iin^{\gamma_\lambda^{\ast} \to q \bar q} ( \xt_{01}, z_i)
\left[\widetilde \psi_\iin^{\gamma_\lambda^{\ast} \to q \bar q}( \xt_{\ov 0 \ov 1},z_{\ov i})\right]^\ast 
=2\frac{\aem}{\pi} Q^2  e_f^2 \mathcal{F}_\lambda(z_0, z_1, z_{\ov 0}, z_{\ov 1})
\end{equation}
with the function  $\mathcal{F}_\lambda$  defined in Eq.~\eqref{eq:lo_photon_wf_part}.

We can now focus on calculating the function $F$.
This is simplest to do by considering each row in Fig.~\ref{fig:final_state_NLO} separately.
The first three rows are finite in $\varepsilon \to 0$ so we can calculate the transverse integrals in two dimensions. This allows us to write them as
\begin{equation}
\label{eq:F_d}
    \begin{split}
        F_{d}=& -16\pi^2 \times  \frac{ z_0 z_1 + z_{\ov 0} z_{\ov 1}}{[  z_0 z_1 z_{\ov 0} z_{\ov 1}]^{3/2}}  
        \times \int \frac{\dd[2]{\Kt_{01}} \dd[2]{\Kt_{\ov 0 \ov 1}}}{(2\pi)^{4}} e^{i \Kt_{\ov 0 \ov 1} \vdot \xt_{\overline 0 \overline 1}-i\Kt_{01} \vdot \xt_{0 1}} 
         (z_{\ov 0} \Kt_{01}-z_0 \Kt_{\ov 0 \ov 1}) \vdot (z_{\ov 1} \Kt_{01}- z_1 \Kt_{\ov 0 \ov 1})
         \times \mathcal{F}_{d}
    \end{split}
\end{equation}
where the index $d$ refers to the name of the diagram, with
\begin{align}
    \mathcal{F}_{\azero} &= \frac{ \theta(z_0 - z_{\ov 0} - \alpha)}{(z_0- z_{\ov 0})^3}
    \frac{ \delta(M_X^2-M_0^2) }{[M_0^2-M_{\aone}^2-i \delta] [M_0^2-M_2^2-i \delta]} \\
    \mathcal{F}_{\aone} &=\frac{ \theta(z_0 - z_{\ov 0} - \alpha)}{(z_0- z_{\ov 0})^3}
    \frac{\delta(M_X^2-M_{\aone}^2) }{[M_{\aone}^2-M_0^2+i \delta] [M_{\aone}^2-M_2^2-i \delta]} \\
    \mathcal{F}_{\atwo} &= \frac{ \theta(z_0 - z_{\ov 0} - \alpha)}{(z_0- z_{\ov 0})^3}
    \frac{ \delta(M_X^2-M_2^2)}{[M_2^2-M_0^2+i \delta] [M_2^2-M_{\aone}^2+i \delta]} \\
    \mathcal{F}_{\bzero} &= \frac{\theta(z_{\ov 0} - z_{0} - \alpha)}{(z_{\ov 0}- z_0)^3}
  \frac{ \delta(M_X^2-M_0^2) }{[M_0^2-M_{\bone}^2-i \delta] [M_0^2-M_2^2-i \delta]} \\
    \mathcal{F}_{\bone} &= \frac{\theta(z_{\ov 0} - z_{0} - \alpha)}{(z_{\ov 0}- z_0)^3}
  \frac{ \delta(M_X^2-M_{\bone}^2)}{[M_{\bone}^2-M_0^2+i \delta] [M_{\bone}^2-M_2^2-i \delta]}\\
    \mathcal{F}_{\btwo} &= \frac{\theta(z_{\ov 0} - z_{0} - \alpha)}{(z_{\ov 0}- z_0)^3}
\frac{  \delta(M_X^2-M_2^2) }{[M_2^2-M_0^2+i \delta] [M_2^2-M_{\bone}^2+i \delta]}.
\end{align}
These have been written in terms of the momenta
\begin{equation}
    \begin{aligned}
    \Kt_{01} &= z_1 \kt_0 - z_0 \kt_1 \\
    \Kt_{\ov 0 \ov 1} &= z_{\ov 1} \kt_{\ov 0} - z_{\ov 0} \kt_{\ov 1}
    \end{aligned}
\end{equation}
which are the natural momentum variables as they correspond to the relative transverse momenta of the quark-antiquark pair going to the shockwave in the amplitude and the complex conjugate as shown in Fig.~\ref{fig:final_state_NLO}.
These momenta appear from Eqs.~\eqref{eq:a2_final-state} and \eqref{eq:d2} by noting that for Diagrams~$\azero$ and $\bzero$ we have $\Pt_{01} = \Kt_{01}$ and for Diagrams~$\atwo$ and $\btwo$ we have $\Pt_{01} = \Kt_{\ov{01}}$. The remaining $\Kt_{01}$ or $\Kt_{\ov{01}}$ integral then comes from the Fourier transform of $\widetilde \psi_\oout^{q \bar q \to q \bar q}$.
For Diagrams~$\aone$ and $\bone$ the momentum integral in Eq.~\eqref{eq:d2} can be written in terms of $\Kt_{01}$ and $\Kt_{\ov{01}}$ by noting that 
\begin{equation}
    \begin{aligned}
        \aone : 
        \Kt_{01} &= \Pt_{01} - \Pt_{12} ,
        &
        \Kt_{\ov{01}} &=  \Pt_{01} + \Pt_{02} ,\\
        \bone : 
        \Kt_{01} &= \Pt_{01} + \Pt_{02},
        &
        \Kt_{\ov{01}} &=  \Pt_{01} - \Pt_{12}.
    \end{aligned}
\end{equation}

The terms $M_d^2$ correspond to invariant masses of different Fock states. The naming scheme is such that for a diagram $d$ the quantity $M_d^2$ is the invariant mass of the Fock state at the final state cut. For some other diagram $d'$ the invariant mass $M_d^2$ might be that of some intermediate state, thus the same quantity $M_d^2$ might appear in the expression for several diagrams. In terms of the momenta we can write the energy denominators and the final state invariant mass restrictions for the diagrams on the first two lines as:
\begin{align}
    M_0^2=M_{\azero}^2=M_{\bzero}^2&=\frac{1}{z_0 z_1} \Kt_{01}^2  \\
    M_{\aone}^2&= \frac{1}{z_0 -z_{\ov 0}} \left[ \frac{z_{\ov 1}}{z_1} \Kt_{01}^2 + \frac{z_0}{z_{\ov 0}} \Kt_{\ov 0 \ov 1}^2 - 2\Kt_{01} \vdot \Kt_{\ov 0 \ov 1} \right] \\
    M_{\bone}^2&= \frac{1}{z_{\ov 0}- z_0} \left[ \frac{z_1}{z_{\ov 1} } \Kt_{\ov 0 \ov 1}^2 + \frac{z_{\ov 0}}{z_0} \Kt_{01}^2 - 2\Kt_{01} \vdot \Kt_{\ov 0 \ov 1} \right] \\
    M_2^2=M_{\atwo}^2=M_{\btwo}^2&= \frac{1}{z_{\ov 0} z_{\ov 1}} \Kt_{\ov 0 \ov 1}^2.
\end{align}

Here it is convenient to get rid of the delta functions and rearrange the diagrams with a trick. In some sense this is the inverse of calculating a cut of a diagram by writing the imaginary part of a propagator as a delta function: here we write a delta function that is a part of the definition of the physical observable as an imaginary part of a propagator. Indeed, thanks to the Sochocki-Plemelj relation
\begin{equation}
    \frac{1}{x-i\delta} =i \pi \delta(x) + \text{PV} \frac{1}{x},
\end{equation}
we have 
\begin{equation} \label{eq:deltaimag}
2\pi \delta(x)  =  i\left( \frac{1}{x+i\delta} -\frac{1}{x+i\delta} \right).
\end{equation}
This trick is particularly useful because the terms in each row in Fig.~\ref{fig:final_state_NLO} have a very similar structure and differ only in the final-state delta function and the energy dominators, allowing us to sum them easily.
First,  consider the sum of $\mathcal{F}_{\azero} + \mathcal{F}_{\aone} + \mathcal{F}_{\atwo}$. We encounter the following combination, where using \eqref{eq:deltaimag} turns the sum of three terms in to six, which cancel so that only two are left:
\begin{equation}
\label{eq:Beuf_trick}
\begin{split}
    &\frac{\delta(M_0^2-M_X^2)}{(M_0^2-M_{\aone}^2-i\delta)(M_0^2-M_2^2-i\delta)} + 
    \frac{\delta(M_{\aone}^2-M_X^2)}{(M_{\aone}^2-M_0^2+i\delta)(M_{\aone}^2-M_2^2-i\delta)}\\
    &+
    \frac{\delta(M_2^2-M_X^2)}{(M_2^2-M_0^2+i\delta)(M_2^2-M_{\aone}^2+i\delta)}\\
    &\quad \quad= \frac{1}{2\pi i} \left[\frac{1}{(M_X^2-M_0^2-i\delta)(M_X^2-M_{\aone}^2-i\delta)(M_X^2-M_2^2-i\delta)}\right.\\
    & \quad \quad\quad \quad  \left.-\frac{1}{(M_X^2-M_0^2+i\delta)(M_X^2-M_{\aone}^2+i\delta)(M_X^2-M_2^2+i\delta)}\right],
\end{split}
\end{equation}
That trick allows us to remove the delta function so that we only have energy denominators (differences of the invariant masses) left. 
Note that the correct signs of the infinitesimals $i\delta$ are crucial for this trick to work. 
This relation is also very general, as it works for any quadruplet of invariant masses $(M_0^2, M_1^2, M_2^2, M_X^2)$.
Thus, a similar relation can also be found for the sum $\mathcal{F}_{\bzero} + \mathcal{F}_{\bone} + \mathcal{F}_{\btwo}$ with the substitution $M_{\aone}^2 \to M_{\bone}^2$.

Another simplification can be done by rewriting the inner product in the numerator in Eq.~\eqref{eq:F_d}, coming from the transverse momenta in the gluon emission and absorption vertices, in terms of the invariant masses:
\begin{equation}
\begin{split}
    &\!\!\!\!\!\!\!\!\!\!\!\!\!\!\!\!\!\!
    \left(z_{\ov 0} \Kt_{01}-z_{0}\Kt_{\ov 0 \ov 1} \right) \vdot \left(z_{\ov 1}\Kt_{01}-z_{1}\Kt_{\ov 0 \ov 1} \right) \\
    =& \frac{1}{2}(z_0 - z_{\ov 0}) \biggl[- (z_{\ov 0} z_1 + z_0 z_{\ov 1}) (M_X^2 - M_{\aone}^2) + z_0 z_{\ov 1} (M_X^2 - M_0^2) 
    + z_0 z_{\ov 1} (M_X^2 - M_2^2) - (z_0 -z_{\ov 0}) M_X^2 \biggr] \\
    =& \frac{1}{2}(z_{\ov 0} - z_{ 0}) \biggl[- (z_{\ov 0} z_1 + z_0 z_{\ov 1}) (M_X^2 - M_{\bone}^2) + z_{\ov 0} z_{ 1} (M_X^2 - M_0^2) + z_{\ov 0} z_{1} (M_X^2 - M_2^2) - (z_{\ov 0} -z_{ 0}) M_X^2 \biggr].
\end{split}
\end{equation}
This allows us to cancel some of the energy denominators from Eq.~\eqref{eq:Beuf_trick}, and we find
\begin{equation}
\label{eq:F_A012}
\begin{split}
    & \!\!\!\!\!\!\!\!\!\!\!\!\!\!\!\!\!\! \left(z_{\ov 0} \Kt_{01}-z_{0}\Kt_{\ov 0 \ov 1} \right) \vdot \left(z_{\ov 1}\Kt_{01}-z_{1}\Kt_{\ov 0 \ov 1} \right)  \times
    [\mathcal{F}_{\azero} + \mathcal{F}_{\aone} + \mathcal{F}_{\atwo}]\\
    &=  \frac{\theta(z_0 - z_{\ov 0 } - \alpha)}{2(z_0 - z_{\ov 0})^2}
     \left[
        z_0 z_{\ov 1} D_{\adiag 01} +z_0 z_{\ov 1} D_{\adiag 12} -(z_{\ov 0} z_1 + z_0 z_{\ov 1}) D_{\adiag 02} -(z_0 -z_{\ov 0}) M_X^2 D_{\adiag 012}
        \right]
\end{split}
\end{equation}
and
\begin{equation}
\label{eq:F_B012}
\begin{split}
    & \!\!\!\!\!\!\!\!\!\!\!\!\!\!\!\!\!\! \left(z_{\ov 0} \Kt_{01}-z_{0}\Kt_{\ov 0 \ov 1} \right) \vdot \left(z_{\ov 1}\Kt_{01}-z_{1}\Kt_{\ov 0 \ov 1} \right)  \times 
    [\mathcal{F}_{\bzero} + \mathcal{F}_{\bone} + \mathcal{F}_{\btwo}]\\
    &=  \frac{\theta(z_{\ov 0} - z_{ 0 } - \alpha)}{2(z_{\ov 0 }-z_0)^2}
      \left[
        z_{\ov 0} z_{1} D_{\bdiag01} +z_{\ov 0} z_{1} D_{\bdiag 12} -(z_{\ov 0} z_1 + z_0 z_{\ov 1}) D_{\bdiag02} -(z_{\ov 0} - z_0) M_X^2 D_{\bdiag 012}
        \right].
\end{split}
\end{equation}
We have used the following short-hand notation for the energy dominators
\begin{equation}
    \begin{split}
    D_{\adiag ij} &= \frac{1}{2\pi i} \left\{\frac{1}{[M_X^2-M_{\adiag i}^2-i\delta][M_X^2-M_{\adiag j}^2-i\delta]}-\frac{1}{[M_X^2-M_{\adiag i}^2+i\delta][M_X^2-M_{\adiag j}^2+i\delta]} \right\} \\
    &= \frac{1}{\pi} \Im \left\{ \frac{1}{[M_X^2-M_{\adiag i}^2-i\delta][M_X^2-M_{\adiag j}^2-i\delta]} \right\} \\
    D_{\adiag ijk} &= \frac{1}{\pi} \Im \left\{\frac{1}{[M_X^2-M_{\adiag i}^2-i\delta][M_X^2-M_{\adiag j}^2-i\delta][M_X^2-M_{\adiag k}^2-i\delta]}\right\}
    \end{split}
\end{equation}
with a similar notation used for Diagrams~$\bdiag$.

At this point, we can see the purpose of combining Diagrams~$\adiag$ and Diagrams~$\bdiag$.
First, we do not have delta functions anymore but only terms with two or three denominators. It is then easier to calculate the Fourier transforms in Eq.~\eqref{eq:F_d} by writing the denominators in terms of Schwinger or Feynman parameters.
Second, the most complicated terms $D_{\adiag 012}$ and $D_{\bdiag 012}$ involving three energy denominators are actually finite.
Roughly speaking, the energy denominators with $M_{\aone}^2$ or $M_{\bone}^2$ behave like 
$z_2' \log z_2'$ after the Fourier transforms, where $z_2' = \abs{z_0 - z_{\ov 0}}$ is the momentum fraction of the exchanged gluon,
 which is enough to cancel the divergence at $z_2' = 0$. 
This also means that the terms $D_{\adiag 01}$, $D_{\adiag 12}$, $D_{\bdiag 01}$ and $D_{\bdiag 12}$ have a double logarithmic divergence $\log^2 \alpha$, while the terms $D_{\adiag 02}$ and $D_{\bdiag 02}$ have a power divergence $1/\alpha$.
The power divergences will cancel with Diagrams~$\cdiag$ corresponding to the instantaneous gluon exchange as we will show next.

The contributions from Diagrams~$\czero$ and $\ctwo$ can be written as
\begin{equation}
\begin{split}
    F_{\czero}
    &= 16 \pi^2 \times
      \frac{ 2z_0 z_1 z_{\ov 0} z_{\ov 1} }{[  z_0 z_1 z_{\ov 0} z_{\ov 1}]^{3/2}}
      \int \frac{\dd[2]{\Kt_{01}} \dd[2]{\Kt_{\ov 0 \ov 1}}}{(2\pi)^{4}} e^{i \Kt_{\ov 0 \ov 1} \vdot \xt_{\overline 0 \overline 1}-i\Kt_{01} \vdot \xt_{0 1}}
        \frac{\theta(z_0 -z_{\ov 0} - \alpha)+\theta(z_{\ov 0} -z_{ 0} - \alpha)}{(z_0 -z_{\ov 0})^2}
         \frac{\delta(M_X^2-M_0^2)}{[M_2^2-M_0^2+i \delta] }
\end{split}
\end{equation}
and
\begin{equation}
\begin{split}
    F_{\ctwo}
    &= 16 \pi^2 \times
      \frac{ 2z_0 z_1 z_{\ov 0} z_{\ov 1} }{[  z_0 z_1 z_{\ov 0} z_{\ov 1}]^{3/2}}
       \int \frac{\dd[2]{\Kt_{01}} \dd[2]{\Kt_{\ov 0 \ov 1}}}{(2\pi)^{4}} e^{i \Kt_{\ov 0 \ov 1} \vdot \xt_{\overline 0 \overline 1}-i\Kt_{01} \vdot \xt_{0 1}}
      \frac{\theta(z_0 -z_{\ov 0} - \alpha)+\theta(z_{\ov 0} -z_{ 0} - \alpha)}{(z_0 -z_{\ov 0})^2}
         \frac{ \delta(M_X^2-M_2^2)}{[M_0^2-M_2^2-i \delta] }
\end{split}
\end{equation}
where we have kept the dependence on the cut-off $\alpha$ explicit.
These can be combined with the similar trick as in Eq.~\eqref{eq:Beuf_trick},
\begin{equation}
\begin{split}
    &\frac{\delta(M_X^2-M_2^2)}{M_0^2-M_2^2-i \delta}+\frac{\delta(M_X^2-M_0^2)}{M_2^2-M_0^2+i \delta}\\
    =& \frac{1}{2\pi i} \left[
    \frac{1}{\left(M_X^2-M_0^2+i\delta \right)\left(M_X^2-M_2^2+i\delta \right)} 
    -\frac{1}{\left(M_X^2-M_0^2-i\delta \right)\left(M_X^2-M_2^2-i\delta \right)} \right]\\
    =& -D_{02},
\end{split}
\end{equation}
and hence
\begin{equation}
\label{eq:F_C02}
\begin{split}
    F_{\czero} + F_{\ctwo}
    =& -16 \pi^2 \times
      \frac{ 2z_0 z_1 z_{\ov 0} z_{\ov 1} }{[  z_0 z_1 z_{\ov 0} z_{\ov 1}]^{3/2}}
      \times \int \frac{\dd[2]{\Kt_{01}} \dd[2]{\Kt_{\ov 0 \ov 1}}}{(2\pi)^{4}} e^{i \Kt_{\ov 0 \ov 1} \vdot \xt_{\overline 0 \overline 1}-i\Kt_{01} \vdot \xt_{0 1}}
       \frac{\theta(z_0 -z_{\ov 0} - \alpha)+\theta(z_{\ov 0} -z_{ 0} - \alpha)}{(z_0 -z_{\ov 0})^2}
        D_{02}. 
\end{split}
\end{equation}
As $D_{\adiag 02} =  D_{\bdiag 02}  = D_{02} $, it is natural to combine the terms containing these energy denominators in Eqs.~\eqref{eq:F_A012}, \eqref{eq:F_B012}, and \eqref{eq:F_C02}.
The terms differ only in their dependence on the momentum fractions $z_i$ and $z_{\ov i}$, and we can write their sum in a way where the power law divergence at $z_0=z_{\ov 0}$ cancels as
\begin{equation}
\begin{split}
    F^{02}
    =&  8\pi^2 \times  \frac{z_0 z_{\ov 0} + z_{1} z_{\ov 1}}{[  z_0 z_1 z_{\ov 0} z_{\ov 1}]^{3/2}} 
        \times \int \frac{\dd[2]{\Kt_{01}} \dd[2]{\Kt_{\ov 0 \ov 1}}}{(2\pi)^{4}} e^{i \Kt_{\ov 0 \ov 1} \vdot \xt_{\overline 0 \overline 1}-i\Kt_{01} \vdot \xt_{0 1}} D_{02} \\
    =& \frac{\pi}{2} \frac{z_0 z_{\ov 0} + z_{1} z_{\ov 1}}{\sqrt{  z_0 z_1 z_{\ov 0} z_{\ov 1}}}
    \left[
    J_0 \left( M_X \abs{\xt_{01}} \sqrt{z_0 z_1} \right)
    Y_0 \left( M_X \abs{\xt_{\ov 0\ov 1}} \sqrt{z_{\ov 0 }z_{\ov 1}} \right)
    +
    Y_0 \left( M_X \abs{\xt_{01}} \sqrt{z_0 z_1} \right)
    J_0 \left( M_X \abs{\xt_{\ov 0\ov 1}} \sqrt{z_{\ov 0 }z_{\ov 1}} \right)
    \right].
\end{split}
\end{equation}
Note that this expression is finite.

The terms involving three energy denominators in Eqs.~\eqref{eq:F_A012} and \eqref{eq:F_B012} are finite and can be calculated using Schwinger or Feynman parametrizations. Evaluating the integrals in this way is not unique, and different forms can be numerically more demanding than others. 
Our calculation for these terms involving three energy denominators, denoted by $ F_{\adiag}^{012}$ and $F_{\bdiag}^{012}$, leads to the forms in Eqs.~\eqref{eq:F_A012_term} and \eqref{eq:F_B012_term} of the final result.

The remaining $\log^2 \alpha$ divergences in Eqs.~\eqref{eq:F_A012} and \eqref{eq:F_B012} are more complicated to cancel, as they will combine with Diagrams~$\done$ and $\eone$ which have a different structure in the transverse momentum integrals. 
For this reason, we choose to Fourier transform the remaining terms first and then show that we can extract the divergences from the integrated expressions.
The different $D_{ij}$ terms give us:
\begin{equation}
\begin{split}
    F_{\adiag 01}=  
    &  \frac{\theta(z_0 -z_{\ov 0} - \alpha)}{z_0- z_{\ov 0}} \frac{z_0 z_1 + z_{\ov 0} z_{\ov 1}}{\sqrt{  z_0 z_1 z_{\ov 0} z_{\ov 1}}} 
    \times \frac{1}{2} \int_0^1 \frac{\dd{t}}{t}
    J_0 \left(M_X \sqrt{ z_0 z_1 \left( \frac{z_{\ov 0}}{z_0} \xt_{\ov 0 \ov 1} -\xt_{01} \right)^2 + \frac{z_0 -z_{\ov 0}}{t} \frac{z_{\ov 0}}{z_0} \xt_{\ov 0 \ov 1}^2 } \right),
\end{split}
\end{equation}
\begin{equation}
\begin{split}
    F_{\adiag 21}=
     &  \frac{\theta(z_0 -z_{\ov 0} - \alpha)}{z_0- z_{\ov 0}} \frac{z_0 z_1 + z_{\ov 0} z_{\ov 1}}{\sqrt{  z_0 z_1 z_{\ov 0} z_{\ov 1}}} \times \frac{1}{2} \int_0^1 \frac{\dd{t}}{t}
    J_0 \left(M_X \sqrt{ z_{\ov 0} z_{\ov 1} \left( \xt_{\ov 0 \ov 1} - \frac{z_1}{z_{\ov 1}} \xt_{01} \right)^2 + \frac{z_0 -z_{\ov 0}}{t} \frac{z_1}{z_{\ov 1}} \xt_{ 0  1}^2 } \right),
\end{split}
\end{equation}
\begin{equation}
\begin{split}
    F_{\bdiag 01}=
     &\frac{\theta(z_{\ov 0} -z_{0} - \alpha)}{z_{\ov 0}- z_{0}} \frac{z_0 z_1 + z_{\ov 0} z_{\ov 1}}{\sqrt{  z_0 z_1 z_{\ov 0} z_{\ov 1}}} \times \frac{1}{2} \int_0^1 \frac{\dd{t}}{t}
    J_0 \left(M_X \sqrt{ z_0 z_1 \left( \frac{z_{\ov 1}}{z_1} \xt_{\ov 0 \ov 1} -\xt_{01} \right)^2 + \frac{z_{\ov 0} - z_0}{t} \frac{z_{\ov 1}}{z_1} \xt_{\ov 0 \ov 1}^2 } \right),
\end{split}
\end{equation}
and
\begin{equation}
\begin{split}
    F_{\bdiag 21}=
    &\frac{\theta(z_{\ov 0} -z_{0} - \alpha)}{z_{\ov 0}- z_{0}} \frac{z_0 z_1 + z_{\ov 0} z_{\ov 1}}{\sqrt{  z_0 z_1 z_{\ov 0} z_{\ov 1}}} \times \frac{1}{2} \int_0^1 \frac{\dd{t}}{t}
    J_0 \left(M_X \sqrt{ z_{\ov 0} z_{\ov 1} \left( \xt_{\ov 0 \ov 1} - \frac{z_0}{z_{\ov 0}} \xt_{01} \right)^2 + \frac{z_{\ov 0}- z_0}{t} \frac{z_0}{z_{\ov 0}} \xt_{ 0  1}^2 } \right).
\end{split}
\end{equation}
To extract the $\alpha$ divergences from $F_{\adiag}^{01} $, $F_{\adiag}^{21} $, $F_{\bdiag}^{01} $, and $F_{\bdiag}^{21} $ we use one of the following identities (see Appendix~\ref{sec:divergence})
\begin{equation}
\begin{split}
\label{eq:simplifying_identity}
    \int_0^1 \frac{\dd{t}}{t} J_0 \left( \sqrt{a^2 + \frac{1}{t}b^2} \right) 
    =&\, 
    -\int_{0}^{1} \frac{\dd{t}}{t} \qty[J_0 \qty( \sqrt{a^2 + t b^2}) - J_0(a) ]
    - J_0( a ) \log( \frac{b^2 e^{\gamma_E}}{2a} )
    - \frac{\pi}{2} Y_0(a)
    \\
    =&\,
    2\int_0^1 \frac{\dd{t}}{t} \left[ 1 - J_0(tb) \right] J_0\left(a \sqrt{1-t^2}\right) 
    - J_0( a ) \log( \frac{b^2 e^{\gamma_E}}{2a} )
    - \frac{\pi}{2} Y_0(a)
\end{split}    
\end{equation}
which works for positive $a$ and $b$.
This allows us to write
\begin{equation}
    F_{\adiag 01} + F_{\adiag 21} + F_{\bdiag 01}  + F_{\bdiag 21} 
    = \Ft + \Fsub + \Fsoft,
\end{equation}
where the finite parts $\Ft$ and $\Fsub$ are defined in Eqs.~\eqref{eq:Ft} and \eqref{eq:Fsub}. The part $\Ft$ contains the $t$-integral in Eq.~\eqref{eq:simplifying_identity}, and the rest of Eq.~\eqref{eq:simplifying_identity} is divergent in the limit $\alpha \to 0$. To make canceling the divergences easier, we subtract from this divergent part the finite term $\Fsub$. 
The divergences are then defined to be part of $\Fsoft$ which reads
\begin{equation}
\begin{split}
    \Fsoft =&  -4 \delta(z_0 -z_{\ov 0}) \int_\alpha^1 \frac{\dd{z_2'}}{z_2'} 
    \Bigg\{
    J_0(M_X \abs{\xt_{01} - \xt_{\ov 0 \ov 1}} \sqrt{z_0 z_1})
    \log(\frac{z_2' M_X \abs{\xt_{01}} \abs{ \xt_{\ov 0 \ov 1}}e^{\gamma_E}}{2
    \abs{\xt_{01} - \xt_{\ov 0 \ov 1}} \sqrt{z_0 z_1}}) \\
    &+ \frac{\pi}{2} Y_0 \left(M_X \abs{\xt_{01} - \xt_{\ov 0 \ov 1}} \sqrt{z_0 z_1} \right)
    \Bigg\}\\
    =&  4 \delta(z_0 -z_{\ov 0}) \log \alpha
    \Bigg\{
    J_0(M_X \abs{\xt_{01} - \xt_{\ov 0 \ov 1}} \sqrt{z_0 z_1})
    \left[
    \log(\frac{M_X\abs{\xt_{01}} \abs{\xt_{\ov 0 \ov 1}}e^{\gamma_E}}{2 \abs{\xt_{01} - \xt_{\ov 0 \ov 1}} \sqrt{z_0 z_1}}) +\frac{1}{2}  \log \alpha    
    \right]\\
    &+ \frac{\pi}{2} Y_0 \left(M_X \abs{\xt_{01} - \xt_{\ov 0 \ov 1}} \sqrt{z_0 z_1} \right)
    \Bigg\}.
\end{split}
\end{equation}
The division between $\Fsub$ and $\Fsoft$ is not unique, and
we have done the division at the integral level where the gluon plus-momentum fraction $z_2' = \abs{z_0 - z_{\ov 0}}$ has not been integrated over. 

To cancel the remaining logarithmic divergences we need to consider Diagrams~$\done$ and $\eone$.
They contain collinear divergences and thus the momentum integrals have to be done in $2-2\varepsilon$ dimensions. Fortunately, the momentum integrals are quite straightforward when written in terms of the natural variables of the diagrams. For the diagram $\done$ these are the relative momentum of the quark and gluon $\Pt_{02}$, and the relative momentum of the quark and antiquark before the gluon emission $\Kt_{01} = \Pt_{01} - \Pt_{12}$. Similarly, for the diagram $\eone$ the natural variables are the relative momentum of the antiquark and gluon $\Pt_{12}$, and the relative momentum of the quark and antiquark before the gluon emission $\Kt_{01} = \Pt_{01} + \Pt_{02}$.
In terms of these variables, the integrals can be written as
\begin{equation}
    \begin{split}
        F_{\done}
        =&  8\pi^2  \delta(z_0- z_{\overline 0}) \int_0^1 \dd{ z_{0p}} \dd{z_{1p}} \int_\alpha^1 \dd{ z_{2p}}  \delta\left(z_1 -z_{1p} \right) \delta(1-z_{0p}-z_{1p}-z_{2p})
        \times \frac{1}{z_0^{2-2\varepsilon}} \times \frac{2}{z_0 z_1 z_{2p}} \left[ z_0^2 +z_{0p}^2 -\varepsilon \deltas  z_{2p}^2 \right] \\
        & \times \int \frac{\dd[2-2\varepsilon]{\Pt_{02}} \dd[2-2\varepsilon]{ \Kt_{01}}}{(2\pi)^{2(2-2\varepsilon)}} 
        e^{i \Kt_{01} \vdot (\xt_{\overline 0 \overline 1}-\xt_{0 1})} 
        \times  \mu^{2\varepsilon} 
        \delta\left(M_X^2 - M_{\done}^2 \right) 
        \frac{1}{z_0 z_{0p} z_{2p}}
        \frac{M_{\done}^2-M_0^2}{[M_{\done}^2-M_0^2-i \delta][M_{\done}^2-M_0^2+i \delta]}\\
        =&   \delta(z_0- z_{\overline 0}) \int_0^1 \dd{ z_{0p}} \dd{z_{1p}} \int_\alpha^1 \dd{ z_{2p}}  \delta\left(z_1 -z_{1p} \right) \delta(1-z_{0p}-z_{1p}-z_{2p})
        \times 
        \frac{-1}{\varepsilon}  \frac{1}{z_{2p} z_0^2}
        \left[ z_0^2 +z_{0p}^2 -\varepsilon \deltas  z_{2p}^2 \right] \\
        & \times \left( \frac{M_X^2}{4\pi^2 \abs{\xt_{01}- \xt_{\ov 0 \ov 1}}^2 \mu^2} \frac{z_{0p} z_{2p}}{z_0} \right)^{-\varepsilon} J_{-2\varepsilon} \left(  \abs{\xt_{01}- \xt_{\ov 0 \ov 1}} M_X \sqrt{z_0 z_1} \right) \\
        =&   \delta(z_0- z_{\overline 0})
        \frac{-1}{\varepsilon}
        \left( \frac{M_X^2}{4\pi^2 \abs{\xt_{01}- \xt_{\ov 0 \ov 1}}^2 \mu^2} \right)^{-\varepsilon} J_{-2\varepsilon} \left(  \abs{\xt_{01}- \xt_{\ov 0 \ov 1}} M_X \sqrt{z_0 z_1} \right) \\
        &\times \frac{1}{2} \left[ -3 - 4 \log \frac{\alpha}{z_0}  +\varepsilon
        \left(-\deltas -6 +\frac{2}{3} \pi^2 +3\log z_0 -2 \log^2 z_0 + 2\log^2\alpha \right) \right]
    \end{split}
\end{equation}
and
\begin{equation}
    \begin{split}
        F_{\eone}
        =&  8\pi^2 \delta(z_0- z_{\overline 0}) \int_0^1 \dd{ z_{0p}} \dd{z_{1p}} \int_\alpha^1 \dd{ z_{2p}}  \delta\left(z_0 - z_{0p} \right) \delta(1-z_{0p}-z_{1p}-z_{2p})
        \times \frac{1}{z_1^{2-2\varepsilon}} \times  \frac{2}{z_0 z_1 z_{2p}} \left[ z_1^2 +z_{1p}^2 -\varepsilon \deltas  z_{2p}^2 \right] \\
        & \times \int \frac{\dd[2-2\varepsilon]{\Kt_{01}} \dd[2-2\varepsilon]{\Pt_{12}}}{(2\pi)^{2(2-2\varepsilon)}} e^{i \Kt_{01} \vdot (\xt_{\overline 0 \overline 1}-\xt_{0 1})}
        \times  \mu^{2\varepsilon} 
        \delta\left(M_X^2 - M_{\eone}^2 \right) 
        \frac{1}{z_1 z_{1p} z_{2p}}\frac{M_{\eone}^2-M_0^2}{[M_{\eone}^2-M_0^2 -i \delta][M_{\eone}^2-M_0^2+i \delta]} \\
        =&  \delta(z_0- z_{\overline 0}) \int_0^1 \dd{ z_{0p}} \dd{z_{1p}} \int_\alpha^1 \dd{ z_{2p}}  \delta\left(z_0 -z_{0p} \right) \delta(1-z_{0p}-z_{1p}-z_{2p})
        \times 
        \frac{-1}{\varepsilon} \frac{1}{z_{2p} z_1^2}
        \left[ z_1^2 +z_{1p}^2 -\varepsilon \deltas  z_{2p}^2 \right] \\
        &\times \left( \frac{M_X^2}{4\pi^2 \abs{\xt_{01}- \xt_{\ov 0 \ov 1}}^2 \mu^2} \frac{z_{1p} z_{2p}}{z_1} \right)^{-\varepsilon} J_{-2\varepsilon} \left(  \abs{\xt_{01}- \xt_{\ov 0 \ov 1}} M_X \sqrt{z_0 z_1} \right) \\
        =&  \delta(z_0- z_{\overline 0})
        \frac{-1}{\varepsilon}
        \left( \frac{M_X^2}{4\pi^2 \abs{\xt_{01}- \xt_{\ov 0 \ov 1}}^2 \mu^2} \right)^{-\varepsilon} J_{-2\varepsilon} \left(  \abs{\xt_{01}- \xt_{\ov 0 \ov 1}} M_X \sqrt{z_0 z_1} \right) \\
        &\times \frac{1}{2} \left[ -3 - 4 \log \frac{\alpha}{z_1}  +\varepsilon
        \left(-\deltas -6 +\frac{2}{3} \pi^2 +3\log z_1 -2 \log^2 z_1 + 2\log^2\alpha \right) \right].
    \end{split}
\end{equation}
Here the energy denominators are written in terms of the invariant masses
\begin{align}
    M_{\done}^2 &= \frac{\Kt_{01}^2}{z_0 z_1} + \frac{\Pt_{02}^2}{z_0 z_{0p} z_{2p}},
    &
    M_{\eone}^2 &= \frac{\Kt_{01}^2}{z_0 z_1} + \frac{\Pt_{12}^2}{z_1 z_{1p} z_{2p}},
\end{align}
and substituting these leads to the following simple expressions:
\begin{align}
    M_{\done}^2 - M_0^2 &=  \frac{\Pt_{02}^2}{z_0 z_{0p} z_{2p}},
    &
    M_{\eone}^2 - M_0^2 &= \frac{\Pt_{12}^2}{z_1 z_{1p} z_{2p}}.
\end{align}
This simple form allows us to perform the momentum integrals analytically.

We note that the other terms that contain divergences in $\varepsilon$, i.e. Eqs.~\eqref{eq:cross_section_photon_wf_nlo} and \eqref{eq:UV_divergence_cross_section},  can be written as the leading-order part times a divergent NLO part.
To cancel these divergences,
it is convenient to expand $F_{\done}$ and $F_{\eone}$ in terms of $\varepsilon$ in such a way that we have a similar structure as in Eq.~\eqref{eq:LO_cross_section}.
This leads to
\begin{equation}
\label{eq:doneeone}
    \begin{split}
        F_{\done} + F_{\eone}
        =&\delta(z_0- z_{\overline 0}) \times  \Bigg\{
        \frac{1}{\varepsilon} \left( 3 + 2 \log \frac{\alpha^2}{z_0 z_1}\right) 
        \left( \frac{\mxbar }{2\pi \abs{\xt_{01}- \xt_{\ov 0 \ov 1}}} \right)^{-\varepsilon} J_{-\varepsilon} \left( \mxbar \abs{\xt_{01}- \xt_{\ov 0 \ov 1}} \right) \\
        &- \left(3 + 2 \log \frac{\alpha^2}{z_0 z_1}\right) 
        \left[   \log( \frac{ M_X }{2\pi \mu^2 \abs{\xt_{01}-\xt_{\ov 0 \ov 1}} \sqrt{z_0 z_1} } )
         J_0 \left( \mxbar  \abs{\xt_{01}- \xt_{\ov 0 \ov 1}}  \right)
        + \frac{\pi}{2} Y_0\left( \mxbar \abs{\xt_{01}-\xt_{\ov 0 \ov 1}} \right) \right]
        \\
        &
        +
        J_0 \left( \mxbar  \abs{\xt_{01}- \xt_{\ov 0 \ov 1}} \right)
        \left[ \deltas + 6 -\frac{2}{3} \pi^2 -\frac{3}{2}\log(z_0 z_1) + \log^2 z_0 +\log^2 z_1
        - 2\log^2\alpha 
        \right]
        \Bigg\}
    \end{split}
\end{equation}
where in the first row we have not expanded all of the $\varepsilon$-dependence in anticipation of canceling this with the other divergent terms.
It should be noted that the $\varepsilon$-divergences here are collinear divergences.
If we treated collinear and UV divergences separately, we would notice that the collinear divergences in the renormalization coefficients $Z_i$ would cancel these collinear divergences exactly, effectively converting them to UV divergences.
Thus, we can consistently combine the $\varepsilon$-divergences in this expression with the rest of the calculation where the $\varepsilon$-divergences have their origin in the UV region.

This concludes calculating the final-state corrections in Fig.~\ref{fig:final_state_NLO}.
In total, we can write
\begin{align}
    F &= F^\text{fin} + F^\text{div}, \\
    F^\text{fin} &=\Ft + \Fsub + F^{02}+ F_\adiag^{012}+ F_\bdiag^{012},  \\
    F^\text{div} &= F_{\done} + F_{\eone} + \Fsoft,
\end{align}
where $F^\text{div}$ contains the divergent terms and the remaining finite term $F^\text{fin}$ gives us Eq.~\eqref{eq:f_final_state} in the final result.
The divergent part contributes to the cross section by
\begin{equation}
\label{eq:final_state_divergence}
    \begin{split}
       &\left[ \frac{\dd{ \sigma^\textrm{D}_{\gamma^*_\lambda +A}}}{\dd[2]{\Deltat}\dd{M_X^2}} \right]^{\text{final-state divergences}}_{\abs{\ref{diag:gamma_qq_qq}}^2+\abs{\ref{diag:gamma_qq_qqg}}^2}\\ 
       &=\sum_f
       \frac{\nc}{(4\pi)^2} \frac{\as \cf}{2\pi} 
       \int_0^1 \dd{z_0} \dd{z_1} \delta(1-z_0-z_1)
       \int 
       \dd[2-2\varepsilon]{\xt_{01}}
       \dd[2-2\varepsilon]{\xt_{\ov{01}}}
       \dd[2-2\varepsilon]{\bt}
       \dd[2-2\varepsilon]{\ov \bt}\\
       &\times 
\widetilde \psi_\iin^{\gamma_\lambda^{\ast} \to q \bar q} (\xt_{01},z_{i})
\left[\widetilde \psi_\iin^{\gamma_\lambda^{\ast} \to q \bar q}( \xt_{\overline 0\overline 1},z_{i})\right]^\ast 
\left(1-\dipole_{01}\right)
\left(1-\dipole_{\ov 0 \ov 1}\right)^\dag 
\times\frac{1}{(2\pi)^{2-2\varepsilon}} e^{i \Deltat \vdot (\ov \bt -\bt)} \\
& \times 
\Biggl\{
\left( \frac{\mxbar}{2\pi \abs{\xt_{\ov{01}}-\xt_{01}}}  \right)^{-\varepsilon} J_{-\varepsilon}\left( \mxbar \abs{\xt_{\ov{01}}-\xt_{01}}\right)
\left[ \frac{1}{\varepsilon} + \log(\pi \mu^2 \abs{\xt_{01}} \abs{\xt_{\ov{01}}}e^{\gamma_E} )\right]
\left( 3 + 2 \log \frac{\alpha^2}{z_0 z_1}\right)
\\
&+\left(2 \log(z_0 z_1)-3 \right)\left[
\log(\frac{M_X e^{\gamma_E}  \abs{\xt_{01}}\abs{\xt_{\ov 0\ov 1}} }{2 \abs{\xt_{01}-\xt_{\ov 0\ov 1}} } ) 
 J_{0}\left( \mxbar \abs{\xt_{\ov{01}}-\xt_{01}}\right)
 +
\frac{\pi}{2} Y_0\left( \mxbar \abs{\xt_{\ov{01}}-\xt_{01}}\right) \right] \\
&+J_{0}\left(\mxbar \abs{\xt_{\ov{01}}-\xt_{01}}\right)
\left[\deltas +6 -\frac{2}{3}\pi^2  -2 \log z_0 \log z_1 \right]
\Biggr\}.
    \end{split}
\end{equation}
We note that the $\log^2 \alpha$ divergences between $ F_{\done}$, $F_{\eone}$,  and  $\Fsoft$ have canceled in this expression.
Let us now show that the $\varepsilon$-divergences cancel with the rest of the calculation. 
The sum of all remaining $\varepsilon$-divergences,
Eqs.~\eqref{eq:UV_divergence_cross_section}, \eqref{eq:cross_section_photon_wf_nlo} and \eqref{eq:final_state_divergence}, is
\begin{equation}
\begin{split}
    & \left[\frac{\dd[]{\sigma^\textrm{D}_{\gamma^*_\lambda +A}}}{\dd[2]{\Deltat}\dd{M_X^2}} \right]_{\ref{diag:gamma_qqg_qq}_\text{UV} \times \ref{diag:gamma_qq_qq}^* + 
    \ref{diag:gamma_qq_qq} \times \ref{diag:gamma_qqg_qq}_\text{UV}^*
    }
    +\left[\frac{\dd[]{\sigma^\textrm{D}_{\gamma^*_\lambda +A}}}{\dd[2]{\Deltat}\dd{M_X^2}} \right]_{\abs{\ref{diag:gamma_qq_qq}}^2}^{\text{$\gamma^* \to q\bar q$ at NLO}}
    + \left[ \frac{\dd{ \sigma^\textrm{D}_{\gamma^*_\lambda +A}}}{\dd[2]{\Deltat}\dd{M_X^2}} \right]^{\text{final-state  divergences}}_{\abs{\ref{diag:gamma_qq_qq}}^2+\abs{\ref{diag:gamma_qq_qqg}}^2}\\
    =&\sum_f
    \frac{\nc}{(4\pi)^2} \frac{\as \cf}{2\pi} \int \dd[2]{\xt_{01}}\dd[2]{\xt_{\overline 0 \overline 1}}\dd[2]{\bt}\dd[2]{\ov \bt} \int_0^1  \dd[]{z_{0}} \dd[]{z_{1}} \delta(1-z_0 -z_1)\\
    &\times \widetilde \psi_\iin^{\gamma^*_\lambda \to q \bar q }(\xt_{01}, z_i) \left[\widetilde \psi_\iin^{\gamma^*_\lambda \to q \bar q }(\xt_{\overline 0 \overline 1},z_i) \right]^*  
\left(1-\dipole_{01}\right)
\left(1-\dipole_{\ov 0 \ov 1}\right)^\dag 
\times\frac{1}{(2\pi)^{2}}e^{i \Deltat \vdot (\ov \bt -\bt)}
\\
    &\times 
    \Biggl\{
    \left(2\log(z_0 z_1)-3 \right)
    \left[\log(\frac{M_X e^{\gamma_E}  \abs{\xt_{01}}\abs{\xt_{\ov 0\ov 1}} }{2 \abs{\xt_{01}-\xt_{\ov 0\ov 1}} }) 
    J_{0} \left( \mxbar |\xt_{01}-\xt_{\overline 0 \overline 1}| \right)
    +\frac{\pi}{2} 
    Y_{0} \left( \mxbar |\xt_{01}-\xt_{\overline 0 \overline 1}| \right)
    \right]\\
    &+\left[11 - \pi^2 +\log^2 z_0 + \log^2 z_1 - 4\log z_0 \log z_1 \right]
     J_{0} \left( \mxbar  |\xt_{01}-\xt_{\overline 0 \overline 1}| \right)
    \Biggr\}
\end{split}
\end{equation}
which is finite: both the $1/\varepsilon$ ad the $\log \alpha$ have canceled. Note that the dependence on the regularization scheme (i.e. on $\deltas$) has also canceled.
Substituting the photon wave functions and using Eq.~\eqref{eq:wave_function_sum} gives us Eq.~\eqref{eq:f_UV}.

\subsection{JIMWLK evolution}
\label{sec:JIMWLK}

While we have now computed all of the contributions to the final result, there is still one divergence left in the dip-trip term~\eqref{eq:diptrip} corresponding to the limit $z_2 \to 0$.
This is a well-known rapidity divergence that has to be absorbed into the JIMWLK evolution of the Wilson lines.
The JIMWLK equation for a general Wilson line operator $\hat{\mathcal{O}}$ reads as~\cite{Jalilian-Marian:1996mkd,Jalilian-Marian:1997jhx,Jalilian-Marian:1997qno,Iancu:2001md, Ferreiro:2001qy, Iancu:2001ad, Iancu:2000hn,Mueller:2001uk} 
\begin{equation}
    \label{eq:JIMWLK_gen}
    \partial_Y \left\langle \hat{\mathcal{O}} \right\rangle = -  \left\langle \hjimwlk \hat{\mathcal{O}} \right\rangle,
\end{equation}
where $\hjimwlk$ is the JIMWLK Hamiltonian.
For diffractive DIS, the relevant operator is either $\hat{\mathcal{O}} = \dipole_{01}$ or $\hat{\mathcal{O}} = \dipole_{01} \dipole_{\ov 0 \ov 1}$ depending on whether we want the total or  coherent diffractive cross section as explained in Sec.~\ref{sec:final_results}.
In both cases the rapidity evolution simplifies to the following form
\begin{multline}
\label{eq:JIMWLK}
    \partial_Y \left[\left(1-\dipole_{01}\right)\left(1-\dipole_{\ov 0 \ov 1}\right)^\dag   \right] 
    =-  \hjimwlk \left[ \left(1-\dipole_{01}\right)\left(1-\dipole_{\ov 0 \ov 1}\right)^\dag  \right] \\
    = - \left[ \hjimwlk  \left(1-\dipole_{01}\right)\right]\left(1-\dipole_{\ov 0 \ov 1}\right)^\dag  
    -
     \left(1-\dipole_{01}\right)\left[ \hjimwlk\left(1-\dipole_{\ov 0 \ov 1}\right)^\dag \right], 
\end{multline}
where one should insert the average $\langle \dots \rangle$ over the target color configuration at the appropriate level.
Essentially, this means that we can  consider separately the JIMWLK evolution of the dipole amplitude in the invariant amplitude and its complex conjugate. 
The JIMWLK operator acts on the dipole amplitude as
\begin{equation}
   - \hjimwlk  \left(1-\dipole_{01}\right)
    = \frac{\as \cf}{\pi^2} \int \dd[2]{\xt_{2}} \frac{\xt_{10}^2}{\xt_{20}^2\xt_{21}^2} 
   \left[ \dipole_{01} -  \tripole_{012} \right].
\end{equation}
We now wish to show that this cancels the remaining rapidity divergence.

The divergent part of the dip-trip term can be found by taking the limit $z_2 \to 0$ in  $ \mathcal{G}^{\nlo}_{\lambda,\text{dip-trip}}$ and $ \mathcal{G}^{\nlo}_{\lambda,\text{dip sub}}$.
For $ \mathcal{G}^{\nlo}_{\lambda,\text{dip-trip}}$ we find
\begin{equation}
    \lim_{z_2 \to 0} \mathcal{G}^{\nlo}_{\lambda,\text{dip-trip}}
    = \frac{4}{z_2} 
\frac{\xt_{01}^2}{\xt_{20}^2 \xt_{21}^2} \mathcal{G}^{\lo}_{\lambda,\text{dip}},
\end{equation}
and for $ \mathcal{G}^{\nlo}_{\lambda,\text{dip sub}}$ we get
\begin{equation}
    \lim_{z_2 \to 0} \mathcal{G}^{\nlo}_{\lambda,\text{dip sub}}
    = \frac{4}{z_2} 
    \left[ 
    \frac{1}{\xt_{20}^2} \exp( - \frac{\xt_{20}^2}{\xt_{01}^2 e^{\gamma_E}} )
    +
    \frac{1}{\xt_{21}^2} \exp( - \frac{\xt_{21}^2}{\xt_{01}^2 e^{\gamma_E}} )
    \right]
    \mathcal{G}^{\lo}_{\lambda,\text{dip}}.
\end{equation}
Using the integral~\cite{Hanninen:2017ddy}
\begin{equation}
    \int \dd[2]{\xt_2} \left[ \frac{1}{\xt_{20}^2} \exp( - \frac{\xt_{20}^2}{\xt_{01}^2 e^{\gamma_E}} )
    +
    \frac{1}{\xt_{21}^2} \exp( - \frac{\xt_{21}^2}{\xt_{01}^2 e^{\gamma_E}} )
    - \frac{\xt_{01}^2}{\xt_{20}^2 \xt_{21}^2} \right] =0
\end{equation}
we note that the divergent part of the dip-trip term reduces to
\begin{equation}
\label{eq:diptrip-divergence}
\begin{split}
      \biggl [ \frac{\dd[]{ \sigma^{\text{D}}_{\gamma^*_{\lambda} + A} }}{\dd[2]{\Deltat} \dd{M_X^2}}  \biggr ]_{\text{dip-trip}}^{z_2 \to 0} 
      =& 
         2 \pi\aem  \nc \sum_f e_f^2
      \int [\mathrm{dPS}]_{\text{dip-trip}} \frac{\as \cf}{2\pi}
       \frac{4}{z_2} 
\frac{\xt_{01}^2}{\xt_{20}^2 \xt_{21}^2} \mathcal{G}^{\lo}_{\lambda,\text{dip}}
      \biggl \{ 2\Re\left(\dipole_{01}-\tripole_{012}\right)
\left(1-\dipole_{\ov 0 \ov 1 }\right)^\dag  \biggr \}
 \\
      =& 
         2 \pi\aem  \nc \sum_f e_f^2
      \int [\mathrm{dPS}]_{\text{dip}}\mathcal{G}^{\lo}_{\lambda,\text{dip}}
      \times\frac{\as \cf}{\pi^2} \int \dd[2]{\xt_2} 
       \int\frac{\dd{z_2}}{z_2} 
\frac{\xt_{01}^2}{\xt_{20}^2 \xt_{21}^2} 
      \biggl \{ 2\Re\left(\dipole_{01}-\tripole_{012}\right)
\left(1-\dipole_{\ov 0 \ov 1 }\right)^\dag  \biggr \}
 \\
      =& 
         2 \pi\aem  \nc \sum_f e_f^2
      \int [\mathrm{dPS}]_{\text{dip}}\mathcal{G}^{\lo}_{\lambda,\text{dip}}
       \int\frac{\dd{z_2}}{z_2} 
\partial_Y \left[\left(1-\dipole_{01}\right)\left(1-\dipole_{\ov 0 \ov 1}\right)^\dag   \right] 
.
\end{split}
\end{equation}
Noting that $Y = \log 1/z_2$ we then see that this term gives exactly the rapidity evolution of the Wilson lines in the leading-order term of Eq.~\eqref{eq:dip}.
This means that this rapidity divergence should be absorbed to the rapidity dependence of the dipole amplitudes, rendering the cross section finite.
The exact procedure of doing this is, however, scheme dependent and therefore we leave the choice of the scheme for numerical applications.

\subsection{Recap on the cancellation of divergences}

Here we present a brief recap on the structure of different divergences present in the calculation, and how they combine to yield a finite result.
In total, the different divergences we have correspond to:
\begin{itemize}
    \item Divergences in the initial state coming from gluon loops before the shockwave. These are contained in the NLO photon wave function $ \psi_\nlo^{\gamma^*_\lambda \to q \bar q}$, and they are part of the contribution  $\abs{\ref{diag:gamma_qq_qq}}^2$.
    Both UV divergences in $\varepsilon$ and divergences in the gluon plus momentum of the form  $\log \alpha$ are present.
    \item Divergences from gluons crossing the shockwave that are not present in the final state are part of the  $\ref{diag:gamma_qq_qq} \times \ref{diag:gamma_qqg_qq}^*$ and $\ref{diag:gamma_qqg_qq} \times \ref{diag:gamma_qq_qq}^*$ contributions.
    These contain $\log \alpha$ divergences, and also UV divergences in $\varepsilon$ if the gluon is emitted and absorbed by the same particle.
    \item Divergences from the renormalization of the Wilson lines.
    These have a $\log \alpha$ rapidity divergence related to the JIMWLK evolution of the Wilson lines.
    \item Divergences from self-energy corrections for the quark and the antiquark. These appear in the renormalization constant but vanish in dimensional regularization with one $\varepsilon$ for IR and UV.
    If one were to consider IR and UV divergences separately, the IR divergences would cancel with the final-state corrections whereas the UV divergences would cancel with the rest of the calculation.
    \item Divergences in the final state where gluon emission happens after the shockwave.
    These contain contributions from virtual gluons ($\abs{\ref{diag:gamma_qq_qq}}^2$) and also from real gluons in the final state ($\abs{\ref{diag:gamma_qq_qqg}}^2$).
    In total, the final-state corrections have collinear divergences in $\varepsilon$ along with $\log \alpha$ divergences.
\end{itemize}
If we combine the final-state divergences with the divergences from the quark and antiquark self-energy corrections, the $\varepsilon$ divergences are effectively turned into UV divergences, and then this combined divergence structure is exactly the same as in the initial state (compare Eqs.~\eqref{eq:cross_section_photon_wf_nlo} and \eqref{eq:final_state_divergence}).
This is not a coincidence: 
if we were to replace the final state with a photon (such as in deeply virtual Compton scattering), we would only need to replace the final-state corrections with the photon wave function and the rest of the calculation (and the divergences) would remain the same.
Thus, the divergence structure of the final-state corrections has to be the same as that of the  photon wave function  $ \psi_\nlo^{\gamma^*_\lambda \to q \bar q}$ for both processes to be finite.

We can also study the divergence structure of the final-state corrections further by separately considering the different diagrams in Fig.~\ref{fig:final_state_NLO} that contribute to the final state:
\begin{itemize}
    \item Regular gluon exchanges (Diagrams $\adiag$ and $\bdiag$): divergences of the form $1/\alpha$, along with $\log^2 \alpha$ and $\log \alpha$.
    \item Instantaneous gluon exchanges (Diagrams $\cdiag$): power divergences $1/\alpha$. These cancel with the ones from Diagrams $\adiag$ and $\bdiag$ as they are an artifact of doing the calculation using LCPT.
    \item Gluons in the final state that are emitted by the same particle in both the amplitude and the complex conjugate (Diagrams $\done$ and $\eone$).
    These have collinear divergences in $\varepsilon$, along with $\log^2 \alpha$ and $\log \alpha$ divergences.
    These diagrams have the structure of a ``cut self-energy corrections''; however, the cut regulates the UV region and thus only collinear divergences in $\varepsilon$ are present, as opposed to the actual self-energy corrections. 
    Combining these with Diagrams $\adiag$ and $\bdiag$ cancels  the $\log^2 \alpha$  divergences.
\end{itemize}
In total, the $1/\alpha$ and $\log^2 \alpha$ divergences that are present in the separate parts of the final-state corrections cancel when summed together, leaving only collinear $\varepsilon$ divergences and $\log \alpha$ divergences to be canceled with the other parts of the calculation.

\section{Conclusions}

We have in this paper calculated, for the first time, the diffractive DIS structure functions for both transverse and longitudinal photons to NLO accuracy in the dipole formalism. These results are fully general in the ``exact eikonal kinematics'', meaning that the interaction with the target is treated in the eikonal limit, but the partonic structure of the probe is treated exactly. In particular, our result is not just valid at high $Q^2$ or for high-$\ptt$ jets, but can be used in the region $Q^2\sim \qs^2$ that is most relevant for the physics of gluon saturation. Our result also allows a completely general impact parameter dependence. It includes as a subset the recent result \cite{Beuf:2022kyp} for the $q\bar{q}g$ component of the NLO cross section, which again generalizes earlier results in approximated kinematics already used in phenomenological studies. We show how the BK/JIMWLK evolution of the target color fields arises in the limit of the gluon momentum fraction becoming small. While we have not explicitly written our result in a way where this rapidity divergence would already be factorized into an evolution of the target, this is straightforward to do following the procedure developed for proton-nucleus collisions in Ref.~\cite{Iancu:2016vyg} and already adapted to the total DIS cross section in Refs.~\cite{Ducloue:2017ftk,Beuf:2020dxl}  and exclusive vector meson production in Refs.~\cite{Mantysaari:2021ryb,Mantysaari:2022bsp,Mantysaari:2022kdm}.

Our full result is presented in Sec.~\ref{sec:final_results} in an explicit form. However, due to the multidimensional coordinate integration involved, moving to a numerical evaluation and confrontation with HERA data still requires some work. The natural starting point will be to assume a factorizable impact parameter profile for a proton. Even then, the numerical integrations are more challenging than for the earlier approximate kinematics results, because there are still more coordinates to integrate over. One aspect that makes the diffractive cross section numerically more challenging than the total one is the oscillatory nature of the integrals: the wave function corresponding to a fixed invariant mass is an oscillatory function of the transverse coordinates. Nevertheless, we believe that these are merely technical problems that can be overcome with sufficient effort, and that the result of this work can become a crucial part of a global systematical understanding of small-$x$ DIS data in the saturation regime.

\begin{acknowledgments}
We thank F. Salazar for the discussions.
We are also grateful to
Y. Mulian for many  discussions that have prompted us to better clarify crucial aspects of this work.
T.L, H.M  and J.P have been supported by the Research Council of Finland, the Centre of Excellence in Quark Matter (project 346324) and projects 338263, 346567 and 321840.
J.P is also supported by the Finnish Cultural Foundation, by the National Science Foundation under grant No. PHY-1945471, and by the U.S. Department of Energy, Office of Science, Office of Nuclear Physics, within the framework of the Saturated Glue (SURGE) Topical Theory Collaboration. R.P is supported by the
Research Council of Finland projects 347499 and 353772.
G.B is supported in part by the National Science Centre (Poland) under the research grant no. 2020/38/E/ST2/00122 (SONATA BIS 10).
G.B  acknowledges financial support from MSCA RISE 823947 ``Heavy ion collisions: collectivity and precision in saturation physics'' (HIEIC).
This work was also supported under the European Union’s Horizon 2020 research and innovation programme by the European Research Council (ERC, grant agreement No. ERC-2018-ADG-835105 YoctoLHC) and by the STRONG-2020 project (grant agreement No. 824093).
The content of this article does not reflect the official opinion of the European Union and responsibility for the information and views expressed therein lies entirely with the authors. 
\end{acknowledgments}

\appendix

\section{Wave functions}
\label{sec:wave_functions}

In this Appendix we list all of the light-cone wave functions needed for the calculation. As a general notation, we denote the momenta for the initial-state photon by $q$, for asymptotic-state quarks and gluons by $p_i$, and for partons at $x^+=0$ by $k_i$.
Note that the transverse and plus momenta are conserved in LCPT meaning that $\qt = \sum_i \pt_i = \sum_i \kt_i$ and $q^+ = \sum_i p_i^+ = \sum_i k_i^+$, and these relations are implicitly assumed for the momenta in the wave functions.
The plus-momentum conservation makes it useful to use momentum fractions $z_{ik} = k^+_i / q^+$ and $z_{ip} = p^+_i / q^+$ which allows us to write plus-momentum conservation as $\sum_i z_{ip} = \sum_i z_{ik} = 1 $.
When these wave functions are used in the main text, we denote $z_{ik} \equiv z_i$ for simplicity if there is no fear of confusion.

\subsection{Photon wave functions}

The photon wave functions in the massless quark case have been calculated previously in Refs.~\cite{Beuf:2016wdz, Beuf:2017bpd, Hanninen:2017ddy}. 
The leading-order wave function $\gamma^* \to q \bar q$ for longitudinal photons reads
\begin{equation}
    \psi_\lo^{\gamma^*_L \to q \bar q}(\xt_{01}, z_{ik}) 
    = \frac{2 e e_f Q}{2\pi} z_{0k} z_{1k} \left( \frac{Q\sqrt{z_{0k} z_{1k} }}{2 \pi \abs{\xt_{01}}} \right)^{-\varepsilon}
    K_{-\varepsilon}\left( \abs{\xt_{01}} Q \sqrt{z_{0k} z_{1k}} \right)
    \frac{1}{2q^+}\bar u(k_0^+) \gamma^+ v(k_1^+),
\end{equation}
and for transverse photons it reads
\begin{equation}
\begin{split}
    \psi_\lo^{\gamma^*_T \to q \bar q}(\xt_{01}, z_{ik})
    =&\frac{e e_f}{ 2\pi} \left( \frac{Q\sqrt{z_{0k} z_{1k} }}{2 \pi \abs{\xt_{01}}} \right)^{-\varepsilon}
    \frac{i \xt_{01}^i}{\abs{\xt_{01}}} \sqrt{z_{0k} z_{1k}} Q K_{1-\varepsilon}\left( \abs{\xt_{01}} Q \sqrt{z_{0k} z_{1k}} \right)
    \varepsilon_\lambda^j \\
    &\times \frac{1}{2q^+}\bar u(k_0^+) \gamma^+ \left\{
    (z_{0k} - z_{1k}) \delta^{ij} 
    + \frac{1}{2} \left[ \gamma^i, \gamma^j \right]
    \right\} v(k_1^+).
\end{split}
\end{equation}
The NLO corrections to these wave functions can be written as
\begin{equation}
    \psi_\nlo^{\gamma^*_\lambda \to q \bar q}(\xt_{01}, z_{ik})
    = \frac{\as \cf}{2\pi}\, K^{\gamma^*}(\xt_{01}, z_{ik}) \times \psi_\lo^{\gamma^*_\lambda \to q \bar q}(\xt_{01}, z_{ik})
\end{equation}
where the function
\begin{equation}
\label{eq:K}
\begin{split}
    K^{\gamma^*}(\xt_{01}, z_{ik}) =& \left[ \frac{3}{2}+ \log(\frac{\alpha}{z_{0k}})+\log(\frac{\alpha}{z_{1k}}) \right] \left\{ \frac{1}{\varepsilon}+ \gamma_E + \log(\pi \xt_{01}^2 \mu^2) \right\} \\
    &+ \frac{1}{2} \log^2 \left( \frac{z_{0k}}{z_{1k}} \right) - \frac{\pi^2}{6} + \frac{5}{2} 
    +\frac{1}{2} \deltas
\end{split}
\end{equation}
is independent of the photon polarization.

The wave functions for the case $\gamma^* \to q \bar qg$ are given by
\begin{equation}
\begin{split}
    &\psi^{\gamma^*_L \to q \bar q g}(\xt_{ij}, z_{ik}) 
    = 4 e e_f Q g z_{0k} z_{1k} \varepsilon^{* j}_\sigma \\
    &\times \frac{1}{2 q^+} \bar u(k_0^+) \gamma^+
    \left\{ 
    \left[ 
    \left(1 + \frac{z_{2k}}{2 z_{0k}} \right) \delta^{ij}
    - \frac{z_{2k}}{4z_{0k}} \left[\gamma^i, \gamma^j \right]
    \right]
    \mathcal{I}_0^i
    - 
    \left[ 
    \left(1 + \frac{z_{2k}}{2 z_{1k}} \right) \delta^{ij}
    + \frac{z_{2k}}{4z_{1k}} \left[\gamma^i, \gamma^j \right]    
    \right]
    \mathcal{I}_1^i
    \right\}
    v(k_1^+)
\end{split}
\end{equation}
for longitudinally polarized photons and
\begin{equation}
\begin{split}
    &\psi^{\gamma^*_T \to q \bar q g}(\xt_{ij}, z_{ik}) 
    = e e_f g \varepsilon^l_\lambda \varepsilon_\sigma^{* j} \Sigma^{lj}
\end{split}
\end{equation}
for transversely polarized photons, where\footnote{We have corrected typos in Eq.~(173) of Ref.~\cite{Beuf:2022ndu} that corresponds to swapping the indices $\mathcal{I}_0^{ik} \to \mathcal{I}_0^{ki}$ and $\mathcal{I}_1^{ik} \to \mathcal{I}_1^{ki}$. }
\begin{equation}
    \begin{split}
        \Sigma^{lj}= \frac{1}{2 q^+} \bar u(k_0^+) \gamma^+
        \Bigg\{&
            -\frac{1}{z_{0k} + z_{2k}} 
            \left[ 
                (2z_{0k} + z_{2k}) \delta^{ij}- \frac{z_{2k}}{2} \left[ \gamma^i, \gamma^j \right]
            \right]
            \left[ 
                (2z_{1k} - 1) \delta^{kl}- \frac{1}{2} \left[ \gamma^k, \gamma^l \right]
            \right]
            \mathcal{I}^{ki}_0 \\
            &
            -\frac{1}{z_{1k} + z_{2k}} 
            \left[ 
                (2z_{0k} -1) \delta^{kl}+ \frac{1}{2} \left[ \gamma^k, \gamma^l \right]
            \right]
            \left[ 
                (2z_{1k} + z_{2k}) \delta^{ij}+ \frac{z_{2k}}{2} \left[ \gamma^i, \gamma^j \right]
            \right]
            \mathcal{I}^{ki}_1 \\
            &+ \frac{z_{0k} z_{2k}}{(z_{0k} + z_{2k})^2} \gamma^j \gamma^l \mathcal{I}_0
            - \frac{z_{1k} z_{2k}}{(z_{1k} + z_{2k})^2}  \gamma^l \gamma^j\mathcal{I}_1 
        \Bigg\}
        v(k_1^+).
    \end{split}
\end{equation}
The following short-hand notations have been used here:
\begin{equation}
\begin{aligned}
    \mathcal{I}_0 &= \mathcal{I}\left(\xt_{0+2;1}, \xt_{20}, \overline Q^2_{0}, \omega_{0} \right), & 
    \mathcal{I}_1 &= \mathcal{I}\left(\xt_{0;1+2}, \xt_{21}, \overline Q^2_{1}, \omega_{1}\right) 
\end{aligned}
\end{equation}
where $\xt_{0+2;1}$, $\xt_{0;1+2}$, $\Qbar[i]$, $\omega_i$ are given in Eqs.~\eqref{eq:x021}, \eqref{eq:x012}, and \eqref{eq:shorthand_notations} with $z_i \equiv z_{ik}$.
The different functions $\mathcal{I}$, $\mathcal{I}^i$, $\mathcal{I}^{ki}$ are defined as
\begin{equation}
\begin{split}
    \mathcal{I}\left(\bt, \rt, \overline Q^2, \omega\right) 
    =& \frac{1}{(2\pi)^{2-2\varepsilon}}
     \left( \frac{\mu}{\omega} \right)^\varepsilon
     \left( \frac{\overline Q}{\sqrt{\bt^2 + \omega \rt^2}} \right)^{1-2\varepsilon}
    K_{1-2\varepsilon} \left( \overline Q \sqrt{\bt^2 + \omega \rt^2} \right) \\
    \overset{\varepsilon=0}{=}& \frac{1}{(2\pi)^2} \left( \frac{\overline Q}{\sqrt{\bt^2 + \omega \rt^2}} \right)
    K_1 \left( \overline Q \sqrt{\bt^2 + \omega \rt^2} \right),
\end{split}
\end{equation}
\begin{equation}
\begin{split}
    \mathcal{I}^i\left(\bt, \rt, \overline Q^2, \omega\right) 
    =&
    \frac{i \mu^{\varepsilon}}{2 (4\pi)^{2-2\varepsilon}}
    \rt^i
    \int_0^\infty \dd{u} u^{-1+\varepsilon} e^{-u \overline Q^2 - \frac{\bt^2}{4u}}
    \int_0^{u/\omega} \dd{t} t^{-2+\varepsilon} e^{-\frac{\rt^2}{4t}}
    \\
    \overset{\varepsilon=0}{=}& \frac{i}{(2\pi)^2} \frac{\rt^i}{\rt^2} 
    K_0 \left( \overline Q \sqrt{\bt^2 + \omega \rt^2} \right),
\end{split}
\end{equation}
and
\begin{equation}
\begin{split}
    \mathcal{I}^{ij}\left(\bt, \rt, \overline Q^2, \omega\right) 
    =&
    -\frac{\mu^{\varepsilon}}{4 (4\pi)^{2-2\varepsilon}}
    \bt^i \rt^j
    \int_0^\infty \dd{u} u^{-2+\varepsilon} e^{-u \overline Q^2 - \frac{\bt^2}{4u}}
    \int_0^{u/\omega} \dd{t} t^{-2+\varepsilon} e^{-\frac{\rt^2}{4t}}
    \\
    \overset{\varepsilon=0}{=}&- \frac{1}{(2\pi)^2}
    \frac{\bt^i \rt^j}{\rt^2}
    \left( \frac{\overline Q}{\sqrt{\bt^2 + \omega \rt^2}} \right)
    K_1 \left( \overline Q \sqrt{\bt^2 + \omega \rt^2} \right).
\end{split}
\end{equation}

\subsection{Wave function for $q \bar q \to q \bar q$ in the final state}

\begin{figure*}[tbp!]
\newdiag{diag:qqbar_to_qqbar}
\centerline{\includegraphics[width=4cm]{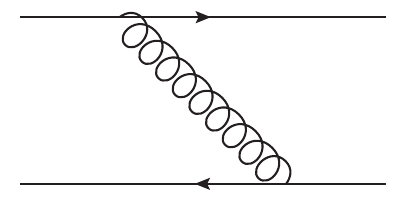}
\begin{tikzpicture}[overlay]
\node[anchor=south west] at (-3cm,-0.5cm) {\namediag{diag:qqbar_qqbar_a}};
            \draw [to-to](-0.7cm,0.2cm) -- (-0.7cm,1.7cm);
            \draw [to-to](-3.5cm,0.2cm) -- (-3.5cm,1.7cm);
         \node[anchor=south east] at (-3.5cm,0.8cm) {$\Pt_{01}$};
         \node[anchor=south east] at (-0.7cm,0.8cm) {$\Kt_{01}$};
         \node[anchor=south east] at (-3.6cm,1.8cm) {$p_{0}$};
         \node[anchor=south east] at (-3.6cm,0.1cm) {$p_{1}$};
         \node[anchor=south east] at (-0.1cm,1.8cm) {$k_{0}$};
         \node[anchor=south east] at (-0.1cm,0.1cm) {$k_{1}$};
\end{tikzpicture}
\rule{2em}{0pt}
\includegraphics[width=4cm]{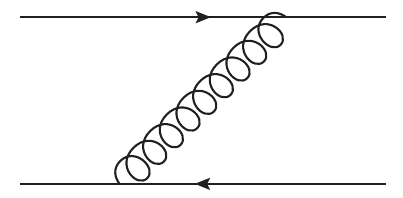}
\begin{tikzpicture}[overlay]
\node[anchor=south west] at (-3cm,-0.5cm) {\namediag{diag:qqbar_qqbar_b}};
\end{tikzpicture}
\rule{2em}{0pt}
\includegraphics[width=4cm]{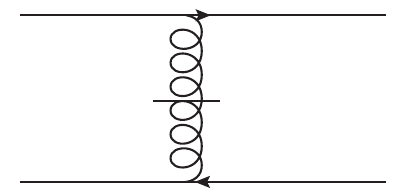}
\begin{tikzpicture}[overlay]
\node[anchor=south west] at (-3cm,-0.5cm) {\namediag{diag:qqbar_qqbar_c}};
\end{tikzpicture}
}
\caption{
Diagrams contributing to the $q\bar q \to q \bar q$ wave function.
}\label{fig:qqbar_to_qqbar}
\end{figure*}
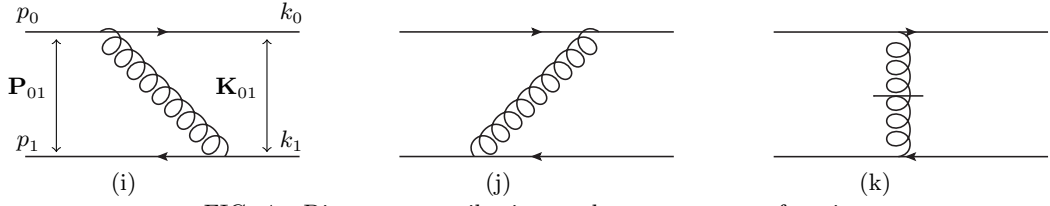

Here we list the parts of the wave function for $q(p_0) \bar q(p_1) \to q(k_0) \bar q(k_1)$ that are needed in the calculation.
At the order required in this work, this wave function has three contributions corresponding to the Feynman diagrams in Fig.~\ref{fig:qqbar_to_qqbar}.
Note that this wave function corresponds to final-state interactions as described in Sec.~\ref{sec:computational_setup}.

The first diagram~\ref{diag:qqbar_qqbar_a} gives us the contribution
\begin{equation}
\begin{split}
\Psi^{q(p_0)\bar q(p_1) \to q(k_0) \bar q(k_1)}_{\oout,\ref{diag:qqbar_qqbar_a}} =&
-g^2 \mu^{2\varepsilon} t^{a}_{\alpha_{0k} \alpha_{0p}} t^{a}_{\alpha_{1p} \alpha_{1k}} 
\frac{\theta(z_{0p} - z_{0k})}{(z_{0p} - z_{0k})^2}  \frac{1}{z_{1k} z_{1p}}
\times S_0^{ij}(z_{0k},z_{0p})  S_1^{kj}(z_{1p},z_{1k})
\\
&\times 
\frac{
 \left(z_{0p} \Kt_{01} - z_{0k} \Pt_{01} \right )^i 
\left (z_{1p} \Kt_{01} - z_{1k} \Pt_{01} \right )^k}{\biggl [\left (z_{0p} \Kt_{01} - z_{0k} \Pt_{01} \right )^2 + i\delta\biggr ] 
\biggl [\frac{\Kt_{01}^2}{z_{0k} z_{1k}} - \frac{\Pt_{01}^2}{z_{0p} z_{1p}} + i\delta\biggr ]
}\\
=&-2g^2 \cf \frac{1}{\nc} \delta^{\alpha_{0k}}_{\alpha_{1k}}\delta^{\alpha_{0p}}_{\alpha_{1p}}
\delta^{h_{0k}}_{  h_{0p}} \delta^{h_{1k}}_{ h_{1p}}
\frac{\theta(z_{0p} - z_{0k})}{(z_{0p} - z_{0k})^3} 
\frac{z_{0k} z_{1k} +  z_{0p}  z_{1p} }{\sqrt{z_{0k} z_{1k}  z_{0p}  z_{1p}}}\\
&\times \frac{
 \left( z_{0p} \Kt_{01} - z_{0k} \Pt_{01} \right ) \vdot 
\left ( z_{1p} \Kt_{01} - z_{1k} \Pt_{01} \right )}{\biggl [\frac{1}{(z_{0p} - z_{0k}) z_{0k}  z_{0p}}\left ( z_{0p} \Kt_{01} - z_{0k} \Pt_{01} \right )^2 + i\delta\biggr ] 
\biggl [\frac{\Kt_{01}^2}{z_{0k} z_{1k}} - \frac{\Pt_{01}^2}{ z_{0p}  z_{1p}}  + i\delta\biggr ]
}\\
&+\text{other spin and color components that vanish}+\mathcal{O}(\varepsilon)
\end{split}
\end{equation}
where 
$S_0$, $S_1$ are given in Eqs.~\eqref{eq:spinor0} and \eqref{eq:spinor1}.
Similarly, the contribution from Diagram~\ref{diag:qqbar_qqbar_b} reads
\begin{equation}
\begin{split}
\Psi^{q(p_0)\bar q(p_1) \to q(k_0) \bar q(k_1)}_{\oout,\ref{diag:qqbar_qqbar_b}} =&
-g^2 \mu^{2\varepsilon} t^{a}_{\alpha_{0k}  \alpha_{0p}} t^{a}_{\alpha_{1p} \alpha_{1k}} 
\frac{\theta(z_{0k}-z_{0p})}{(z_{0k}-z_{0p})^2} \frac{1}{z_{0k} z_{0p}}
\times S_0^{ij}(z_{0k},z_{0p})S_1^{kj}(z_{1p},z_{1k})
\\
&\times\frac{
 \left( z_{0p} \Kt_{01} - z_{0k} \Pt_{01} \right )^i 
\left ( z_{1p} \Kt_{01} - z_{1k} \Pt_{01} \right )^k}{\biggl [\left ( z_{1p} \Kt_{01} - z_{1k} \Pt_{01} \right )^2 + i\delta\biggr ] 
\biggl [\frac{\Kt_{01}^2}{z_{0k} z_{1k}} - \frac{\Pt_{01}^2}{ z_{0p}  z_{1p}}  + i\delta\biggr ]
}\\
=& -2g^2 \cf \frac{1}{\nc} \delta^{\alpha_{0k}}_{\alpha_{1k}}\delta^{\alpha_{0p}}_{\alpha_{1p}}
\delta^{h_{0k}}_{  h_{0p}} \delta^{h_{1k}}_{ h_{1p}}
\frac{\theta(z_{0k}-z_{0p})}{(z_{0k}-z_{0p})^3} \frac{z_{0k} z_{1k} +  z_{0p}  z_{1p}}{\sqrt{z_{0k} z_{1k}  z_{0p}  z_{1p}}} 
\\
&\times \frac{
 \left( z_{0p} \Kt_{01} - z_{0k} \Pt_{01} \right ) \vdot 
\left ( z_{1p} \Kt_{01} - z_{1k} \Pt_{01} \right )}{\biggl [\frac{1}{(z_{0k}-z_{0p}) z_{1k}  z_{1p}}\left ( z_{1p} \Kt_{01} - z_{1k} \Pt_{01} \right )^2 + i\delta\biggr ] 
\biggl [\frac{\Kt_{01}^2}{z_{0k} z_{1k}} - \frac{\Pt_{01}^2}{ z_{0p}  z_{1p}}  + i\delta\biggr ]
}\\
&+\text{other color components that do not contribute}+\mathcal{O}(\varepsilon),
\end{split}
\end{equation}
and Diagram~\ref{diag:qqbar_qqbar_c} gives us
\begin{equation}
\begin{split}
\Psi_{\oout,\ref{diag:qqbar_qqbar_c}}^{q(p_0)\bar q(p_1) \to q(k_0) \bar q(k_1)}
 = &
4g^2 \mu^{2\varepsilon} t^{a}_{\alpha_{0k} \alpha_{0p}} t^{a}_{\alpha_{1p} \alpha_{1k}}
\frac{1}{(z_{0k} - z_{0p})^2}
\frac{1}{\left[\frac{\Kt_{01}^2}{z_{0k} z_{1k}} - \frac{\Pt_{01}^2}{z_{0p} z_{1p}} + i\delta \right]}
\\
&\times \frac{1}{(2q^+)^2}
\biggl [\bar u(k_0^+)\gamma^+ u(p_0^+) \biggr ] \biggl [\bar v(p_1^+) \gamma^+ v(k_1^+) \biggr ] \\
=& 4g^2 \cf \frac{1}{\nc} \delta^{\alpha_{0k}}_{\alpha_{1k}}\delta^{\alpha_{0p}}_{\alpha_{1p}}
\delta^{h_{0k}}_{  h_{0p}} \delta^{h_{1k}}_{ h_{1p}}
\frac{1}{(z_{0k} - z_{0p})^2}
\frac{\sqrt{z_{0k} z_{1k} z_{0p} z_{1p}}}{\left[\frac{\Kt_{01}^2}{z_{0k} z_{1k}} - \frac{\Pt_{01}^2}{z_{0p} z_{1p}} + i\delta \right]}\\
&+\text{other color components that  do not contribute}+\mathcal{O}(\varepsilon).
\end{split}
\end{equation}
Here we have kept only the color-singlet contribution, and also contributions proportional to the quark helicity have been dropped as they vanish when summing over the helicities.

These wave functions can be transformed to the coordinate space by a transverse Fourier transform
\begin{equation}
\begin{split}
&\!\!\!\!\!\!\!\!\!\!\!\!\!\!\!\!\!    \widetilde \Psi_{\oout}^{q(p_0)\bar q(p_1) \to q(\xt_0,z_{0k}) \bar q(\xt_1,z_{1k})}
\\
    &= \int \frac{\dd[2-2\varepsilon]{\kt_0} \dd[2-2\varepsilon]{\kt_1}}{(2\pi)^{2(2-2\varepsilon)}} e^{i \kt_0 \vdot \xt_0 + i \kt_1 \vdot \xt_1} (2\pi)^{2-2\varepsilon} \delta^{(2-2\varepsilon)}(\kt_0 + \kt_1 - (\pt_0 + \pt_1)  ) \Psi_{\oout}^{q(p_0)\bar q(p_1) \to q(k_0) \bar q(k_1)} \\
    &= e^{ i \bt \vdot \sum_i \pt_i} \int \frac{\dd[2-2\varepsilon]{\Kt_{01}}}{(2\pi)^{2-2\varepsilon}} e^{i \Kt_{01} \vdot \xt_{01} }
    \Psi_{\oout}^{q(p_0)\bar q(p_1) \to q(k_0) \bar q(k_1)}.
\end{split}
\end{equation}
While we do not need the coordinate-space wave functions because of the way we calculate final-state corrections in Sec.~\ref{sec:final-state_corrections}, we list the Fourier transforms for completeness.
For Diagram~\ref{diag:qqbar_qqbar_a} this evaluates to
\begin{equation}
\begin{split}
\widetilde \Psi^{q(p_0)\bar q(p_1) \to q(\xt_0,z_{0k}) \bar q(\xt_1,z_{1k})}_{\oout,\ref{diag:qqbar_qqbar_a}} 
=&
-e^{ i \bt \vdot \sum_i \pt_i} g^2 \mu^{2\varepsilon} t^{a}_{\alpha_{0k} \alpha_{0p}} t^{a}_{\alpha_{1p} \alpha_{1k}} 
\frac{\theta(z_{0k} - z_{0p})}{(z_{0k} - z_{0p})^2}  \frac{1}{z_{1k} z_{1p}}
\times S_0^{ij}(z_{0k},z_{0p})  S_1^{kj}(z_{1p},z_{1k})
\\
&\times 
(z_{0p}i \partial_{\xt_{01}} + z_{0k} \Pt_{01} )^i
(z_{1p}i \partial_{\xt_{01}} + z_{1k} \Pt_{01} )^k
\mathcal{F}_{\ref{diag:qqbar_qqbar_a}}
\end{split}
\end{equation}
where
\begin{equation}
    \mathcal{F}_{\ref{diag:qqbar_qqbar_a}}
    = \frac{z_{0k} z_{1k}}{8 z_{0p}^2} \left( \frac{4\pi^2 \xt_{01}^2}{ \Pt_{01}^2}\right)^{\varepsilon/2} 
    \int_0^1 \dd{t} \exp(i (1-t) \frac{z_{0k}}{z_{0p}} \xt_{01} \vdot \Pt_{01})
    \frac{1}{\xi_{\ref{diag:qqbar_qqbar_a}}^{1+\varepsilon}}  
    H^{(2)}_{-1 -\varepsilon}(\abs{\xt_{01}} \abs{\Pt_{01}} \xi_{\ref{diag:qqbar_qqbar_a}})
\end{equation}
and
\begin{equation}
    \xi_{\ref{diag:qqbar_qqbar_a}} = \sqrt{ t \frac{z_{0k} z_{1k}}{z_{0p} z_{1p}} \left( 1 - (1-t) \frac{z_{0k} z_{1p}}{z_{0p} z_{1k}} \right)}  .
\end{equation}
Here $H^{(2)}_{-1 -\varepsilon}$ is the Hankel function of the second kind.
Similarly, for Diagram~\ref{diag:qqbar_qqbar_b} we get
\begin{equation}
\begin{split}
\widetilde \Psi^{q(p_0)\bar q(p_1) \to q(\xt_0,z_{0k}) \bar q(\xt_1,z_{1k})}_{\oout,\ref{diag:qqbar_qqbar_b}} 
=&
-e^{ i \bt \vdot \sum_i \pt_i} g^2 \mu^{2\varepsilon} t^{a}_{\alpha_{0k}  \alpha_{0p}} t^{a}_{\alpha_{1p} \alpha_{1k}} 
\frac{\theta(z_{0p}-z_{0k})}{(z_{0k}-z_{0p})^2} \frac{1}{z_{0k} z_{0p}}
\times S_0^{ij}(z_{0k},z_{0p})S_1^{kj}(z_{1p},z_{1k})
\\
&\times 
(z_{0p}i \partial_{\xt_{01}} + z_{0k} \Pt_{01} )^i
(z_{1p}i \partial_{\xt_{01}} + z_{1k} \Pt_{01} )^k
\mathcal{F}_{\ref{diag:qqbar_qqbar_b}}
\end{split}
\end{equation}
where
\begin{equation}
    \mathcal{F}_{\ref{diag:qqbar_qqbar_b}}
    = \frac{z_{0k} z_{1k}}{8 z_{1p}^2} \left( \frac{4\pi^2 \xt_{01}^2}{ \Pt_{01}^2} \right)^{\varepsilon/2}
    \int_0^1 \dd{t} \exp(i (1-t) \frac{z_{1k}}{z_{1p}} \xt_{01} \vdot \Pt_{01})
    \frac{1}{\xi_{\ref{diag:qqbar_qqbar_b}}^{1+\varepsilon}} 
    H^{(2)}_{-1 -\varepsilon}(\abs{\xt_{01}} \abs{\Pt_{01}} \xi_{\ref{diag:qqbar_qqbar_b}})
\end{equation}
and
\begin{equation}
    \xi_{\ref{diag:qqbar_qqbar_b}} = \sqrt{ t \frac{z_{0k} z_{1k}}{z_{0p} z_{1p}} \left( 1 - (1-t) \frac{z_{1k} z_{0p}}{z_{1p} z_{0k}} \right)}  .
\end{equation}
Finally, for Diagram \ref{diag:qqbar_qqbar_c} the Fourier-transformed wave function is given by
\begin{equation}
\begin{split}
\widetilde \Psi_{\oout,\ref{diag:qqbar_qqbar_c}}^{q(p_0)\bar q(p_1) \to q(\xt_0,z_{0k}) \bar q(\xt_1,z_{1k})} 
 = &
4e^{ i \bt \vdot \sum_i \pt_i} g^2 \mu^{2\varepsilon} t^{a}_{\alpha_{0k} \alpha_{0p}} t^{a}_{\alpha_{1p} \alpha_{1k}}
\frac{1}{(z_{0k} - z_{0p})^2}
\times \frac{1}{(2q^+)^2}
\biggl [\bar u(k_0^+)\gamma^+ u(p_0^+) \biggr ] \biggl [\bar v(p_1^+) \gamma^+ v(k_1^+) \biggr ]\\
&\times (-i)\frac{z_{0k}z_{1k}}{4} \left( \frac{4\pi^2 \xt_{01}^2}{ \Pt_{01}^2}  \frac{z_{0p}z_{1p}}{z_{0k} z_{1k}}\right)^{\varepsilon/2}
\times  H^{(2)}_{-\varepsilon} \left( \abs{\xt_{01}} \abs{\Pt_{01}} \sqrt{\frac{z_{0k} z_{1k}}{z_{0p} z_{1p}}}  \right).
\end{split}
\end{equation}
Note these wave functions are finite when $\varepsilon \to 0$ but contain divergences in $\alpha$ when integrated over $z_{ik}$ and $z_{ip}$.

These results can to some extent be compared to the results in Ref.~\cite{Caucal:2021ent}, although an exact comparison is not possible because our calculation is structured in a different way than in  Ref.~\cite{Caucal:2021ent}. To the extent that we have been able to make a direct comparison, in particular the parts related to the Hankel function $H^{(2)}_{-n-\varepsilon}$,
our results agree\footnote{Note that the Hankel function $H^{(2)}$ can be related to the Bessel $K$ function with an imaginary argument by
$H^{(2)}_{n-\varepsilon}(x) = \frac{2}{\pi} e^{i \frac{\pi}{2} (1+n-\varepsilon)}  K_{n-\varepsilon}(ix)$ when $x > 0$.
}.

\subsection{Wave function for $q \bar q \to q \bar q g$ in the final state}

\begin{figure*}[tbp!]
\newdiag{diag:qqbar_to_qqbarg}
\centerline{\includegraphics[width=4cm]{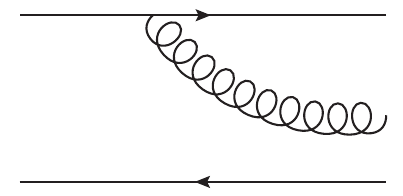}
\begin{tikzpicture}[overlay]
\node[anchor=south west] at (-3cm,-0.5cm) {\namediag{diag:qqbar_qqbarg_a}};
            \draw [to-to](-3.5cm,0.2cm) -- (-3.5cm,1.7cm);
         \node[anchor=south east] at (-3.6cm,0.8cm) {$\Pt_{01}$};
         \node[anchor=south east] at (-3.6cm,1.8cm) {$p_{0}$};
         \node[anchor=south east] at (-3.6cm,0.1cm) {$p_{1}$};
         \node[anchor=south east] at (-0.1cm,1.8cm) {$k_{0}$};
         \node[anchor=south east] at (-0.1cm,0.1cm) {$k_{1}$};
         \node[anchor=south east] at (-0.1cm,1.0cm) {$k_{2}$};
\end{tikzpicture}
\rule{2em}{0pt}
\includegraphics[width=4cm]{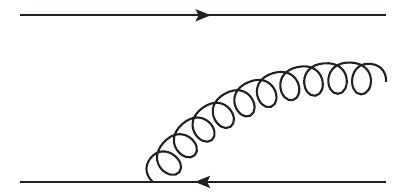}
\begin{tikzpicture}[overlay]
\node[anchor=south west] at (-3cm,-0.5cm) {\namediag{diag:qqbar_qqbarg_b}};
\end{tikzpicture}
}
\caption{
Diagrams contributing to the $q\bar q \to q \bar q g$ wave function.
}\label{fig:qqbar_to_qqbarg}
\end{figure*}

We also need the wave functions for $q(p_0) \bar q(p_1) \to q(k_0) \bar q(k_1) g(k_2)$ in the final state. The relevant Feynman diagrams are shown in Fig.~\ref{fig:qqbar_to_qqbarg} and their contributions read
\begin{equation}
\begin{split}
    \Psi^{q(p_0) \bar q(p_1) \to q(k_0) \bar q(k_1) g(k_2)}_{\oout,\ref{diag:qqbar_qqbarg_a}} = &(2\pi)^{2-2\varepsilon} \delta^{(2-2\varepsilon)}(\pt_1- \kt_1) 4\pi z_{1p} \delta(z_{1p}-z_{1k}) \delta_{\alpha_{1k}}^{\alpha_{1p}}\delta_{h_{1k}}^{h_{1p}}\\
    &\times 
    g \mu^{\varepsilon} t^{a}_{\alpha_{0k} \alpha_{0p}}\varepsilon^{j \ast}_{\sigma}  \frac{ \left(z_{0p} \kt_2-z_{2k} \pt_0 \right)^i }{\left( z_{0p} \kt_2 -z_{2k} \pt_0 \right)^2+i\delta}
    \times S_0^{ij}(z_{0k}, z_{0p})
\end{split}
\end{equation}
and
\begin{equation}
\begin{split}
    \Psi^{q(p_0) \bar q(p_1) \to q(k_0) \bar q(k_1)g(k_2) }_{\oout,\ref{diag:qqbar_qqbarg_b}} 
    =& (2\pi)^{2-2\varepsilon} \delta^{(2-2\varepsilon)}(\pt_0- \kt_0) 4\pi z_{0p} \delta(z_{0p}-z_{0k}) \delta_{\alpha_{0k}}^{\alpha_{0p}}\delta_{h_{0k}}^{h_{0p}}\\
    &\times 
     (- g) \mu^{\varepsilon} t^{a}_{\alpha_{1p} \alpha_{1k}} \varepsilon^{j \ast}_{\sigma} \frac{ \left( z_{1p}\kt_2-z_{2k} \pt_1 \right)^i}{\left(z_{1p} \kt_2 -z_{2k} \pt_1 \right)^2+i\delta} 
     \times S_1^{ij}(z_{1p},z_{1k}).
\end{split}
\end{equation}

The corresponding coordinate-space wave functions are also needed. In this case, the Fourier transform is given by
\begin{equation}
\begin{split}
    &\widetilde \Psi^{q(p_0) \bar q(p_1) \to q(\xt_0,z_{0k}) \bar q(\xt_1,z_{1k}) g(\xt_2,z_{2k})}_\oout \\
    &= \int \frac{\dd[2-2\varepsilon]{\kt_0} \dd[2-2\varepsilon]{\kt_1} \dd[2-2\varepsilon]{\kt_2} }{(2\pi)^{3(2-2\varepsilon)}} 
    e^{i \sum_i \kt_i \vdot \xt_i} (2\pi)^{2-2\varepsilon} \delta^{(2-2\varepsilon)}\left(\sum_i \kt_i - \sum_i \pt_i  \right) 
    \Psi_{\oout}^{q(p_0)\bar q(p_1) \to q(k_0) \bar q(k_1) g(k_2)}.
\end{split}
\end{equation}
Summing the contributions from Diagrams~\ref{diag:qqbar_qqbarg_a} and \ref{diag:qqbar_qqbarg_b}, the Fourier transform evaluates to
\begin{multline}
\label{eq:wf_qq_to_qqg}
    \widetilde \Psi^{q(p_0) \bar q(p_1) \to q(\xt_0,z_{0k}) \bar q(\xt_1,z_{1k}) g(\xt_2,z_{2k})}_\oout 
    =
    e^{i  \bt \vdot \sum_i \pt_i }g \varepsilon^{j\ast}_{\sigma} \frac{i}{2\pi} \Gamma\left( 1-\varepsilon \right)\\
    \times \Bigg[ 
    4\pi z_{1p} \delta(z_{1p}-z_{1k}) \delta_{h_{1k}}^{h_{1p}}
     \delta_{\alpha_{1k}}^{\alpha_{1p}}  t^{a}_{\alpha_{0k} \alpha_{0p}}
    e^{ i \Pt_{01} \vdot \xt_{0+2;1 }
    } 
      \frac{1}{z_{0p}}  \frac{\xt_{20}^i}{\xt_{20}^2} (\mu \pi \xt_{20}^2)^{\varepsilon}
      \times S_0^{ij}(z_{0k}, z_{0p})\\
    -4\pi z_{0p} \delta(z_{0p}-z_{0k}) \delta_{h_{0k}}^{h_{0p}}
    \delta_{h_{0k}}^{h_{0p}}   t^{a}_{\alpha_{1p} \alpha_{1k}}
    e^{i \Pt_{01} \vdot \xt_{0;1+2 }
    } 
     \frac{1}{z_{1p}} \frac{\xt_{21}^i}{\xt_{21}^2} (\mu \pi \xt_{21}^2)^{\varepsilon}
     \times S_1^{ij}(z_{1p},z_{1k})
    \Bigg]
\end{multline}
where $\xt_{0+2;1 } = \frac{z_{0k} \xt_0 + z_{2k} \xt_2}{z_{0k}+z_{2k}}- \xt_1$ and 
$\xt_{0;1+2 } = \xt_0 - \frac{z_{1k} \xt_1 + z_{2k} \xt_2}{z_{1k}+z_{2k}}$
are the transverse separations of the $q \bar q$ pair before the gluon emission.
The reduced wave function can then be written as
\begin{multline}
\label{eq:wf_qq_to_qqg_red}
   \widetilde \psi^{q \bar q \to q \bar q g}_\oout(\Pt_{01}, z_{ip}; \xt_{ij}, z_{ik}) 
    =
    g \varepsilon^{j\ast}_{\sigma} \frac{i}{2\pi} \Gamma\left( 1-\varepsilon \right)
   \\
    \times \Bigg[ 
    4\pi z_{1p} \delta(z_{1p}-z_{1k}) \delta_{h_{1k}}^{h_{1p}}
    e^{ i \Pt_{01} \vdot \xt_{0+2;1}
    } 
      \frac{1}{z_{0p}}  \frac{\xt_{20}^i}{\xt_{20}^2} (\mu \pi \xt_{20}^2)^{\varepsilon}
      \times S_0^{ij}(z_{0k}, z_{0p})\\
    -4\pi z_{0p} \delta(z_{0p}-z_{0k}) \delta_{h_{0k}}^{h_{0p}}
    e^{i \Pt_{01} \vdot \xt_{0;1+2}
    } 
     \frac{1}{z_{1p}} \frac{\xt_{21}^i}{\xt_{21}^2} (\mu \pi \xt_{21}^2)^{\varepsilon}
     \times S_1^{ij}(z_{1p},z_{1k})
    \Bigg].
\end{multline}

\subsection{Wave function for $q \bar q g \to q \bar q$ in the final state}

\begin{figure*}[tbp!]
\newdiag{diag:qqbarg_to_qqbar}
\centerline{\includegraphics[width=4cm]{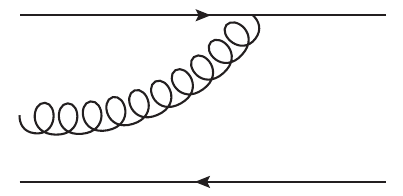}
\begin{tikzpicture}[overlay]
\node[anchor=south west] at (-3cm,-0.5cm) {\namediag{diag:qqbarg_qqbar_a}};
            \draw [to-to](-3.4cm,1.0cm) -- (-3.4cm,1.7cm);
            \draw [to-to](-3.4cm,0.2cm) -- (-3.4cm,0.6cm);
         \node[anchor=south east] at (-2.6cm,1.0cm) {$\Pt_{02}$};
         \node[anchor=south east] at (-2.6cm,0.1cm) {$\Pt_{12}$};
         \node[anchor=south east] at (-3.6cm,1.8cm) {$p_{0}$};
         \node[anchor=south east] at (-3.6cm,0.1cm) {$p_{1}$};
         \node[anchor=south east] at (-0.1cm,1.8cm) {$k_{0}$};
         \node[anchor=south east] at (-0.1cm,0.1cm) {$k_{1}$};
         \node[anchor=south east] at (-3.6cm,1.0cm) {$p_{2}$};
\end{tikzpicture}
\rule{2em}{0pt}
\includegraphics[width=4cm]{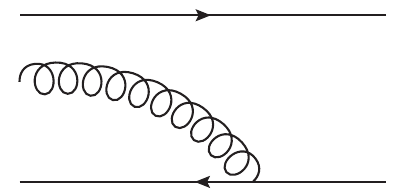}
\begin{tikzpicture}[overlay]
\node[anchor=south west] at (-3cm,-0.5cm) {\namediag{diag:qqbarg_qqbar_b}};
\end{tikzpicture}
}
\caption{
Diagrams contributing to the $q\bar q g\to q \bar q $ wave function.
}\label{fig:qqbarg_to_qqbar}
\end{figure*}

The final type of wave functions we need is for the case $q(p_0) \bar q(p_1) g(p_2) \to q(k_0) \bar q(k_1)$. This wave function has two contributions shown in Fig.~\ref{fig:qqbarg_to_qqbar}, and they read
\begin{equation}
    \begin{split}
\Psi^{q(p_0) \bar q(p_1) g(p_2) \to q(k_0) \bar q(k_1) }_{\oout,\ref{diag:qqbarg_qqbar_a}}
= &(2\pi)^{2-2\varepsilon} \delta^{(2-2\varepsilon)}(\pt_1- \kt_1) 4\pi z_{1p} \delta(z_{1k}-z_{1p}) \delta_{\alpha_{1k}}^{\alpha_{1p}}\delta_{h_{1k}}^{h_{1p}}
    \\
    & \times g \mu^\varepsilon \varepsilon_\sigma^j t^{a}_{\alpha_{0k} \alpha_{0p} }  \frac{\left( z_{2p}\kt_0 -z_{0k}\pt_2
    \right)^i}{\left( z_{2p} \kt_0 - z_{0k} \pt_2
    \right)^2 -i\delta} 
    \times S_0^{ij}(z_{0k},z_{0p})
    \end{split}
\end{equation}
and
\begin{equation}
    \begin{split}
\Psi^{q(p_0) \bar q(p_1) g(p_2) \to q(k_0) \bar q(k_1) }_{\oout,\ref{diag:qqbarg_qqbar_b}}
=& (2\pi)^{2-2\varepsilon} \delta^{(2-2\varepsilon)}(\pt_0- \kt_0) 4\pi z_{0p} \delta(z_{0k} - z_{0p}) \delta_{\alpha_{0k}}^{\alpha_{0p}}\delta_{h_{0k}}^{h_{0p}}  \\
    &\times
    (-g) \mu^\varepsilon \varepsilon_\sigma^j t^{a}_{\alpha_{1p} \alpha_{1k} } 
      \frac{\left( z_{2p} \kt_1 - z_{1k} \pt_2
    \right)^i}{\left( z_{2p} \kt_1 - z_{1k} \pt_2
    \right)^2 -i\delta} 
    \times S_1^{ij}(z_{1p},z_{1k})
    \end{split}
\end{equation}

The Fourier transform of the $q \bar q g \to q \bar q$ wave function is given by
\begin{multline}
    \widetilde \Psi_{\oout}^{q(p_0) \bar q(p_1) g(p_2) \to q(\xt_0,z_{0k}) \bar q(\xt_1,z_{1k})}\\
    = \int \frac{\dd[2-2\varepsilon]{\kt_0} \dd[2-2\varepsilon]{\kt_1}}{(2\pi)^{2(2-2\varepsilon)}} e^{i \kt_0 \vdot \xt_0 + i \kt_1 \vdot \xt_1} (2\pi)^{2-2\varepsilon} \delta^{(2-2\varepsilon)}(\kt_0 + \kt_1 - (\pt_0 + \pt_1 +\pt_2)  ) \Psi_{\oout}^{q(p_0) \bar q(p_1) g(p_2) \to q(\xt_0,z_{0k}) \bar q(\xt_1,z_{1k})} 
\end{multline}
which evaluates to
\begin{multline}
    \widetilde \Psi^{q(p_0) \bar q(p_1) g(p_2) \to q(\xt_0,z_{0k}) \bar q(\xt_1,z_{1k}) }_{\oout} 
    =g \mu^\varepsilon \varepsilon_\sigma^j e^{i \bt \vdot \sum_i \pt_i }
    \\
     \times  \Bigg[4\pi z_{1p} \delta( z_{1k} - z_{1p} ) \delta_{\alpha_{1k}}^{\alpha_{1p}}\delta_{h_{1k}}^{h_{1p}} t^{a}_{\alpha_{0k} \alpha_{0p} } 
    e^{i \xt_{01} \vdot ( \Pt_{01} -\Pt_{12}) }
      \frac{\Pt_{02}^i}{\Pt_{02}^2}
      \times S_0^{ij}(z_{0k},z_{0p})\\
    - 4\pi z_{0p} \delta( z_{0k} - z_{0p} ) \delta_{\alpha_{0k}}^{\alpha_{0p}}\delta_{h_{0k}}^{h_{0p}}t^{a}_{\alpha_{1p} \alpha_{1k} }
    e^{i \xt_{01} \vdot ( \Pt_{01} + \Pt_{02}) }
    \frac{ \Pt_{12}^i}{ \Pt_{12}^2  } 
    \times S_1^{ij}(z_{1p},z_{1k})
    \Bigg].
\end{multline}
The corresponding reduced wave function is then
\begin{multline}
\label{eq:wf_qqg_to_qq_red}
     \widetilde \psi_\oout^{q \bar qg \rightarrow q \bar q}(\Pt_{ij},z_{ip} ;  \xt_{01},z_{ik}) 
     =g \mu^\varepsilon \varepsilon_\sigma^j 
     \times  \Bigg[4\pi z_{1p} \delta( z_{1k} - z_{1p} ) \delta_{h_{1k}}^{h_{1p}}
    e^{i \xt_{01} \vdot (\Pt_{01}- \Pt_{12}) }
      \frac{\Pt_{02}^i}{\Pt_{02}^2}
      \times S_0^{ij}(z_{0k},z_{0p})\\
    - 4\pi z_{0p} \delta( z_{0k} - z_{0p} ) \delta_{h_{0k}}^{h_{0p}}
    e^{i \xt_{01} \vdot (\Pt_{01} + \Pt_{02}) }
    \frac{ \Pt_{12}^i}{ \Pt_{12}^2 } 
    \times S_1^{ij}(z_{1p},z_{1k})
    \Bigg].
\end{multline}

\subsection{UV-divergent part of a general $X \to q \bar q \to q \bar q g$ wave function}
\label{sec:UV_divergence}

\begin{figure*}[tbp!]
\newdiag{diag:X_qqg}
\centerline{\includegraphics[width=4cm]{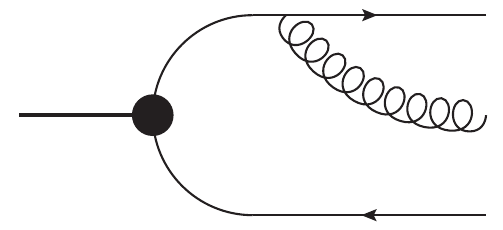}
\begin{tikzpicture}[overlay]
\node[anchor=south west] at (-3cm,-0.5cm) {\namediag{diag:X_q_qg}};
         \node[anchor=south east] at (-3.5cm,0.9cm) {$q$};
         \node[anchor=south east] at (-2.0cm,1.7cm) {$k_{0}'$};
         \node[anchor=south east] at (-2.0cm,0.2cm) {$k_{1}'$};
         \node[anchor=south east] at (-0.1cm,1.7cm) {$k_{0}$};
         \node[anchor=south east] at (-0.1cm,0.1cm) {$k_{1}$};
         \node[anchor=south east] at (-0.1cm,1.0cm) {$k_{2}$};
\end{tikzpicture}
\rule{2em}{0pt}
\includegraphics[width=4cm]{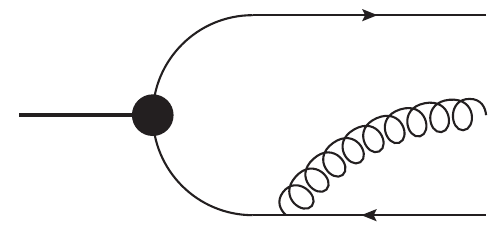}
\begin{tikzpicture}[overlay]
\node[anchor=south west] at (-3cm,-0.5cm) {\namediag{diag:X_qbar_qbarg}};
\end{tikzpicture}
}
\caption{
Diagrams contributing to the UV-divergent part of the $X \to q \bar q g $ wave function.
}\label{fig:X_qqg}
\end{figure*}
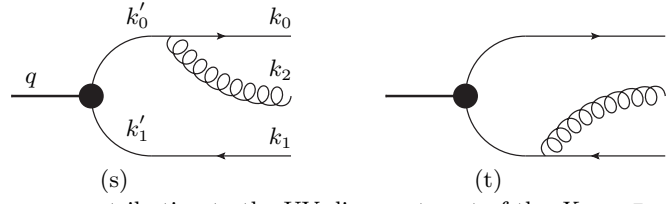

In this Section, we consider a general way to extract the most UV-divergent part of a wave function $X \to q \bar q \to q \bar q g$.
This UV divergence arises from the emitted gluon having a large transverse momentum or, equivalently, having a small transverse separation from the emitting particle.
The gluon can be emitted either from the quark or the antiquark, as shown in Fig.~\ref{fig:X_qqg}.

Let us consider a general $X \to q \bar q$ wave function $\Psi^{X(q) \to q(k_0) \bar q(k_1)}$, and then a gluon emission from this state. If the quark emits the gluon, the resulting wave function can be written as
\begin{equation}
    \begin{split}
        \Psi^{X(q) \to q(k_0) \bar q(k_1) g(k_2)}_{\iin,\ref{diag:X_q_qg}}
        =&
        \int_0^1 \dd{z_{0k}'} \dd{z_{1k}'} \delta(1-z_{0k}'-z_{1k}') \delta(z_{1k} - z_{1k}') 
         \Psi^{X(q) \to q(k_0') \bar q(k_1)}_\iin \\
         &\times g \mu^\varepsilon \frac{-1}{z_{0k}z_{2k}(1-z_{1k})^2} \varepsilon_{\sigma}^{*j} t^a_{\alpha_{k} \alpha_{0k}'}
        \times S_0^{ij}(z_{0k}, z_{0k}') \\
        &\times \frac{\Kt_{20}^i}
        { q^2 
        -\frac{1}{z_{0k} z_{2k}(1-z_{1k})} \Kt_{20}^2 - \frac{1}{z_{1k}(1-z_{1k})} \Kt_{01}^{\prime \, 2} + i \delta} \\
        \overset{\Kt_{20}^2 \text{ large}}{\approx}&
        \int_0^1 \dd{z_{0k}'} \dd{z_{1k}'} \delta(1-z_{0k}'-z_{1k}')\delta(z_{1k} - z_{1k}')  
        \Psi^{X(q) \to q(k_0') \bar q(k_1)}_\iin \\
        &\times g \mu^\varepsilon \frac{1}{1-z_{1k}} \varepsilon_{\sigma}^{*j} t^a_{\alpha_{0k} \alpha_{0k}'}
        \frac{\Kt_{20}^i}{\Kt_{20}^2} 
        \times S_0^{ij}(z_{0k}, z_{0k}').
    \end{split}
\end{equation}
The momenta in the intermediate state are denoted by primed variables $\kt_i', z_i'$.
Additionally, were we have denoted $\Kt_{01}^{\prime \, 2} = z_{1k}' \kt_{0}' -z_{0k}' \kt_1'$, and used the momentum conservation $z_{0k}' + z_{1k}' = z_{0k} + z_{1k} + z_{2k} = 1$ and $\kt_0'+\kt_1' = \kt_0 + \kt_1 + \kt_2$.
One thing to note is that this approximation also works in the limit $z_{2k} \to 0$, which means that this subtraction also gives the correct rapidity-divergent part that goes into the JIMWLK contribution.
We note that the large transverse momentum approximation $\Kt_{20}^2 \gg Q^2, \Kt_{01}^{\prime \, 2}$ is actually not needed at this point as what we want is the limit $\xt_{20} \to 0$. This approximation does not affect the end result but somewhat simplifies the derivation, by removing the dependence of the $\gamma^* \to q\bar{q}$ splitting momentum $\Kt_{01}^{\prime \, 2}$ from the energy denominator. This allows one to factorize the gluon emission part from the earlier $\gamma^* \to q\bar{q}$  splitting already before the Fourier transform to mixed space, while without this approximation the same factorization appears in mixed space as a rather nontransparent property of nested Fourier transforms~\cite{Beuf:2017bpd,Hanninen:2017ddy}.

Taking the Fourier transform of this, we get:
\begin{equation}
    \begin{split}
        & \! \! \! \! \! \! \! \! \! \! \! \! \! \! \! \! \! \! \! \! \! \!
        \widetilde \Psi^{X(q) \to q(\xt_0, z_{0k}) \bar q(\xt_1, z_{1k}) g(\xt_2, z_{2k})}_{\iin,\ref{diag:X_q_qg}} 
        =\int \frac{\dd[2-2\varepsilon]{\kt_0}\dd[2-2\varepsilon]{\kt_1}\dd[2-2\varepsilon]{\kt_2}}{(2\pi)^{3(2-2\varepsilon)}} e^{i \xt_0 \vdot \kt_0+i \xt_1 \vdot \kt_1+i \xt_2 \vdot \kt_2} \\
        &\times (2\pi)^{2-2\varepsilon} \delta^{(2-2\varepsilon)}(\qt - \kt_0-\kt_1-\kt_2)
        \Psi^{X(q) \to q(k_0) \bar q(k_1) g(k_2)}_{\iin,\ref{diag:X_q_qg}} \\
        \approx & 
        \int_0^1 \dd{z_{0k}'} \dd{z_{1k}'} \delta(1-z_{0k}'-z_{1k}')\delta(z_{1k} - z_{1k}')\\
        &\times \int \frac{\dd[2-2\varepsilon]{\kt_0'}\dd[2-2\varepsilon]{\kt_1}}{(2\pi)^{2(2-2\varepsilon)}} e^{i 
        \frac{z_{0k} \xt_0 + z_{2k} \xt_2}{z_{0k}+z_{2k}} \vdot \kt_0'+i \xt_1 \vdot \kt_1} 
        (2\pi)^{2-2\varepsilon} \delta^{(2-2\varepsilon)}(\qt - \kt_0'-\kt_1) \Psi^{X(q) \to q(k_0') \bar q(k_1)}_\iin \\
        &\times g \mu^\varepsilon \frac{1}{(1-z_{1k})^{3-2\varepsilon}} \varepsilon_{\sigma}^{*j} t^a_{\alpha_{0k} \alpha_{0k}'}
        \times S_0^{ij}(z_{0k}, z_{0k}')
        \times  \int \frac{\dd[2-2\varepsilon]{\Kt_{20}}}{(2\pi)^{2-2\varepsilon}} e^{i \frac{1}{z_{0k}+z_{2k}}\xt_{20} \vdot \Kt_{20}}  \frac{\Kt_{20}^i}{\Kt_{20}^2}\\
        \overset{\xt_{20}^2  \text{ small}}{\approx}& \int_0^1 \dd{z_{0k}'} \dd{z_{1k}'} \delta(1-z_{0k}'-z_{1k}')\delta(z_{1k} - z_{1k}')
        \times  \widetilde \Psi^{X(q) \to q(\xt_0,z_{0k}') \bar q(\xt_1, z_{1k})}_\iin \\
        &\times g \mu^\varepsilon \frac{1}{(1-z_{1k})^2} \varepsilon_{\sigma}^{*j} t^a_{\alpha_{0k} \alpha_{0k}'}
        \times S_0^{ij}(z_{0k}, z_{0k}')
        \times \frac{i}{2\pi} \frac{ \xt_{20}^i}{\xt_{20}^2} \Gamma(1-\varepsilon) (\pi \xt_{20}^2)^\varepsilon
    \end{split}
\end{equation}
The limit $ \xt_{20}^2 \to 0$ causes the coordinate combination $\frac{z_{0k} \xt_0 + z_{2k} \xt_2}{z_{0k}+z_{2k}}$ to reduce to just $\xt_0$, removing the effect of the gluon emission on the $X\to q\bar{q}$ splitting and leading to the factorization  of  the wave function $\widetilde{\Psi}^{X(q) \to q(\xt_0,z_{0k}') \bar q(\xt_1, z_{1k})}_\iin$ without gluon emission from the process. The limit of a high transverse momentum for the gluon means that the gluon in coordinate space is close to its emitter. In general in coordinate space a later gluon emission has a recoil effect on the earlier $X\to q\bar{q}$ splitting, but perhaps paradoxically this recoil effect vanishes in the limit of high transverse momentum or small plus-momentum fraction gluon emissions.
A similar expression can be derived for the antiquark emitting the gluon:
\begin{multline}
        \widetilde \Psi^{X(q) \to q(\xt_0, z_{0k}) \bar q(\xt_1, z_{1k}) g(\xt_2, z_{2k})}_{\iin,\ref{diag:X_qbar_qbarg}} \\
        \overset{\xt_{21}^2 \text{ small}}{\approx} 
        -\int_0^1 \dd{z_{0k}'} \dd{z_{1k}'} \delta(1-z_{0k}'-z_{1k}')\delta(z_{0k} - z_{0k}')
        \times  \widetilde \Psi^{X(q) \to q(\xt_0,z_{0k}') \bar q(\xt_1.z_{1k}')}_\iin
        \\
        \times g \mu^\varepsilon \frac{1}{(1-z_{0k})^2} \varepsilon_{\sigma}^{*j} t^a_{\alpha_{1k}' \alpha_{1k}}
        \times S_1^{ij}(z_{1k}', z_{1k})
        \times
        \frac{i}{2\pi} \frac{ \xt_{21}^i}{\xt_{21}^2} \Gamma(1-\varepsilon) (\pi \xt_{21}^2)^\varepsilon.
\end{multline}
The corresponding reduced wave functions are then
\begin{equation}
    \label{eq:wf_UV_quark}
    \begin{split}
        \widetilde \psi^{X \to q \bar q g}_{\iin,\ref{diag:X_q_qg},\text{UV}}(\xt_{ij},z_{ik})
        =& \int_0^1 \dd{z_{0k}'} \dd{z_{1k}'} \delta(1-z_{0k}'-z_{1k}')\delta(z_{1k} - z_{1k}')
        \times  \widetilde \psi^{X\to q \bar q}_\iin(\xt_{01},z_{ik}')\\
        &\times g \mu^\varepsilon \frac{1}{(1-z_{1k})^2} \varepsilon_{\sigma}^{*j} 
        \times S_0^{ij}(z_{0k}, z_{0k}')
        \times
        \frac{1}{2\pi} \frac{i \xt_{20}^i}{\xt_{20}^2} \Gamma(1-\varepsilon) (\pi \xt_{20}^2)^\varepsilon
    \end{split}
\end{equation}
and
\begin{equation}
\label{eq:wf_UV_antiquark}
    \begin{split}
        \widetilde \psi^{X \to q \bar q g}_{\iin,\ref{diag:X_qbar_qbarg},\text{UV}}(\xt_{ij},z_{ik})
        =&   -\int_0^1 \dd{z_{0k}'} \dd{z_{1k}'} \delta(1-z_{0k}'-z_{1k}')\delta(z_{0k} - z_{0k}')
        \times  \widetilde \psi^{X\to q \bar q}_\iin(\xt_{01},z_{ik}')\\
        &\times g \mu^\varepsilon \frac{1}{(1-z_{0k})^2} \varepsilon_{\sigma}^{*j} 
        \times S_1^{ij}(z_{1k}', z_{1k})
        \times
        \frac{1}{2\pi} \frac{i \xt_{21}^i}{\xt_{21}^2} \Gamma(1-\varepsilon) (\pi \xt_{21}^2)^\varepsilon.
    \end{split}
\end{equation}
These are the equations that will be used when subtracting the UV-divergent part in Sec.~\ref{sec:details}.
These wave functions also have the correct behavior in the limit $z_2 \to 0$ which means that when subtracting the UV divergence we also subtract the rapidity divergence.
This is useful as then we do not need to do any additional subtractions to get the JIMWLK evolution for the Wilson lines.
We also note that this subtraction procedure is independent of the wave function $\psi^{X\to q \bar q}$ which makes canceling the divergences more straightforward and independent of the photon polarization.

\section{Deriving Eq.~\eqref{eq:simplifying_identity}}
\label{sec:divergence}

We wish the extract the divergent part of the integral
\begin{equation}
\label{eq:Jab}
\mathcal{J}(a,b)=
    \int_0^1 \frac{\dd{t}}{t} J_0 \left( \sqrt{a^2 + \frac{1}{t}b^2} \right) 
    = \int_{b^2}^\infty \frac{\dd{u}}{u} J_0 \qty( \sqrt{a^2 + u} )
\end{equation}
in the limit $b^2 \to 0$.
Subtracting and adding the term 
\begin{equation}
\mathcal{J}_0(a,b)= \int_{0}^{b^2} \frac{\dd{u}}{u} \qty[J_0 \qty( \sqrt{a^2 + u}) - J_0(a) ]
\end{equation}
we can write Eq.~\eqref{eq:Jab} as
\begin{equation}
\label{eq:Jab2}
\mathcal{J}(a,b)=
     \int_{0}^\infty \frac{\dd{u}}{u} \qty[ J_0 \qty( \sqrt{a^2 + u} ) - \theta \qty( b^2 -u ) J_0(a)]  - \mathcal{J}_0(a,b).
\end{equation}
Note that $\mathcal{J}_0(a,b)\propto b^2$ in the limit $b^2 \to 0$ and the divergence is contained in the first term of Eq.~\eqref{eq:Jab2}.
This can be extracted by adding a regulator $\eta > 0$, as then the different parts of the integrals can be evaluated as
\begin{equation}
\begin{split}
   \int_{0}^\infty \frac{\dd{u}}{u^{1-\eta}}  J_0 \qty( \sqrt{a^2 + u} ) 
   =&a^{2\eta} \int_{1}^\infty  \dd{v} (v-1)^{\eta-1}  J_0 \qty( a\sqrt{v} ) 
   = \Gamma(\eta) (2a)^\eta J_{-\eta}(a) \\
   =& \qty[ \frac{1}{\eta} - \gamma_E + \log(2a) ] J_0(a) - \frac{\pi}{2} Y_0(a) + \mathcal{O}(\eta)
\end{split}
\end{equation}
and
\begin{equation}
   \int_{0}^\infty \frac{\dd{u}}{u^{1-\eta}}   \theta \qty( b^2 -u ) %
   = \frac{1}{\eta} + \log(b^2) +  \mathcal{O}(\eta).
\end{equation}
Inserting these two relations into  Eq.~\eqref{eq:Jab2}  yields
\begin{equation}
\label{eq:Jab3}
\mathcal{J}(a,b)=
     -\log(\frac{b^2 e^{\gamma_E}}{2a})  J_0(a) - \frac{\pi}{2} Y_0(a)
     - \mathcal{J}_0(a,b).
\end{equation}
Finally, we note that
\begin{equation}
\begin{split}
    \mathcal{J}_0(a,b)=& \int_{0}^{1} \frac{\dd{t}}{t} \qty[J_0 \qty( \sqrt{a^2 + t b^2}) - J_0(a) ]
    =
    \sum_{k=1}^\infty \sum_{l=0}^\infty \frac{(-1)^{k+l}}{4^{k+l} k \Gamma(k+1)\Gamma(l+1)\Gamma(k+l+1)}
    a^{2l} b^{2k}\\
    =&
    -2\int_0^1 \frac{\dd{t}}{t} \left[ 1 - J_0(tb) \right] J_0\left(a \sqrt{1-t^2}\right) 
\end{split}
\end{equation}
which can be proven by expanding the integrands as power series in terms of $a$ and $b$.
This gives us Eq.~\eqref{eq:simplifying_identity}.

\bibliographystyle{JHEP-2modlong.bst}
\bibliography{refs}

\end{document}

%% file: defs.tex
\newcommand{\Pt}{{\mathbf{P}}}

\newcommand{\rt}{{\mathbf{r}}}
\newcommand{\xt}{{\mathbf{x}}}

\newcommand{\Kt}{{\mathbf{K}}}
\newcommand{\bt}{{\mathbf{b}}}

\newcommand{\pt}{{\mathbf{p}}}
\newcommand{\qt}{{\mathbf{q}}}
\newcommand{\kt}{{\mathbf{k}}}

\newcommand{\ptt}{p_T} % scalar
\newcommand{\ktt}{k_T} % scalar
\newcommand{\Deltat}{{\boldsymbol{\Delta}}}

\newcommand{\ov}{\overline}
\newcommand{\azero}{ \hyperref[fig:A0]{\text{A0}} }
\newcommand{\aone}{  \hyperref[fig:A1]{\text{A1}} }
\newcommand{\atwo}{  \hyperref[fig:A2]{\text{A2}} }
\newcommand{\bzero}{ \hyperref[fig:B0]{\text{B0}} }
\newcommand{\bone}{  \hyperref[fig:B1]{\text{B1}} }
\newcommand{\btwo}{  \hyperref[fig:B2]{\text{B2}} }
\newcommand{\czero}{ \hyperref[fig:C0]{\text{C0}} }

\newcommand{\ctwo}{  \hyperref[fig:C2]{\text{C2}} }

\newcommand{\done}{  \hyperref[fig:D1]{\text{D1}} }

\newcommand{\eone}{  \hyperref[fig:E1]{\text{E1}} }

\newcommand{\adiag}{ {\hyperref[fig:A1]{\text{A}}} }
\newcommand{\bdiag}{ {\hyperref[fig:B1]{\text{B}}} }
\newcommand{\cdiag}{ {\hyperref[fig:C0]{\text{C}}} }

\newcommand{\dipole}{ \hat{S}^{(2)} }
\newcommand{\tripole}{ \hat{S}^{(3)} }

\newcommand{\octet}{ \text{c.o.} }

\newcommand{\Upsilonbb}{ \Upsilon^{\abs{(b)}^2}_\text{reg.} }
\newcommand{\Upsiloncc}{ \Upsilon^{\abs{(c)}^2}_\text{reg.} }
\newcommand{\Upsilonbc}{ \Upsilon^{(b)\times (c)}_\text{interf.} }
\newcommand{\Upsilond}{ \Upsilon^{(d)}_\text{inst.} }
\newcommand{\Upsilone}{ \Upsilon^{(e)}_\text{inst.} }

\newcommand{\Upsilonuone}{ \Upsilon^\text{U1}}
\newcommand{\Upsilonutwo}{ \Upsilon^\text{U2} }
\newcommand{\Upsilonuthree}{\Upsilon^\text{U3} }
\newcommand{\Upsilonufour}{ \Upsilon^\text{U4} }

\newcommand{\Upsilonvone}{ \Upsilon^\text{V1}}
\newcommand{\Upsilonvtwo}{ \Upsilon^\text{V2} }
\newcommand{\Upsilonvthree}{\Upsilon^\text{V3} }
\newcommand{\Upsilonvfour}{ \Upsilon^\text{V4} }

\newcommand{\kvec}{{\vec{k}}}
\newcommand{\pvec}{{\vec{p}}}

\newcommand{\kpvec}{{{\vec{k}}{\, '}}}

\newcommand{\qvec}{{\vec{q}}}
\newcommand{\qpvec}{{{\vec{q}}{\, '}}}

\newcommand{\lo}{{\textnormal{LO}}}
\newcommand{\nlo}{{\textnormal{NLO}}}

\newcommand{\ud}{\, \mathrm{d}}

\newcommand{\nc}{N_\mathrm{c}}

\newcommand{\half}{\frac{1}{2}}

\newcommand{\cf}{C_\mathrm{F}}

\newcommand{\qs}{Q_\mathrm{s}}

\newcommand{\as}{\alpha_{\text{s}}}
\newcommand{\aem}{\alpha_{\text{em}}}

\newcommand{\xpom}{{x_\mathbb{P}}}

\newcommand{\iin}{{\text{in}}}
\newcommand{\oout}{{\text{out}}}

\newcommand{\UV}{{\text{UV}}}

\newcommand{\hjimwlk}{\mathcal{H}_\text{JIMWLK}}

\newcommand{\ft}{f_\text{non-soft}}
\newcommand{\Ft}{F_\text{non-soft}}
\newcommand{\Fsoft}{F_\text{soft}}
\newcommand{\Fsub}{F_\text{sub}}
\newcommand{\fsub}{f_\text{sub}}
\newcommand{\mxbar}[1][ ]{\ov M_{\! X #1}}
\newcommand{\Qbar}[1][ ]{\ov Q_{ #1}}

\newcommand{\deltas}{\delta_{(s)}}
\newcommand{\Ds}{D_{(s)}}